\documentclass[11pt,a4paper]{article}
\pdfoutput=1

\usepackage{newunicodechar} % Input encoding
\usepackage[table,xcdraw]{xcolor} % Color support, including table colors
\usepackage[english]{babel} % Language support
\usepackage[a4paper,left=2cm,top=2.25cm,right=2cm,bottom=2.25cm]{geometry} % Page layout
\usepackage[hidelinks]{hyperref} % Hyperlink support for PDF + footnotes

\definecolor{lightturquoise}{rgb}{0.5, 1.0, 1.0} % Light turquoise shade

\hypersetup{
    colorlinks=false,                % Keep link text black
    pdfborder={1 1 1},               % Show border: horz, vert, width
    linkbordercolor=lightturquoise, % TOC, figure/table refs
    citebordercolor=lightturquoise, % Citations
    urlbordercolor=lightturquoise   % External links
}

\usepackage[ddmmyyyy]{datetime} % automatically update date
\usepackage{enumitem}
\usepackage{graphics} % Basic graphics support
\usepackage{graphicx} % Enhanced graphics support
\addto\extrasenglish{

    \setcounter{secnumdepth}{4}
}
\usepackage{subcaption} % Subfigures and subcaptions
\usepackage{parskip}
\usepackage{comment}

\usepackage[none]{hyphenat}
\usepackage{amsmath}  % For aligning equations
\usepackage{amssymb}  % For mathematical symbols
\usepackage{mathtools} % Enhanced math tools
\usepackage{gensymb} % Generic symbols (e.g., degree symbol)
\usepackage{multicol} % Multiple columns
\usepackage{wrapfig} % Wrap text around figures

\usepackage{booktabs} % Enhanced table formatting
\usepackage{multirow} % Multirow cells in tables
\usepackage{float} % Improved float placement
\usepackage{numprint} % decimal points in table

\usepackage[format=plain, labelfont=bf, labelsep=period]{caption} % Caption customization
\usepackage{pdfpages}

\usepackage[numbers]{natbib} % Enhanced bibliography support
\usepackage{notoccite} % Not misnumber reference in captions
\usepackage{appendix}
\usepackage{tikz} % To make simple vector graphic drawings
\tikzset{
    every picture/.append style={
        line cap=round,
    }
}

\usepackage{listings} % Add code
\usepackage{makecell} % add this in the preamble

\usepackage{nomencl}
\nomenclature{symbol}{description}
\makenomenclature
\usepackage{placeins}

\usepackage{fancyhdr}
\usetikzlibrary{shapes.geometric, arrows.meta, positioning}

\tikzstyle{process} = [rectangle, minimum width=3.8cm, minimum height=1cm, text centered, draw=black, font=\scriptsize, fill=blue!10]
\tikzstyle{arrow} = [thick, ->, >=stealth]

\begin{document}
\begin{titlepage}
\begin{figure}[h] 
\includegraphics[scale=.3]{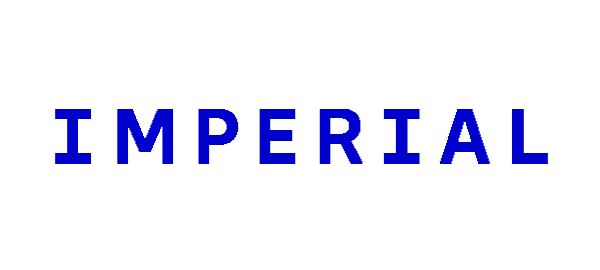}
\end{figure}

\vspace{1cm}

\begin{center}
    \hrule \vspace{1cm} %get a line
    \huge{\bf Set-up and Characterisation of Atmospheric Boundary Layers in the 10'$\times$5' Wind Tunnel} \\
    \vspace{0.5cm}
    \Large{Final Year Project 2024-25} \\
    \Large{Department of Aeronautics} \\
    \vspace{1cm} \hrule  %get a line
\end{center}

\vspace*{1.5cm}

\centering
    \begin{tabular}{rl}
    {\bf Author:} & {Sita Mandakini Nair}
    \\
    {\bf CID:} & {02014568}
    \\
    {\bf Supervisor:} & {Dr. Kevin Gouder}
    \\
    {\bf Second Marker:} & {Prof. Oliver Buxton} 
    \\

    % {\bf Group members:} & {} \\
    % {}                   & {} \\
    % {}                   & {} \\
    % {}                   & {} \\
    % {}                   & {} \\
    % {}                   & {} \\
    {\bf Date:} & 05/06/2025

    \end{tabular}

\vspace{2cm}

\begin{center}
    \includegraphics[scale=0.32]{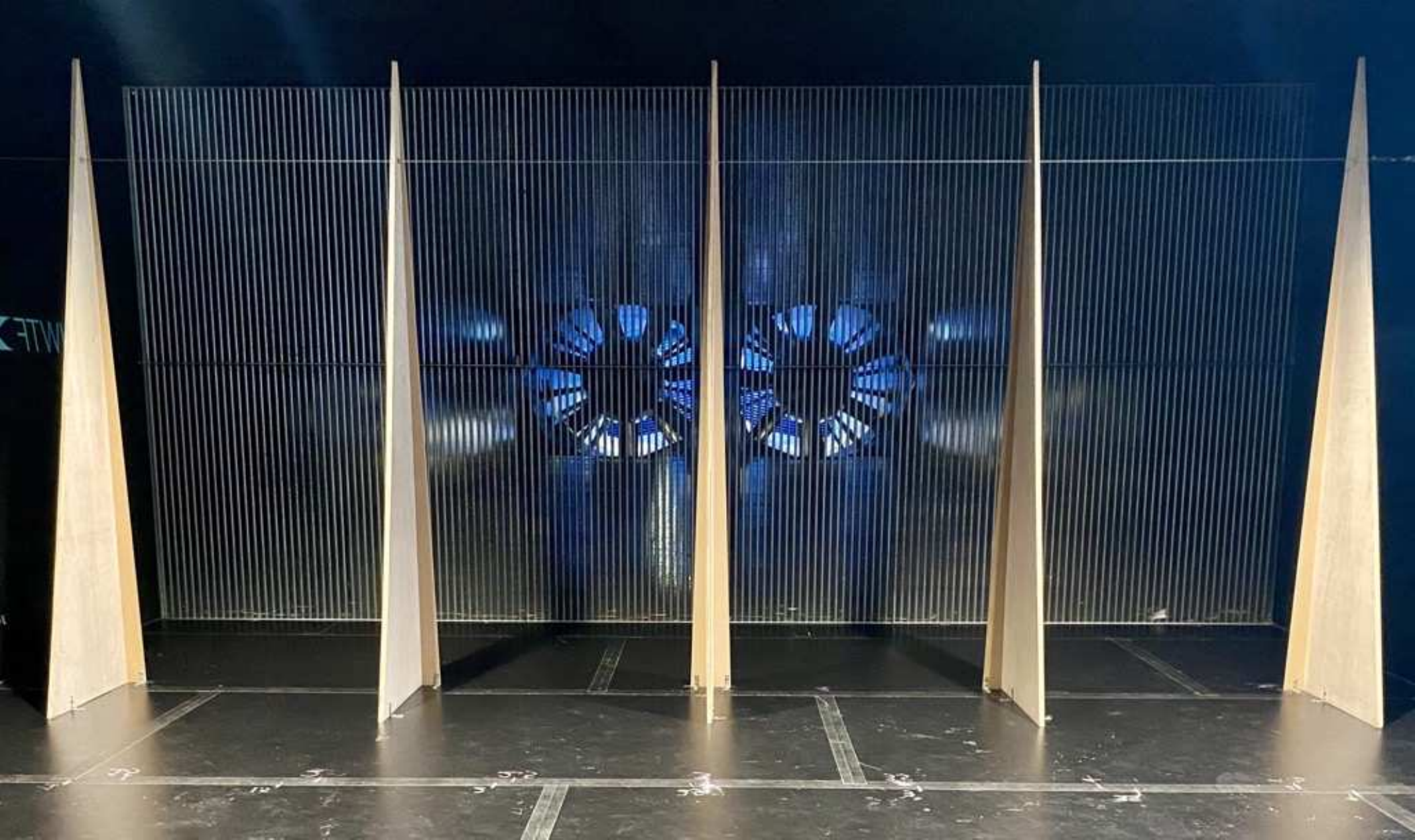} % change to your desired image
\end{center}

% REMEMBER : abstract should summarise report and present key results, INCLUDE NUMBERS
\newpage
\thispagestyle{empty} % Remove page number from the acknowledgment page
\noindent
\vspace*{6cm}
\begin{center}
    \textbf{\large Abstract}
\end{center}
\vspace{0.5cm} % Space between title and text
\noindent The Atmospheric Boundary Layer (ABL) plays a critical role in influencing objects exposed to atmospheric conditions, making its study crucial. Due to the high cost of real-world testing, this thesis focuses on replicating marine ABLs in a wind tunnel environment. Two profiles were developed: one that served as a framework establishing commonality between the various international wind engineering standards (`Profile 1'), and a second profile, which is more suitable for modelling the inflow of wind farms in the English Channel and the North Sea (`Profile 2').

The ABLs were generated using Irwin spires, and no addition of floor roughness elements, with the flow characteristics measured via Laser Doppler Anemometry (LDA), and a multi-hole probe (MHP). MHP showed a reasonably high accuracy when compared to LDA, with less than 1\% deviation in the streamwise velocity, and under 5\% standard deviation in the streamwise, spanwise, and wall-normal velocity components. The profiles achieved a good agreement with target metrics such as normalised velocity and turbulence intensity. Spanwise uniformity and spectral analysis confirmed the robustness of the simulation across varying inflow velocities.

Overall, Irwin spires proved to be a cost-effective and reliable method for simulating marine ABLs, offering valuable insights for optimising offshore wind energy systems. 

\end{titlepage}
\pagenumbering{roman}
% Acknowledgment page, without section numbering or page numbers
\newpage
\thispagestyle{empty} % Remove page number from the acknowledgment page
\begin{center}
    \textbf{\large Acknowledgments}
\end{center}
\vspace{1cm} % Space between title and text
\noindent First and foremost, I would like to express my heartfelt gratitude to everyone who supported me throughout the course of my Final Year Project. 
I am especially grateful to my supervisor, Dr. Kevin Gouder, for his unwavering guidance, encouragement, and invaluable pieces of advice. His dedication and belief in me, particularly during challenging times, kept me motivated and focused; I truly could not have asked for a better mentor. Many thanks to Prof. Oliver Buxton for his constructive feedback during the interim presentation, which helped me refine and develop my work further. 

A sincere thank you goes to Postdoctoral Researcher Dr. Craig Thompson for his continuous support and expertise. His assistance with the multi-hole probe setup, calibration, and data analysis methods, along with his general advice throughout the project, was instrumental to my progress. 
I would also like to wholeheartedly acknowledge the Research Officers at the 10'$\times$5', William McArdle and Ricardo Huerta Cruz, and most importantly Paul Howard, the technician. From day one, along with my supervisor, they have provided tireless support, offering practical insights, verifying experimental approaches, and assisting with all aspects of the experimental setup, including but not limited to LDA calibration and acquisition, quick-turnaround manufacturing, and hands-on problem-solving. 

A big thank you to Alan Smith, for manufacturing all the spires exactly to specification, and to Roland Hutchins, Jordan Farrar, and Mark Thornton for their help with 3D printing. 

Special thanks to Karthik Krishnan for his help with CAD, and for being a constant source of support throughout this project. I am also incredibly grateful to all my dear friends, especially my flatmate Anoushka Khot, for their encouragement and companionship, including the countless late nights spent working together in the library and at my accommodation. 

Finally, I am deeply thankful to my parents and extended family for their unwavering support, love, and belief in me throughout this journey. Their presence, dedication, and constant encouragement gave me the strength to persevere during my time at Imperial, and I truly could not have accomplished this milestone without them. 
\vspace{8.5cm} % Space between title and text

\noindent \textbf{Disclaimer:} I hereby declare that this thesis is the result of my own work. All research, analysis, and writing are original, unless otherwise referenced. I acknowledge the use of large language models (LLMs) for technical clarifications, code debugging, and improving language clarity, in accordance with Imperial College London's guidelines.

\newpage
\thispagestyle{empty} % removes page number

\vspace*{\fill}
\begin{center}
    ``If I have seen further, it is by standing on the shoulders of giants,''\\[1em]
    \textit{Sir Isaac Newton}
\end{center}
\vspace*{\fill}
\newpage

\newpage % After the acknowledgment, normal page numbering starts
\addcontentsline{toc}{section}{Table of Contents}
\tableofcontents
\newpage
\addcontentsline{toc}{section}{List of Figures}
\listoffigures
\newpage
\addcontentsline{toc}{section}{List of Tables}
\listoftables
% \newpage
% \addcontentsline{toc}{section}{Nomenclature}
% \input{Sections/nomenclature}
% \printnomenclature[2cm]
\newpage
% Symbols
\nomenclature[A]{$z_{ref}$}{{Reference height}}
\nomenclature[A]{$U_{ref}$}{{Reference wind speed at $z_{ref}$}}
\nomenclature[A]{$z_0$}{{Terrain roughness parameter}
\nomunit{\SI{}{}}}
\nomenclature[A]{$\delta$}{{Boundary layer height}}
\nomenclature[A]{$U_{\delta}$}{{$U$ velocity at $\delta$}}
\nomenclature[A]{$I_u(z)$}{{Turbulence intensity due to $u$ fluctuations}}
\nomenclature[A]{$L_{u,x}$}{{Turbulent length scale}}
\nomenclature[A]{$\alpha$}{{Power law exponent}}
\nomenclature[A]{$h$}{{Spire height}}
\nomenclature[A]{$b$}{{Spire base (triangular Irwin spire)}}
\nomenclature[A]{$b_1$}{{Spire bottom width (truncated Irwin spire)}}
\nomenclature[A]{$b_2$}{{Spire top width (truncated Irwin spire)}}
\nomenclature[A]{$b_s$}{{Splitter plate base}}

% Abbreviations
\nomenclature[D]{WMO}{World Meteorological Organisation}
\nomenclature[D]{AS/NZS}{Australian/ New Zealand Standards}
\nomenclature[D]{ESDU}{Engineering Sciences Data Unit}
\nomenclature[D]{ASCE}{American Society of Civil Engineers}
\nomenclature[D]{ISO}{International Standard for Standardisation}
\nomenclature[D]{ABL}{Atmospheric Boundary Layer}
\nomenclature[D]{DNH}{Deaves and Harris}
\nomenclature[D]{BL}{Boundary Layer}
\nomenclature[D]{VG}{Vortex Generator}
\nomenclature[D]{TS}{Test Section}
\nomenclature[D]{`Profile 1'}{Framework marine profile, establishing the commonality between various standards}
\nomenclature[D]{`Set 1'}{Spires designed for `Profile 1'}
\nomenclature[D]{`Profile WF'}{Profile more suitable for wind farm (WF) inflow in the English Channel and North Sea}
\nomenclature[D]{`Set 2'}{Spires designed for `Profile WF'}

\begin{thenomenclature} 
\nomgroup{A}
  \item [{$\alpha$}]\begingroup {Power law exponent}\nomeqref {0}\nompageref{1}
  \item [{$\delta$}]\begingroup {Boundary layer height}\nomeqref {0}\nompageref{1}
  \item [{$b$}]\begingroup {Spire base (triangular Irwin spire)}\nomeqref {0}\nompageref{1}
  \item [{$b_1$}]\begingroup {Spire bottom width (truncated Irwin spire)}\nomeqref {0}\nompageref{1}
  \item [{$b_2$}]\begingroup {Spire top width (truncated Irwin spire)}\nomeqref {0}\nompageref{1}
  \item [{$b_s$}]\begingroup {Splitter plate base}\nomeqref {0}\nompageref{1}
  \item [{$h$}]\begingroup {Spire height}\nomeqref {0}\nompageref{1}
  \item [{$I_u(z)$}]\begingroup {Turbulence intensity due to $u$ fluctuations}\nomeqref {0}\nompageref{1}
  \item [{$L_{u,x}$}]\begingroup {Turbulent length scale}\nomeqref {0}\nompageref{1}
  \item [{$U_{\delta}$}]\begingroup {$U$ velocity at $\delta$}\nomeqref {0}\nompageref{1}
  \item [{$U_{ref}$}]\begingroup {Reference wind speed at $z_{ref}$}\nomeqref {0}\nompageref{1}
  \item [{$z_0$}]\begingroup {Terrain roughness parameter}\nomeqref {0}\nompageref{1}
  \item [{$z_{ref}$}]\begingroup {Reference height}\nomeqref {0}\nompageref{1}
\nomgroup{D}
  \item [{`Profile 1'}]\begingroup Framework marine profile, establishing the commonality between various standards\nomeqref {0}\nompageref{1}
  \item [{`Profile WF'}]\begingroup Profile more suitable for wind farm (WF) inflow in the English Channel and North Sea\nomeqref {0}\nompageref{1}
  \item [{`Set 1'}]\begingroup Spires designed for `Profile 1'\nomeqref {0}\nompageref{1}
  \item [{`Set 2'}]\begingroup Spires designed for `Profile WF'\nomeqref {0}\nompageref{1}
  \item [{ABL}]\begingroup Atmospheric Boundary Layer\nomeqref {0}\nompageref{1}
  \item [{AS/NZS}]\begingroup Australian/ New Zealand Standards\nomeqref {0}\nompageref{1}
  \item [{ASCE}]\begingroup American Society of Civil Engineers\nomeqref {0}\nompageref{1}
  \item [{BL}]\begingroup Boundary Layer\nomeqref {0}\nompageref{1}
  \item [{DNH}]\begingroup Deaves and Harris\nomeqref {0}\nompageref{1}
  \item [{ESDU}]\begingroup Engineering Sciences Data Unit\nomeqref {0}\nompageref{1}
  \item [{ISO}]\begingroup International Standard for Standardisation\nomeqref {0}\nompageref{1}
  \item [{TS}]\begingroup Test Section\nomeqref {0}\nompageref{1}
  \item [{VG}]\begingroup Vortex Generator\nomeqref {0}\nompageref{1}
  \item [{WMO}]\begingroup World Meteorological Organisation\nomeqref {0}\nompageref{1}

\end{thenomenclature}

\addcontentsline{toc}{section}{List of Symbols}
\newpage

\pagenumbering{arabic}

\section{Introduction}
% INTRODUCTION

The Atmospheric Boundary Layer (ABL) is the lowest part of the Earth's atmosphere, characterised by high turbulence stemming from its interaction with the Earth's surface, which facilitates the mixing and redistribution of heat, moisture, pollutants, and more. Its depth typically ranges from a few metres to several kilometres, depending on the meteorological conditions \cite{abl_depth}. The dynamics of the atmospheric boundary layer influence various engineering and environmental applications. These include, but are not limited to, weather forecasting, pollutant dispersion analysis, structural wind loading assessments, and wind energy harvesting. 

Wind tunnel simulation and testing of ABLs are indispensable for replicating realistic flow characteristics, since the costs and complexity involved in full-scale testing are immense. The main goal behind these simulations is to capture key flow features such as the vertical variation of mean velocity, turbulence intensity, and integral length scales, assuming neutral atmospheric stability conditions. To complement physical experiments, advanced computational methods, such as Reynolds-Averaged Navier-Stokes (RANS), and Large Eddy Simulations (LES), are utilised to enable detailed analyses of complex ABL flows over varied terrains.

\subsection{Background and Motivation}

Atmospheric boundary layer simulation in the wind tunnel has been highly instrumental in advancing experimental aerodynamics, enabling researchers to replicate and study atmospheric flow phenomena under controlled laboratory conditions. This approach supports investigations of wind-induced effects on structures, vehicles, and numerous other objects exposed to the natural atmosphere. For such experiments to yield reliable and applicable results, the simulated flow must reflect the key characteristics of the real ABL, ensuring that conclusions drawn can be reliably applied to real-world applications. 

Since the 1960s, several methods have been established to simulate the ABL, with the goal of reproducing key flow features. These include near-wall effects, which are characterised by the reduction in flow velocity aligned with an increase in the turbulence intensity closer to the ground. Therefore, to mimic this, a rougher floor is usually installed in laboratory settings in order to obtain the desired flow characteristics. Design of the wind tunnel setup is commonly dictated by the choice of terrain being simulated, which can be broadly classified into marine, rural, suburban, or urban \cite{EN1991}. This involves the deployment of various passive devices such as spires, roughness elements, and mixing devices, specifically formulated to reproduce particular flow attributes. Such approaches typically yield a good representation of mean flow characteristics and turbulence properties essential for engineering analyses. 

These generation techniques are of special interest in the context of wind energy research, since they call for meticulous modelling of the atmospheric wind flow. Wind farms, which are integral to renewable energy strategies, operate within the ABL. Therefore, requiring an accurate simulation of the atmospheric flow is essential for optimising turbine and farm performance, as well as structural safety. This thesis explores methods for accurately simulating the ABL in wind tunnels. Special emphasis is attributed to reproducing realistic flow features, establishing commonality between the various international wind engineering standards, and in addition generating a profile which is more suitable for wind farm inflow conditions in the English Channel, such as in the ICONIC wind farm optimisation project. This project at Imperial seeks to implement coordinated turbine control, optimise wind farm performance by amplifying the energy output, and reduce loads acting on the turbine, in place of conventional greedy control strategies. A critical review of the existing methodologies and experimental setups highlights the optimal approaches and challenges involved in modelling ABL(s) under varied terrain conditions. Furthermore, the research undertaken alongside the techniques developed throughout the course of this project will be instrumental in generating custom wind profiles more efficiently in future experimental campaigns at Imperial College London, especially in projects focused on environmental flow studies, energy, and structural testing. 

\subsection{Objectives}

\begin{itemize}
    \item Review various international wind engineering standards to develop a framework profile that aligns with the common elements shared across all standards;
    \item Generate the marine atmospheric boundary layers in the 10'$\times$5' wind tunnel based on the standard profiles, and design the hardware for the simulation of urban boundary layers; 
    \item Investigate the performance of multi-hole probe (MHP) against Laser Doppler Anemometry (LDA) in the lower test section, to validate MHP's independent use in the upper test section; 
    \item Produce a target wind profile in the upper test section, more suitable for modelling wind farms situated off the coast of English Channel and North Sea;
    \item Compare experimental data to target profiles using various metrics such as mean velocity, turbulence intensity, etc.;
    \item Evaluate the performance and the re-usability of the designed spires and roughness elements for future experiments;
    \item Generate power spectra from the data acquired, and review the various methods to calculate length scales. 
\end{itemize}

\newpage
\section{Literature Review}
Exploring existing literature laid a foundation aiding in the design, development, and validation of the wind tunnel experiments performed in this project, to simulate atmospheric boundary layers. Firstly, various international wind standards were examined, providing a basis for benchmarking experimental results. Thereafter, various techniques for producing the desired ABL(s) in the wind tunnel were evaluated to identify viable approaches and limitations relevant to achieving the objectives of this project. Finally, the importance of estimating turbulent length scales was considered, as well as the different approaches for calculating them from the experimental data.

\subsection{Target Wind Profiles}

It is essential to explore and compare various wind standards that define the variation of wind velocity with height, the effect of terrain roughness, and other environmental factors. This in turn provided a framework to develop the required boundary layer simulation in the wind tunnel, by analysing different international wind standards. These include the Eurocode \cite{EN1991}, British National Annex version of the Eurocode (BS EN) \cite{BSEN}, International Organisation for Standardisation (ISO) \cite{ISO}, American Society of Civil Engineers (ASCE) \cite{asce}, The Engineering Sciences Data Unit (ESDU) \cite{esdu85020}, and the Australian/ New Zealand Standards (AS/NZS) \cite{asnzs}. 

Davenport \cite{davenport} popularised the conventional use of hourly mean wind velocities, due to the availability of such data from meteorological stations and the assumption of statical stationarity over this period. As such, there were employed in harmonising the various standards mentioned below. However, not all wind standards adopt this convention. For example, Eurocode uses a 10-minute mean, which is also recommended by the World Meteorological Organisation (WMO) for marine and offshore conditions, especially during tropical or subtropical cyclones \cite{wmo}. The 10-minute average strikes a good balance between sensitivity to short-lived gusts and smoothing out from longer averages. Empirically, it has been determined that the 10-minute mean velocities are appromximately 5\% higher than the hourly means \cite{ENdocguide}. 

\textbf{Eurocode}

The Eurocode \cite{EN1991} is an extensive European standard that provides detailed terrain categorisation with National Annexes for multiple countries. It also addresses the $z_{min}$ value for all the terrain classifications, which marks the realistic height off the ground from where wind pressure starts varying. It defines terrain categories ranging from Category 0 (open sea) to Category IV (dense urban areas). In Eurocode, the basic wind velocity $v_b$ is defined as the 10-minute mean wind velocity at a height of 10 m given by $v_b = c_{dir} \cdot c_{season} \cdot v_{b,0}$. This accounts for wind direction, and seasons ($c_{dir}=1$, $c_{season}=1$ being the recommended values respectively). Here, $v_{b,0}$ is the characteristic mean velocity at 10 m, independent of wind direction and the time of the year, measured over terrain Category II. 

The mean velocity, $v_m$, at an arbitrary height of $z$ off the ground is derived from Equation \ref{eq:vm}, where the roughness factor, $c_r$, is based on the logarithmic profile, given by $\bar u/U_\delta = ln(z_g/z_0)/ln(\delta/z_0)$ (where $U_\delta$ is the velocity at the edge of the boundary layer $\delta$, $z_0$ is the roughness parameter for the specific terrain, $\bar u$ is the mean velocity calculated at a height $z_g$ above the ground),  and is influenced by the terrain roughness $z_0$, and the terrain factor $k_r$. The orography factor $c_0$ accounted for topographic features such as hills, cliffs, etc. Unless specified, the default value of $c_0$ is typically taken as 1. To calculate $c_r$, Equation \ref{eq:cr_combined} can be followed, where $z_{0}$\textsubscript{,II}=0.05 m is the roughness parameter for terrain II. 

Additionally, Eurocode defines turbulence intensity $I_v(z)$ as in Equations \ref{eq:Iv_combined}, representing the ratio of standard deviation of turbulence to the mean velocity, where $\sigma_v = k_{r} \cdot v_{b} \cdot k_l$. Turbulence length scale $L(z)$ represents the average gust size \cite{EN1991}, given by Equation \ref{eq:Lz_combined}, for heights less than 200 m, where $z_t=$ 200 m is the reference height, $L_t=$ 300 m is the reference length scale, with $\alpha$ defined as $\alpha=0.67+0.05 ln(z_0)$. 
\vspace{-2cm}
{\small
\begin{center}
\begin{minipage}[t]{0.48\textwidth}
\begin{align}
v_m(z) &= c_{r}(z)\cdot c_{0}(z)\cdot v_{b} \label{eq:vm} \\
c_r(z) &= 
\begin{cases}
c_r(z_{\min}) &\text{, } z \leq z_{\min} \\
k_r \cdot \ln\left(\dfrac{z}{z_0}\right) & \text{, } z_{\min} < z \leq z_{\max}
\end{cases} \label{eq:cr_combined} \\
k_r &= 0.19 \cdot \left( \dfrac{z_0}{z_{0,\mathrm{II}}} \right)^{0.07} \label{eq:kr}
\end{align}
\end{minipage}
\hfill
\begin{minipage}[t]{0.48\textwidth}
\begin{align}
I_v(z) &= 
\begin{cases}
I_v(z_{\min}) & \text{, } z \leq z_{\min} \\
\dfrac{k_l}{c_0(z) \cdot \ln\left(\dfrac{z}{z_0}\right)} & \text{, } z_{\min} < z \leq z_{\max}
\end{cases}
\label{eq:Iv_combined}
\end{align}

\begin{equation}
L(z) =
\begin{cases}
L(z_{\min}) & \text{, } z < z_{\min} \\
L_t \cdot \left(\dfrac{z}{z_t}\right)^\alpha & \text{, } z \geq z_{\min}
\end{cases}
\label{eq:Lz_combined}
\end{equation}
\end{minipage}
\end{center}
}

The UK has nationally adopted the Eurocode, resulting in the British Standard BS EN 1991-1-4:2005 \cite{BSEN}, sharing the fundamental methods of modelling velocity using the logarithmic law, and atmospheric physics. However, it is tailored specifically for UK-specific wind conditions and complex terrain, accounting for its irregular coastline and varied topography. The BS EN also uses the 10-minute mean for wind velocities, similar to Eurocode. Despite the complexity involved in the estimations followed by the BS EN, it considers most of the region-specific conditions relevant to the UK, unlike Eurocode. 

\textbf{ASCE}

In contrast, the ASCE \cite{asce} which is the nationally adopted wind loading standard in the USA and around the globe, defines wind speed $U$, at a height $z$, using a power law profile, for $z$ < $z_{min}$, given in Equation \ref{eq:powerlaw}. Altough ASCE defines wind velocities as 3-second gusts at a 10 m height from the ground over Exposure C (which represents open terrain with scattered obstructions), it also acknowledges the use of hourly means for long-term load estimations, via different $\alpha$ values, given by Equation \ref{asce_vel} (0.78 and 1/8 are emperical constants for the marine terrain). 

\noindent
\begin{minipage}{0.48\textwidth}
\begin{equation}
\frac{U}{U_{ref}}=\left(\frac{z}{z_{ref}}\right)^\alpha
\label{eq:powerlaw}
\end{equation}
\end{minipage}
\hfill
\begin{minipage}{0.48\textwidth}
\begin{equation}
U_{hourly}=0.78 \times \left(\frac{z}{z_{ref}}\right)^{1/8} \times U_{ref}
\label{asce_vel}
\end{equation}
\end{minipage}

 The fixed $\alpha$ value also limits the applicability of this standard to terrains like marine or urban, since it does not take into account the change in sea states or urban transitions. Terrains are referred to as exposure Category D to B, where D refers to water surfaces that are flat and unobstructed, and B is representative of urban and suburban exposures. 

Furthermore, ASCE does not define fundamental terrain parameters like $z_0$ and $z_{min}$, unlike Eurocode. ASCE suggests $I_z=0.15(10/z)^{1/6}$ for turbulence intensities at different heights $z$, where $0.15$ is the constant for marine (category D). 

\textbf{ISO}

On the other hand, the ISO \cite{ISO} standard recommends a 1-hour averaging, taking the reference elevation as 10 m off the ground. This is high enough to minimise surface disturbances, and is commonly considered as the standard reference height. It addresses means ranging from 10 minutes to 3 hours, noting that shorter durations yield higher wind speeds and increased spatial variability. 

For marine profiles, ISO adopts the Fr{\o}ya logarithmic model, which provides a more precise method reflecting physical processes and surface characteristics of the open sea. This model forges from the Charnock relationship, linking sea roughness to wind speed and sea state, under the assumption of neutral atmospheric stability \cite{offshore}. Log law is generally better for marine profiles and offshore conditions as given by WMO \cite{marinepaper}, but not for urban/ land-based terrain profiles. For representing turbulence in the marine boundary layer, ISO also incorporates the Kaimal spectrum. The empirical relations stemming from ISO which aids in the calculation of mean velocities and turbulence intensities are given by Equations \ref{uw1h}, and \ref{iu}, with $C=0.0573 \cdot (1+0.15U_{w0})^{1/2}$. These are specifically fine-tuned for offshore conditions, where the log-law provides a more accurate representation of the velocity profile than the power law. Therefore, the target profiles for marine ABL(s), serving as a framework for this project, will be developed in alignment with the ISO standard. 

\noindent
\begin{minipage}[b]{0.48\textwidth}
\begin{equation}
U_{w,1h}(z) = U_{w0} \left(1 + C \ln\left(\dfrac{z}{z_r}\right)\right)
\label{uw1h}
\end{equation}
\end{minipage}
\hfill
\begin{minipage}[b]{0.48\textwidth}
\begin{equation}
I_u(z) = 0.06 \left[1 + 0.043U_{w0}\right] \left(\dfrac{z}{z_r}\right)^{-0.22}
\label{iu}
\end{equation}
\end{minipage}

Here, $U_{w,1h}(z)$ is the one-hour sustained wind speed at a height z, $U_{w0}$ is the same at 10 m in m/s, $z_r=$ 10 m is the reference height, and $I_u(z)$ is the turbulence intensity in $u$ at z. The power law exponents suggested are $\alpha=0.11$, and $\alpha=0.14$ for onshore, and offshore conditions respectively \cite{offshore}.

\textbf{ESDU}

ESDU provides a detailed tool for modelling wind over complex terrains, accounting for the combination of different roughness parameters, by calculating an effective $z_0$ value. ESDU 85020 \cite{esdu85020} supports both log and power law profiles, as well as anisotropic turbulence, which is particularly dominant near the ground where surface effects dominate, hence reflecting the variation in turbulent characteristics in the longitudinal, lateral, and vertical directions. It is based on the Deaves and Harris (DNH) set of measurements for the open fetches, such as the sea \cite{deavesharris}. The DNH model provides a family of wind velocity profiles for strong winds in the ABL that satisfy the boundary conditions at the top and bottom of the boundary layer (BL), offering an improved accuracy compared to the power law. The calculations for ESDU were based off on a pre-built spreadsheet, which provided various key parameters like wind velocity $V_z$, turbulence intensity $I_u$, and turbulence length scales $L_{u,x}$ (distance $x$ over which turbulent eddies of velocity scale $u$ are correlated), and similarly for the $v$, and $w$ components \cite{ESDU01008}. Even though it operates on hourly mean wind speeds with typical gust durations of 3 seconds by default, the final outputs of the mean velocity and turbulence are independent of this assumption, enhancing its flexibility for different applications. 

The model includes a zero-plane displacement height $d$ for urban profiles, indicating that the meaningful velocity data begins only at this height. This is typically set to 0 for marine profiles. The ESDU standard enhances terrain classification via roughness parameter $z_0$ values, with the reference site representing undisturbed terrains like the open sea, and the target site reflecting the terrain of interest.

\textbf{AS/NZS}

AS/NZS 1180.2 standard \cite{asnzs} uses 0.2-second gusts as the basis of wind load assessment, with terrain categories ranging from Category 1 (marine) to Category 4 (urban). The turbulence intensity values are tabulated in the standard, up to a height of $z=$ 200 m for various terrain types. The wind velocity, $V_{des\theta}=M_{z,cat}V_{ref}$, where $V_{ref}$ is the reference velocity at $z_{ref}=$ 10 m, and $M_{z,cat}$ is the terrain/ height multiplier based on the terrain category. Here, marine is referred to as Category 1, whilst large city centres are referred to as Category 4. The AS/NZS also follows a logarithmic mean wind profile based on DNH \cite{asnzsDoc}. 

\newpage

\textbf{`Profile 1'}

\begin{figure}[H]
    \centering  \includegraphics[width=1.04\linewidth, height=2.7cm]{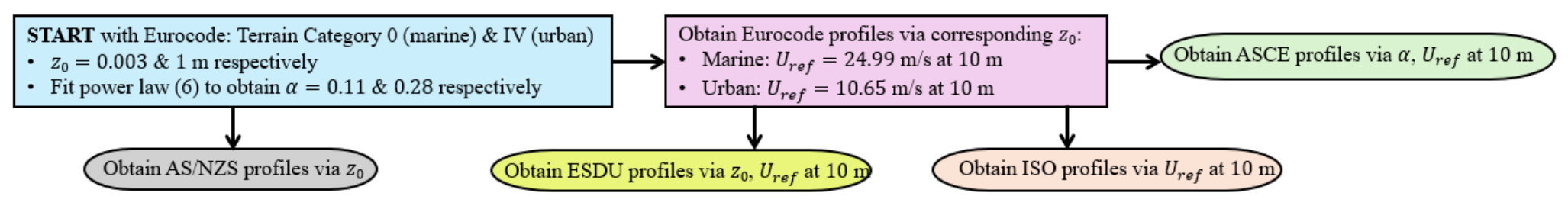}
    \caption{Flowchart to summarise the establishment of `Profile 1'}
    \vspace{-0.3cm}
    \label{fig:flowchart}
\end{figure}

\begin{figure}[H]
    \centering
    \begin{minipage}{.47\linewidth}
        \centering
        \includegraphics[width=1.08\linewidth]{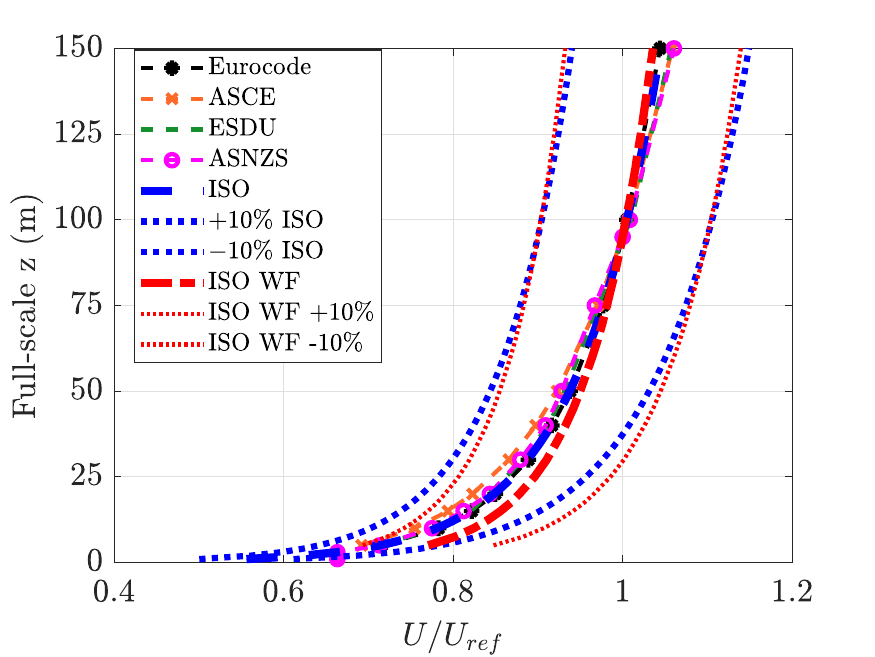}
        \caption{$U/U_{ref}$ from `Profile 1', `Profile WF'}
        \label{fig:standards_U}
    \end{minipage}\hfill
    \begin{minipage}{.47\linewidth}
        \centering
        \includegraphics[width=1.08\linewidth]{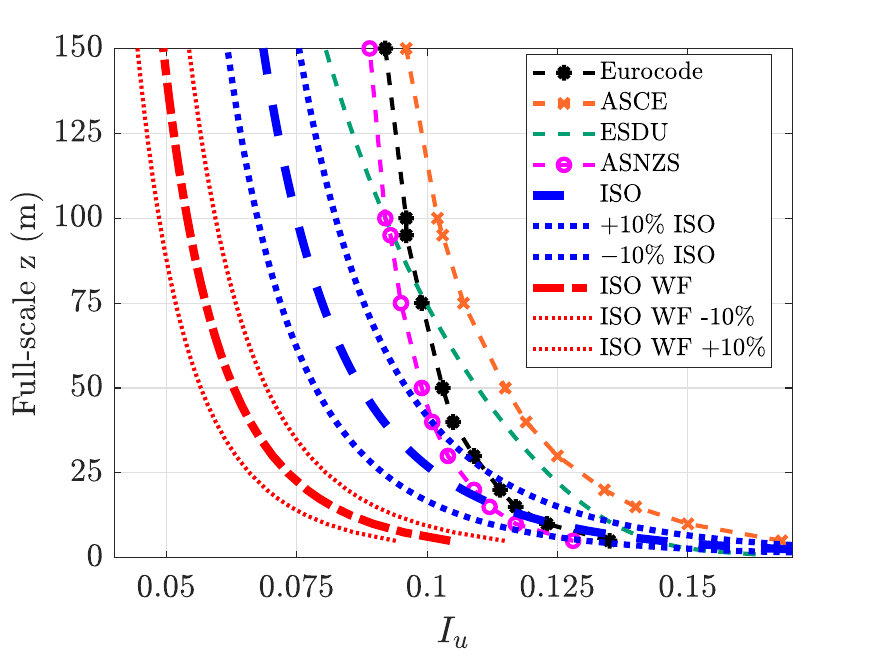}
        \caption{$I_u$ from `Profile 1', `Profile WF'}
        \label{fig:standards_iU}
    \end{minipage}
\end{figure}

\label{unify_standards}
A marine profile was developed to reflect the commonality among all the standards aforementioned, serving as a baseline for comparison. This was performed under a common condition, i.e., a wind speed of 25 m/s at a height of 10 m, which is near the upper end of expected wind speeds. Using this condition, respective profiles were generated from each standard, as outlined in Figure \ref{fig:flowchart}. The resulting normalised mean velocity and turbulence intensity profiles, are based on the reference velocity, $U_{ref}$ at a reference height of 95 m (which is representative of the hub height of turbines in wind farms off the coast of the English Channel/ North Sea). These marine profiles are shown in Figures \ref{fig:standards_U} and \ref{fig:standards_iU} (all except ISO WF). This profile, which represents the commonality between all the various standards, will be referred to as `Profile 1' throughout the rest of this report. `Set 1' spires were designed to simulate this profile, as detailed in Section \ref{spire_design} [For urban profiles, please see Appendix \ref{urbanprofiles}]

\vspace{-0.3cm}

\vspace{-0.3cm}
\textbf{`Profile Wind Farm (WF)'}

Having established a methodology for calculating the desired ABL characteristics from various standards, the procedure is now used to calculate the wind characteristics of a particular marine site located in the North Sea. The ISO methodology was selected for this design, since it takes into account the sea roughness due to surface waves. The profiles (ISO and other standards, except the ones labelled ISO WF) shown in Figures \ref{fig:standards_U} and \ref{fig:standards_iU}, were calculated with a $U_{ref}$ of around 25 m/s at $z_{ref}=$ 10 m, as prescribed by an example site in the standards. For wind farms like the Thornton Bank, which is situated 30 km offshore of Belgium, at 25 m/s, the farm is likely to be on the verge of being shut down for safe preservation. Their mean yearly wind speed at turbine hub height (95 m in full-scale) is typically around 8.5 m/s. However, for the specific turbines, the design hub wind speed is approx. 14 m/s. This led to the generation of a second profile called `Profile WF', which is more suitable for the inflow of such wind farms situated in the English Channel, as they require a lower turbulence intensity. Consequently, the procedure outlined by ISO was applied to calculate the characteristics of a profile with a $U_{ref}=$ 14 m/s at a $z_{ref}=$ 95 m. This adjusted profile, labelled as ISO WF, is presented alongside `Profile 1', in Figures \ref{fig:standards_U} and \ref{fig:standards_iU}. While the differences in the normalised velocities are minimal, the lower $U_{ref}$ value used, results in a lower intensity distribution, uniformly shifting the ISO WF profile to the left. To simulate `Profile WF', `Set 2' spires were designed (Section \ref{spire_design}).

\vspace{-0.4cm}

\subsection{Techniques for Atmospheric Boundary Layer Generation}

Understanding the variation in wind properties with height is rudimentary to simulate realistic atmospheric boundary layer flows. The main purpose of this project is to generate a neutral ABL in the wind tunnel facility at Imperial, specifically in accordance with international wind standards (`Profile 1'), and one specific marine profile typical of wind farms in the English Channel and the North Sea (`Profile WF'). In neutral ABLs, no temperature stratification affects the flow, implying no significant vertical temperature gradient is present. Key parameters which have to be considered include the vertical velocity gradient, turbulence intensity in the longitudinal direction (i.e., in $U$), power spectral density (PSD) of the velocity fluctuations, and the distribution of Reynolds stresses \cite{depaepe}. There are various ABL simulation devices, such as spires, roughness elements, and barriers or fences, which can be installed in the tunnel to artificially replicate real-life ABLs. These are very often considered essential because typical wind tunnels are too short to allow the natural development of the desired neutral ABL profile, which requires a sufficiently long fetch. Hence, the design of specific hardware to simulate ABLs would help to artificially accelerate the formation of a neutral ABL profile. 

While the average wind speed at the hub height, $U_{hub}$, is commonly acknowledged to significantly influence the wind load characteristics, the mean wind loads affected by wind shear have also been shown to play a major role. Higher turbulence in the incoming wind was observed to result in significant fluctuations of the rotational speed of wind turbines, yielding in increased fatigue loads \cite{tian}. This portrays the importance of monitoring and implementing various methodologies to control mean velocity, and turbulence intensities experienced by wind turbines, adhering to their design requirements. 

Furthermore, Tian et al. \cite{tian} underscored the importance of accurately reproducing the required velocity and turbulence intensity at the hub height. Turbulence intensity $I_u$ can therefore be calculated as the ratio of the rms of the velocity fluctuations to the mean velocity $U_{local}$, given by Equation \ref{IU} and is a fundamental metric for assessing aerodynamic loading conditions. It gives a measure of the amplitude of velocity fluctuations in the flow caused by the superposition of eddies transported by the mean flow, and also helps to determine an estimation of the forces and bending moments acting on the turbines \cite{varshney}. 
\vspace{-0.2cm}
 \begin{equation}
    I_u = \sigma_u/U_{local}
    \label{IU}
    \end{equation}
    
\vspace{-0.3cm}

One of the devices which are commonly used for the generation of the ABL is spires, which play a crucial role in influencing the wind profile in the wall-normal direction. They help create an initial momentum deficit, accompanied by vortex shedding, thus contributing to the acceleration of wind shear, and turbulence intensities. There are various spire types that can be considered; for example, Irwin spires with broader bases would be more effective at increasing turbulence due to their shape, avoiding having to use a combination of narrow spires (such as Counihan spires) and fences or barriers. Hancock et al. \cite{hancock01} \cite{hancock02} \cite{hancock03} have extensively studied the iterative process for the design of vortex generators VGs and roughness elements. From their work, it can be deciphered that desired boundary layer characteristics such as mean velocity and turbulence intensity profiles can be achieved in the wind tunnel by appropriate optimisation of spire geometry, its spacing, and roughness elements. 

Counihan et al. \cite{coun01} discussed the generation of neutrally stable ABLs using setup constituting of a barrier, combined with vortex generators. The barriers help in inducing a momentum deficit, whereas the VGs help to recover the velocity profile, and enhance turbulence, as they are placed at alternating angles downstream. Installing coarse grid screens can also help to generate turbulence, but controlling the mean velocity profile using these can be tricky, since turbulent scales depend on the grid spacing. A typical set-up consists of triangular VGs in combination with an upstream wall or barrier, frequently complemented by roughness elements on the floor to enhance the near-wall turbulence, even using simple LEGO blocks \cite{coun05}. Similar implementation has been followed by Bortoli et al. \cite{debortoli}, but for part-depth ABL simulation. Various VG types were tested, out of which, the elliptic VGs were considered more effective. Typical height of the VG has to be about 1/10th of the boundary layer thickness, and the distance between the barrier and the VGs depends on the vertical distribution of turbulence. Further studies performed by Counihan demonstrated that it would take about 4.5 times the boundary layer height downstream of the turbulence generating devices, for the boundary layer to be fully developed \cite{coun03}. Counihan spires use flow generators to provide a thick boundary layer, so longer working sections are not usually necessary \cite{hancock02}. They also require some amount of refining to match with the target standards like ESDU \cite{hohman} \cite{robins1979}. There are also other VGs which are flat triangular, and hence easy to manufacture, providing profiles close to the target ESDU standard, which was set as a benchmark for comparison \cite{hancock03} \cite{irwin} \cite{hancock01}. 

Counihan and Standen et al. \cite{coun02} \cite{standen2} developed roughness, barrier, and mixing method, which is termed to be simple, and works for the full-depth simulation of the ABL \cite{debortoli}. Velocity profiles for urban terrains tend to follow the power law with an exponent $\alpha$ of 0.4, as originally documented by Davenport (1963), but Counihan reports a lower $\alpha$ of 0.28, which aligns with many field observations as given by Wardlaw and Moss (1970). This is in close proximity to the power law exponent value of $\alpha=0.285$ utilised in the unification of wind standards, described in Section \ref{unify_standards}.

To represent the mean velocity profile in an ABL, there are two classical approaches: either by using the log law or by using the power law \cite{seth}. Power law is commonly expressed by relating normalised velocities to normalised heights from above the ground, given by Equation \ref{eq:powerlaw}, where the power law exponent, whose common value can be approximated as $n=1/7$ for fully turbulent flows on open terrains \cite{1/7thlaw}. In various standards, power law is also represented in terms of boundary layer height and velocity at its edge, $\delta$, and  $U_\delta$ respectively, instead of $z_{ref}$, and  $U_{ref}$. WMO recommends a reference height $z_{ref}$ of 10m, which is commonly used my most of the wind engineering standards \cite{zref}. 

Logarithmic law provides a more completed approach to defining the velocity distribution as it takes into account physical parameters such as surface roughness or friction velocity, given by $\bar u/U_\delta = ln(z_g/z_0)/ln(\delta/z_0)$ (where $U_\delta$ is the velocity at the edge of the boundary layer $\delta$, $z_0$ is the roughness parameter for the specific terrains, $\bar u$ is the mean velocity calculated at a height $z_g$ above the ground). Unlike the power law, this model incorporates physically meaningful parameters that are critical for accurate assessment of wind characteristics, and also for design of wind turbines \cite{loredo}. The ABL typically extends for 1-2 km off the ground, with the first 50 m consituting of the surface layer, where shear stress remains roughly constant, but mechanical turbulence dominates \cite{kaimal}. 

The lower 30 to 50 m of the boundary layer is often described by a logarithmic law, under moderate wind and neutrally stable conditions, where the terrain-induced roughness strongly affects the wind profiles \cite{coun06}. However, Avelar et al. \cite{avelar} states that the log law applies from the surface to a height of about 100-150 m. On the other hand, power law is more appropriate for higher altitudes and stronger winds \cite{coun06}. The main drawback of the power law is that it is valid for any reference height, hence not recognising the top of the ABL in the model \cite{loredo}. Abubaker et al. \cite{abubaker} showcased the coexistence of log law and power law in the velocity profiles of ABLs, rendering the inner and outer layer of the boundary layer. 

Power law exponent $\alpha$ varies with terrain, giving lower values for smoother terrains, and higher values for rougher ones. It typically lies in the range rural terrains is 0.143 to 0.167, which aligns well with the value of $\alpha=0.11$ in the unification of standards in Section \ref{unify_standards}, and the value suggested for urban terrains is $\alpha=0.28$, which also matches with the one used in unifying various standards \cite{coun06}. For ocean studies, the acceptable value of $\alpha$ is described to be between 0.11, and 0.15 \cite {hsu} \cite{barb} \cite{bless}. Kozmar at al. \cite{kozmar01} \cite{kozmar2010} also provides detailed terrain-dependent parameters, stating the range of power law exponents to be from 0.16 (rural) to 0.35 (urban), whose spectral analysis was proven to closely align with classical turbulence theories, such as the Kolmogorov's -5/3rd law, and the von K\'arm\'an spectral methods. These values also match with the ones given by the standards, such as ASCE, and Eurocode (by fitting the velocity profile obtained, to the power law - to find the $\alpha$ value), although a discrepency can be noted in the power law exponent values defining urban, and suburban terrains between the standards. The $z_0$ values stated for urban city centres come to around 1 to 3, which is similar to that stated by Eurocode. 

Cook \cite{cook01} explored boundary layer simulation of the lower one-third part of the urban ABL by combining square mesh grids with perforated or solid walls, using plastic cups as roughness elements. The reported power law exponent ranged from 0.26 to 0.35, which indicates that the value of $\alpha=0.28$, stated by Counihan is a reasonable approximation for suburban simulations. Hence, this value was used in unifying various standards, detailed in Section \ref{unify_standards}. Cook \cite{cook05} further described the use of a barrier to create an initial momentum deficit, and mixing devices downstream to develop the turbulent boundary layer. A configuration consisting of a plane wall upstream of the mixing device, combined with elliptic wedge VGs downstream was incorporated, but this resulted in the reduction of the momentum deficit created along the centre-line of each of the VG due to strong turbulence mixing. As a solution, the height of the barrier wall could be increased upstream of the VGs, or castellated walls could be utilised, potentially improving the wake interactions. Standen \cite{standen} combined mixing, and barrier devices with tapering spires, but was only able to simulate a part of the BL, and not the entire BL. This indicates the challenges of fully replicating the target ABLs, in a scientific and laboratory setting. Therefore, it can also be concluded that spires and roughness elements are indispensable for producing realistic flow conditions within the wind tunnel \cite{hohman} \cite{hancock01}. 

One of the most commonly used and easiest ways to simulate the ABL in the wind tunnel is by using Irwin spires. The wind tunnel setup of hardwares for ABL generation using Irwin spires consists of a lateral array of drag devices, which are the spires at the upstream end of the test section, followed by an array of roughness elements, to aid in obtaining the desired wind profile, and sustained turbulence, as represented by boundary layers. Each spire has a triangular front plate which is iscoceles, placed normal to the flow, and a back plate conventionally called as a splitter plate, placed parallel to the flow at the leeward side. The splitter plate joins the front plate at its symmetry plane, and is mainly responsible for providing enough support to the standing spire. This spire design generates large scale turbulent structures, which are crucial for the development of critical flow features. Roughness elements are responsible for sustaining the turbulence generated by the spires, and also affects the value of the power law exponent \cite{ivancoNASA}. The Irwin spires are typically known to reproduce a BL, $\delta$, of around 80\% of the spire height $h$, whereas Counihan spires are more straightforward in generating BLs approximately equal to their full height.  

It has also been reported that approximately triangular shaped spires are important to produce the mean velocity, and turbulence intensity properties downstream of the spires, as the resulting characteristics are commonly insensitive to the spire shape \cite{irwin}. Irwin reports that the fetch length adequate to develop the target BL is approximately 6 times the spire height, $6 \cdot h$, at which the flow develops lateral uniformity, where $h$ is the height of the spire. Various empirical relations have been developed to link the boundary layer characteristics to the design of the spire and roughness elements in the Irwin methodology \cite{irwin}. Equations \ref{h} and \ref{b/h} outline the process to be followed, which has been implemented for designing the spires tested in this project. 

\begin{figure}[H]
  \begin{minipage}{0.48\linewidth}
    \begin{equation}
      h = \frac{1.39\delta}{1 + \alpha/2}
      \label{h}
    \end{equation}
  \end{minipage}\hfill
  \begin{minipage}{0.48\linewidth}
    \begin{equation}
      \frac{b}{h} = 0.5 \left[ \frac{\psi(H/\delta)}{1 + \psi} \right] (1 + \alpha/2)
      \label{b/h}
    \end{equation}
  \end{minipage}
\end{figure}
\vspace{-0.5cm}
where: $\psi = \beta{[2/(1+2\alpha]+\beta-[1.13\alpha/(1+\alpha)(1+\alpha/2)]}/(1-\beta)^2$, $\beta = (\delta/H)\alpha/(1+\alpha)$.

\begin{wrapfigure}{l}{0.6\textwidth}
  \centering
  \vspace{-10pt}
  \includegraphics[width=0.28\textwidth]{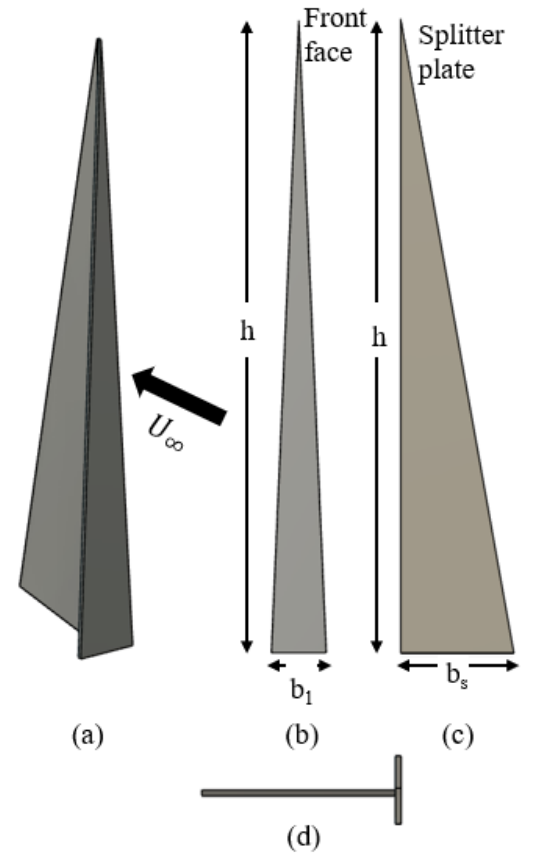}%
  \hspace{0.01\textwidth}%
  \includegraphics[width=0.3\textwidth]{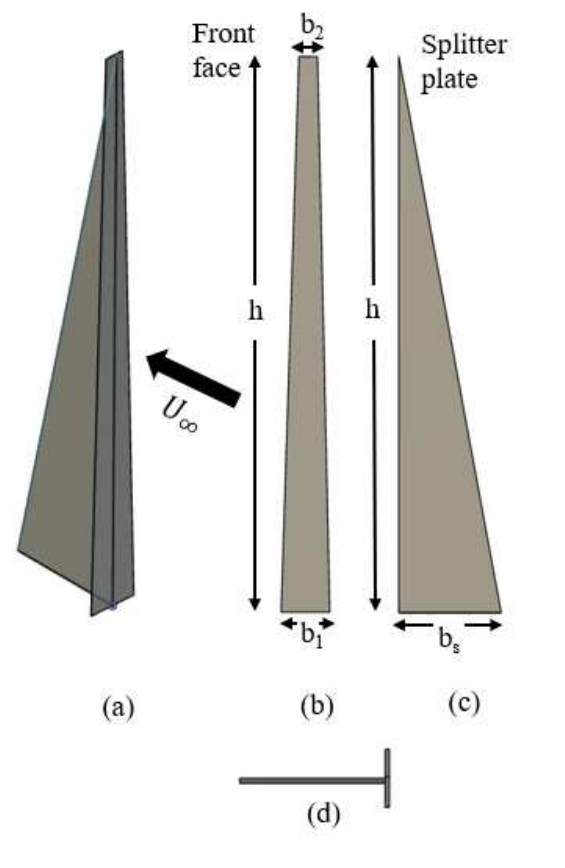}
  \caption{Comparison of a general set of Irwin spires — original (left), truncated (right): (a) 3D view, (b) front view, (c) side view, (d) plan view.}
  \label{fig:irwin_spires_wrap}
  \vspace{-9pt}
\end{wrapfigure}

Here, $\delta$ refers to the boundary layer height in full-scale, $\alpha$ is the power law exponent, and $b$ is the spire base. The equation for $b/h$ includes the aerodynamic drag of the floor roughness, which cannot be ignored though its consequences would be nearly insignificant at a distance of $6 \cdot h$ downstream. Guidlines provided by Irwin recommend a spire spacing of $h/2$ from centre-line to centre-line, where $h$ stands for spire height \cite{standen2}. The recommended splitter plate's base dimension is $h/4$. Figure \ref{fig:irwin_spires_wrap} presents a visual comparison of the Irwin spires, in original and truncated (used in cases where recommended spire height exceeds the tunnel section height) forms, with labelled dimensions of the triangular front plate, and the rear splitter plate. 

The influence of surface roughness on the spatial distribution of velocity, turbulence intensity, and turbulence length scales within the boundary layer, has been underlined by Varhsney et al. \cite{varshney}. Calculations for roughness distribution at this downstream distance can be carried out assuming that the conditions would be near equilibrium. Therefore, the skin friction coefficient $C_f$ can be approximated by Equation \ref{cf}, which quantifies the drag contribution of surface roughness as given by Irwin \cite{irwin}. The ratio of the cube height $k$ to the boundary layer height $\delta$, relates to the cube spacing $D$ via the Equation \ref{k/delta}, which is valid for 30 < $\delta D^2/k^3$ < 2000.
% \clearpage

\vspace{-0.4cm}

\begin{figure}[H]
  \begin{minipage}{0.40\linewidth}
    \begin{equation}
      C_f = 0.136 \left[ \frac{\alpha}{1 + \alpha^2} \right]^2
      \label{cf}
    \end{equation}
  \end{minipage}\hfill
  \begin{minipage}{0.60\linewidth}
    \begin{equation}
      \frac{k}{\delta} = \exp\left( \frac{2}{3} \ln\left( \frac{D}{\delta} \right) - 0.1161 \left[ \left( \frac{2}{C_f} + 2.05 \right)^{1/2} \right] \right)
      \label{k/delta}
    \end{equation}
  \end{minipage}
\end{figure}

\vspace{-0.5cm}

The Irwin and Counihan spires have also been employed to simulate suburban type terrains, as reported by Paepe et al. \cite{depaepe}, who compared the wind tunnel results using Counihan quarter ellipse spires with roughness elements to similar configurations involving Irwin spires. The resulting velocity profiles were evaluated against various international standards, which portrayed a good agreement. This suggests that truncated spires may be considered a suitable final method for suburban flow simulation. 

More importantly, a similar work was done by Ivanco et al. \cite{ivancoNASA} by using using Irwin spires, and roughness elements. Furthermore, another project involving the study of wind turbine wakes interference, and boundary layer generation was extensively studied by Ozby et al. \cite{ozbay}, who stated the use of triangular spires combined with chains. However, no specific design methodology or optimisation criterias were detailed. In addition to this, Pires et al.  \cite{pires} accentuated the use of surface roughness elements using readily accessible materials such as carpets, and so on to help finetune the desired boundary layer characteristics. These aid in promoting the formation of a fully developed boundary layer with uniform velocity and turbulence profiles, which can be achieved by disintegrating the larger vortices, and by facilitating the process of turbulence decay. 

Hancock et al. \cite{hancock01} has also investigated the experimental simulation of the wakes of wind turbines, for which Irwin spires were deployed with an array of roughness elements, with no barrier or fence. A trial and error approach was then followed to obtain the desired profile that matched with the ESDU data, which was used as the criterion for comparison. The classification of various terrain types, and the related values of the power law exponent, given by the ASCE standard has been experimentally validated by Shojaee et al. \cite{shojaee}. The methodology outlined by Irwin has been utilised in this paper to design their spires and roughness elements, reporting the efficacy of Irwin-type spires in reproducing boundary layer characteristics across varied terrain classes. 

The widespread use of Irwin spires in simulating diverse environments, including urban, marine, and wind farm settings, demonstrate their suitability for replicating `Profile 1' (marine profile indicating the commonality between all the standards), and `Profile WF' (an incident marine profile typical ahead of a wind farm in the English Channel or North Sea) in this project. 

\subsection{Power Spectral Density \& Length Scales}
\label{lxmethods}
 % \item when the surface roughness is removed - int scales decrease, natural terrain - more open is the terrain - bigger is the int scales 
 %    \item finding Lu,x by fitting the PSD of the long turb component to ESDU atm spectrum formula with Lux as the fitting parameter - see formula \cite{cookLux}

Power Spectral Density (PSD) is commonly used to represent the power of a signal across various frequencies. PSD of the velocity fluctuations near the surface, up to about a height of 50 m in full scale is shown to follow the von K\'arm\'an spectrum \cite{vonk}, and also exhibit slopes consistent with the theoretical Kolmogorov's -5/3rds law \cite{kolmogorov}. Above this height, transition occurs to the boundary layer, which causes deviation from these theoretical models in reality \cite{farellIy}. 

The von K\'arm\'an spectrum for PSD of the longitudinal velocity fluctuations (in $U$) is given by Equation \ref{PSD} \cite{esdu85020}.
\vspace{-0.1cm}
\begin{equation}
S_u(f) = \frac{{4\sigma_u^2 L_{u,x}}/{U}}{
\left[ 1 + 70.8 \left( \displaystyle\frac{L_{u,x}f}{U} \right)^2 \right]^{5/6} }
\label{PSD}
\end{equation}

\vspace{-0.1cm}
The PSD $S_u(f)$ of longitudinal velocity fluctuations is $\bar u'^2=\int_{0}^{\infty}{S_u(f)df}$, where $u(t)=\bar u+u'(t)$.

Understanding length scales in turbulent flows is fundamental to representing the structure and dynamics of turbulence, especially in the atmospheric boundary layer. Integral length scales are used to measure the typical size of the energy-containing eddies, providing insights into the spatial and temporal coherence of turbulent structures by quantifying the maximum correlation distance or period between two points in the flow \cite{lengthscale} \cite{autocorrelation-lengthscale}.  

Pope \cite{pope2001turbulent} details the physical significance of these length scales within the energy cascade framework, where kinetic energy enters turbulence at large scales, gets transferred via intermediate scales by the inviscid process, and is eventually dissipated by viscosity at the smallest scales. Hence, it is essential to accurately capture the largest scales in turbulence modelling. The size and intensity of these turbulent eddies directly affect the aerodynamic loads on the structures and their fatigue life, highlighting the practical importance of integral length scales. They can be estimated via several methods:
 
 \begin{enumerate}[leftmargin=*, label=\arabic*.]
    \item \textbf{Zero-frequency spectral (y-intercept) method}: This method estimates the integral length scale $L_{u,x}$ from zero-frequency limit of the velocity spectral density, using Equation \ref{L-zerofreq}, where ${S_u{(0)}}$ is the spectral density at zero-frequency from Equation \ref{PSD}, $\sigma_u^2$ is the variance of the velocity fluctuations in $u$, and $U$ is the mean velocity.
\vspace{-0.1cm}
    \begin{equation}
    L_{u,x}=\frac{U}{4} \cdot \frac{S_u{(0)}}{\sigma_u^2}
    \label{L-zerofreq}
    \end{equation}
    However, this method assumes isotropic and homogenous turbulence (which breaks down near boundaries or in complex flow conditions), following the von K\'arm\'an spectrum model. The spectrum is also expected to cover a wide range of frequencies, especially at lower frequencies. Taylor's frozen turbulence hypothesis is used to convert frequency to spatial scale, assuming that only stationary turbulence is advected past the sensor without distortion \cite{pope2001turbulent}. Nevertheless, this method provides a direct spectral measure related to the energy-containing eddies and can also be implemented via spectral estimation functions such as `pwelch' in MATLAB. But it is highly sensitive to low-frequency spectral noise, errors arising from spectral fitting, and also due to poor data quality, which causes significant inaccuracies. Various filters can be used to combat this issue as suggested by \cite{farellIy}.

    \item \textbf{Autocorrelation method}: This is the most robust and physically meaningful method, since it is solely based on the velocity autocorrelation function, $R_{u,x}$, which measures the correlation of a velocity component with itself over spatial or temporal lags. The integral length scale is defined as $L_{u,x}$, calculated via Equation \ref{Lux}, where $R_{u,x}$ is given by Equation \ref{Rux}, and $\bar U$ is the mean velocity of the flow \cite{pope2001turbulent}. 
    Area under $R_{u,x}$ gives the time scale, which when multiplied with the mean velocity $\bar U$ gives the integral length scale, by utilising Taylors frozen turbulence hypothesis \cite{pope2001turbulent}.

    \vspace{-0.1cm}
\noindent
\begin{minipage}{0.35\textwidth}
\begin{equation}
L_{u,x}=\bar U \int_{0}^{T} Ru_x (\tau)d\tau
\label{Lux}
\end{equation}
\end{minipage}%
\hfill
\begin{minipage}{0.55\textwidth}
\begin{equation}
Ru_x=\frac{\overline{[U(t)-\bar U(t)][U(t+\tau)-\bar U(t+\tau)]}}{\bar U^2}
\label{Rux}
\end{equation}
\end{minipage}

    The autocorrelation method relies on several key assumptions, including Taylor's frozen turbulence hypothesis, stationarity, homogeneity, adequate sampling to capture relevant turbulence scales, and minimal flow disturbance by the measurement probe \cite{kozmar2010}. Its advantages include being less sensitive to noise than spectral methods, not requiring fitting to theoretical spectral models, and also directly reflecting the physical correlation properties of turbulence \cite{Lucomparison}. However, this method has its own limitations since Taylor's hypothesis may be invalid near the ground or in complex flows where shear and convection vary, and the practical limits of the integration are also finite, which leads to uncertainty. The common choices for the integration limits are, integrating up to the first zero-crossing, across the entire available domain, stopping where the $R_{u,x}$ reaches its minimum, or using the $1/e$ decay point as a cut-off \cite{trittonLu}. For homogenous isotropic turbulence, the $R_{u,x}$ typically decays smoothly to zero, validating the zero-crossing approach. Additionally, measurement of the domain size relative to the length scale can also affect its value \cite{varshney}. 

    \item \textbf{Central peak frequency method}: Length scales can also be estimated from the dominant spectral frequency $f_{peak}$, after normalising the spectrum by the variance, and then using Equation\ref{peakfreq}, where $U$ is the mean streamwise velocity \cite{Lucomparison}. 
    \vspace{-0.2cm}
    \begin{equation}
    L_{u,x}=\frac{0.146 \times U}{f_{peak}}
    \label{peakfreq}
    \end{equation}    
    This method assumes homogenous turbulence with a distinct spectral peak, requires a high quality, low noise spectral data, and also relies on the validity of Taylor's frozen turbulence hypothesis. Therefore, it is highly sensitive to spectral noise and poor resolution, and the reliability of this method decreases in complex anisotropic or boundary-affected flows, where spectral peaks may be unclear or even absent \cite{Lucomparison}. 

    \item \textbf{von K\'arm\'an fit method}: $L_{u,x}$ can be calculated from the von K\'arm\'an spectrum for PSD of the longitudinal velocity fluctuations in $U$, given by Equation \ref{PSD}. This equation can be fit to the data, having computed its $u_{rms}$, $S_u$, and $f$, alongside the $U$. Hence, the only unknown in the equation would be the length scale $L_{u,x}$, which can then be adjusted for a minimum of a least squares fit. This method assumes that turbulence is homogenous and isotropic locally, and that the data follows the theoretical von K\'arm\'an model, to represent the energy distribution across scales.  

\end{enumerate}

    Lu et al. \cite{Lucomparison} compared different approaches to calculate length scales, highlighting the underlying assumptions and the need for careful interpretation of the results thus obtained. The autocorrelation method is generally preferred for its robustness, and physical relevance, especially when a good quality velocity time series is available. 

    Significant variability exist in the length scale definitions, and values reported by different international wind standards. Standards like Eurocode, ESDU, and others have varied formulas, averaging periods, referencing heights, terrain considerations, and so on, which impacts the length scale calculations/ methods given by them. For instance, the reference height used by Eurocode is 200 m, whilst ESDU's accounts for displacement height $d$, which is a factor not accounted for in Eurocode. Kozmar \cite{kozmar2011wind} highlighted these variations, emphasising the difficulty in simulating all turbulence features of the ABL within wind tunnels due to these inherent difference in basic definitions, and empirical models.

    The above mentioned methods to calculate integral length scales have been utilised to anlayse the results of the test(s) performed in this project. They have also been compared to those given by various international wind standards, which can be found in Section \ref{lux_compare}.
\newpage
\section{Experimental Method and Design}
A comprehensive overview of the experimental methods and design considerations, leading to the final setup in the upper and lower test sections (TS) of Imperial's 10'$\times$5' wind tunnel, is presented in this section. Iterations and refinements of the setup were carried out in the lower test section, enabling the use of both Laser Doppler Anemometry (LDA) and multi-hole probe (MHP) to capture extensive flow characteristics. Valuable insights gained from these preliminary tests facilitated to replicate a similar set-up in the upper test section, where only the MHP technique was deployed for retrieving flow data. 
\vspace{-0.5cm}
\subsection{Spires}
\label{spire_design}
The design of the spires was guided by Irwin's \cite{irwin} methodology, elaborated in Section \ref{spire_design}. To simulate the marine atmospheric boundary layer conditions, two sets of spires were developed for the lower and the upper test sections (TS). Two profiles were tested and implemented, summarised in Table \ref{tab:spire_sets}.

\begin{table}[H]
\centering
\begin{tabular}{lp{10cm}} % Adjust width as needed
\toprule
\textbf{Spire Set} & \textbf{Description} \\ \midrule\hline
\textbf{`Set 1'} & \textbf{`Profile 1'} : framework profile, establishing commonality between standards ($\alpha=0.11$) \\
\textbf{`Set 2'} & \textbf{`Profile WF'} : more suitable for wind farm inflow in English Channel/ North Sea ($\alpha=0.07$) \\
\bottomrule
\end{tabular}
\caption{Description of spire sets, and the corresponding profiles}
\label{tab:spire_sets}
\end{table}

\vspace{-0.7cm}
\begin{wrapfigure}[16]{l}{0.55\textwidth}
    \centering
    \vspace{-10pt}
    \includegraphics[width=1\linewidth]{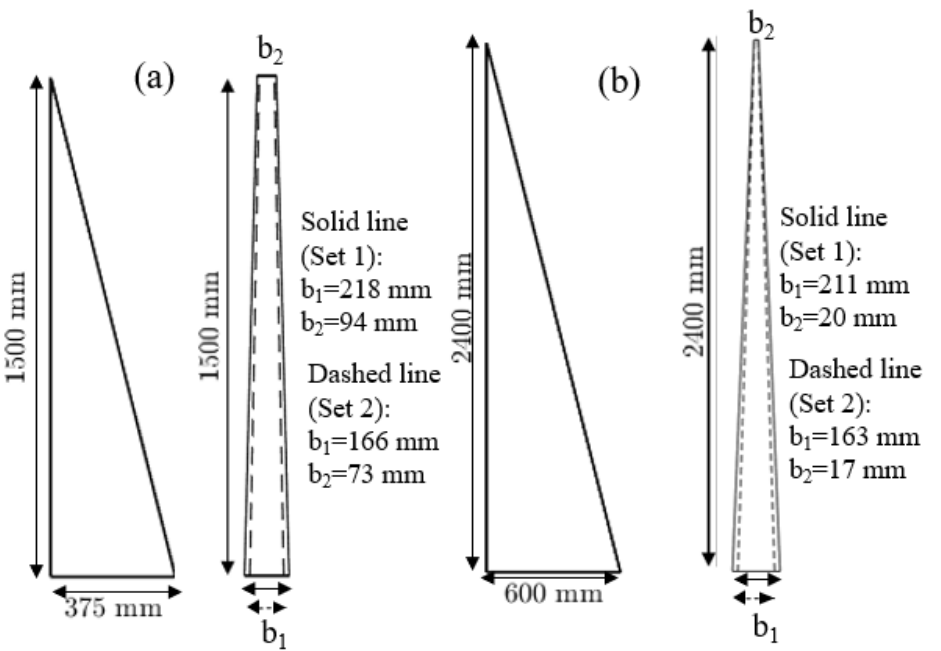}
    \vspace{-0.8cm}
    \caption{(a) Lower TS, (b) Upper TS}
    \label{fig:lowerupperts_spires}
\end{wrapfigure}

The design process was largely dependent on the wind tunnel testing scale (1:250, which is typical for the wind farms in the British Channel or North Sea), full-scale boundary layer height $\delta$, the power law exponent value $\alpha$, and the tunnel test section height $H$. The scale 1:250 is a sweet spot for the wind tunnel facility here at Imperial, since at 1:250, a single 125 m diameter wind turbine can be modelled as a 0.5 m diameter model turbine, with a blockage of 4.3\% in the lower TS. In the upper TS, it allows a small wind farm of 6 turbines to be modelled with a reasonable turbine to turbine spacing. The $\alpha$ value stemmed from the unification of standards, and the specific wind farm profile, discussed in Section \ref{unify_standards}. A conservative estimate of $\delta = 500$\,m was chosen based on Pasquill's suggested boundary layer limit of 600\,m~\cite{pasquill}, where mechanical energy production diminishes. 

\par\vspace{1ex} % end wrapping cleanly and add some vertical spacing
% \vspace{2cm}
\begin{table}[H]
\centering
\scriptsize
\renewcommand{\arraystretch}{1.1}
\begin{tabular}{l|p{4.2cm}|c|c|c|c|c|c|c|c}
\textbf{Dimensions} & \textbf{Description} & \multicolumn{4}{c|}{\textbf{Original Spire}} & \multicolumn{4}{c}{\textbf{Truncated Spiew}} \\ \hline
 & & \multicolumn{2}{c|}{\textbf{Lower TS}} & \multicolumn{2}{c|}{\textbf{Upper TS}} & \multicolumn{2}{c|}{\textbf{Lower TS}} & \multicolumn{2}{c}{\textbf{Upper TS}} \\ \hline
 & & Set 1 & Set 2 & Set 1 & Set 2 & Set 1 & Set 2 & Set 1 & Set 2 \\ \hline\hline
h (m) & Spire height & 2.6 & 2.7 & 2.7 & 2.7 & 1.5 & 1.5 & 2.4 & 2.4 \\
$b_1$ (mm) & Spire bottom width & 218 & 166 & 211 & 163 & 218 & 166 & 211 & 163 \\
$b_2$ (mm) & Spire top width & N/A & N/A & N/A & N/A & 94 & 73 & 20 & 17 \\
$b_s$ (mm) & Splitter Plate base & 375 & 166 & 211 & 163 & 375 & 166 & 211 & 163 \\
x (m) [n] & \begin{tabular}[t]{@{}l@{}}Spacing [spire no.] \\ \multicolumn{1}{c}{} \end{tabular} 
& \multicolumn{4}{c|}{1.3 [2] / 0.7 [3]} & \multicolumn{4}{c}{1.3 [4] / 1.2 [5]} \\
k (mm) & Roughness cube size & 0.094 & 0.016 & 0.094 & 0.016 & 0.094 & 0.016 & 0.094 & 0.016 \\
D (mm) & Roughness spacing & 1 & 1 & 1 & 1 & 1 & 1 & 1 & 1 \\
\end{tabular}
\caption{Summary of spires and roughness elements for target profiles in the lower \& upper TS}
\label{tab:geometry_table}
\end{table}
\clearpage

The `Set 1' spires were designed to replicate `Profile 1', establishing the commonality between the various standards \ref{spire_design} and \ref{unify_standards}. On the other hand, `Set 2' was developed using Equation~\ref{eq:powerlaw} to match the conditions of an incident wind farm profile given by `Profile WF'. The key dimensions of `Set 1' and `Set 2' spires for the lower and upper sections are summarised in Figure \ref{fig:lowerupperts_spires} (same labelling as in Figure \ref{fig:irwin_spires_wrap}), with common splitter plates across both spire configurations. The solid lines in Figure \ref{fig:lowerupperts_spires} represent the `Set 1' spire, whereas the dashed lines show the `Set 2' spire. Table \ref{tab:geometry_table} gives the values of both the original spire (directly given by Irwin's method) and the truncated spire (spire design which needs to be truncated to fit into Imperial's wind tunnel). 

The spire set for the lower TS was truncated to fit within the wind tunnel height. Whereas, spires in the upper TS were truncated to simplify construction. The design ambiguity arose regarding the centre-line spacing recommendation of $h/2$; a lack of clarity in Irwin \cite{irwin} whether $h$ referred to the truncated or the non-truncated spire height (both appear in literature). To resolve this, both interpretations were tested, resulting in four test configurations per test section. 
The measurements were taken 16.3 m downstream ($6 \cdot h$) in the lower TS, and 14.3 m downstream, $5.5 \cdot h$ in the upper test section to avoid interference from the wind tunnel corner vanes. 

All spires were manufactured from 18 mm plywood, offering an optimal balance between weight and structural integrity. M8 inserts were employed to connect the triangular front faces to the rear splitter plates, allowing easy assembly and reconfiguration, allowing splitter plates to be reused across tests in both the upper and lower sections. In the upper test section, spires were secured by turnbuckles connected via steel cable (see Appendix \ref{upperts_cable}).

Spires for generating urban terrain conditions were also developed to complete the establishment of commonality between the standards, given in Appedix \ref{urbanprofiles}. 

% To ensure the spires’ structural stability during wind tunnel operation, vibrational phenomena such as galloping were analyzed. Verification of the Strouhal number and the natural frequencies of the spires was performed to prevent resonance and excessive oscillations under flow conditions. This verification process confirmed that the spire designs were dynamically stable, minimizing flow-induced vibrations that could compromise experimental accuracy.

\subsection{Roughness Elements}
\label{R}
The design of the roughness elements was carried out following the methodology outlined by Irwin \cite{irwin}, ensuring consistency in the approach followed for the design of spires. Table \ref{tab:geometry_table} summarises the final calculated values for the spacing $D$, and the size of the roughness elements $k$, for `Set 1', and `Set 2' spires, which are in the form of cubes as suggested by Irwin's design approach.

Notably, the values denoting the size of the roughness elements are considerably small, indicating that the natural tunnel floor characteristics combined with natural frictional forces are practically sufficient to provide the necessary surface roughness. In a physical sense, this implies that the bare tunnel floor would be adequate, or the addition of a very fine, carpet-like texture affixed to the floor could help replicate the necessary profile that matches with the target boundary layer conditions.

To validate the amount of roughness required, following the results from the testing of `Set 1' spires in the lower section with no external roughness elements, a second iteration of experimental testing was performed using roughness in the form of green mesh material for a distance of 6.84 m (out of the 16.3 m distance between the measurement location and the spires) from the measurement location in the tunnel, to elevate the surface texture (shown in Appendix \ref{greenmesh}. The combination of the two `Set 1' spires and the green mesh provided a more pronounced roughness distribution than what was necessary, which can be observed in the results presented in Section \ref{R_result}.

\subsection{Instrumentation Techniques}

The characterisation of ABLs in the 10'$\times$5' wind tunnel and its comparison against the target velocity profiles were achieved via two complementary measurement techniques: Laser Doppler Anemometry (LDA), and the seven-hole probe, also known as a multi-hole probe (MHP). To measure the flow features, and compare the profiles, these techniques were employed to capture the instantaneous velocity components at specific spatial locations within the wind tunnel. Selecting LDA and MHP as the measurement methods allowed for a reliable verification of the velocity data and turbulent characteristics. LDA and  MHP, both require careful calibration, data acquisition, and data processing to ensure accuracy and consistency in the measurements procured. Both methods possess inherent strengths and limitations. LDA requires complex optical setup and meticulous alignment during the calibration procedure alongside safety precautions. But it provides a non-intrusive point measurement of the velocity components, in contrast with the multi-hole probe that converts pressure measurements into velocity components.  

\subsubsection{Laser Doppler Anemometry}

\begin{wrapfigure}{r}{0.5\textwidth}
    \centering
    \vspace{-10pt} % Adjust vertical spacing as needed
    \includegraphics[width=0.45\textwidth]{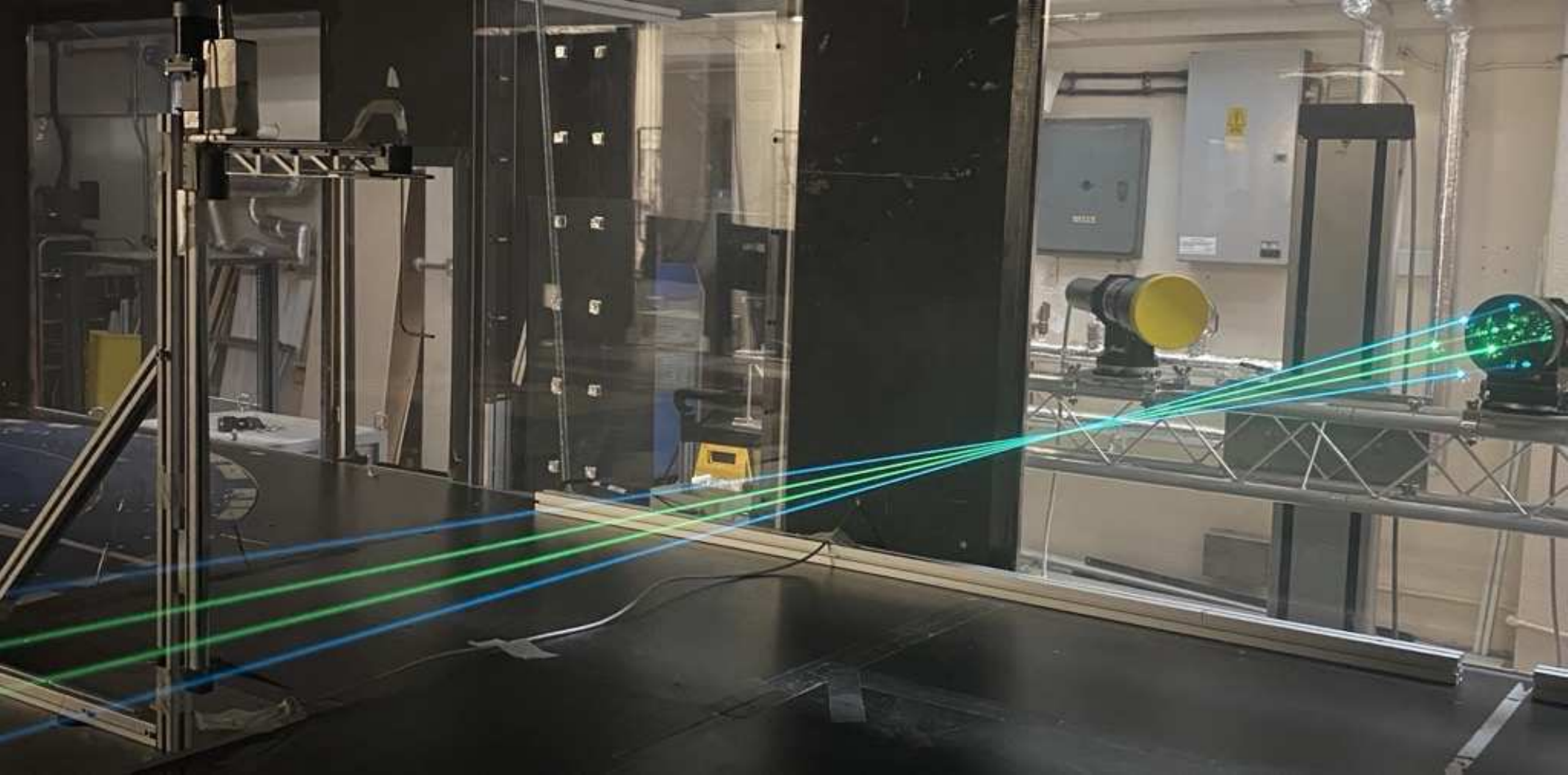}
    \caption{LDA and MHP, lower TS}
    \label{fig:LDA}
    \vspace{-10pt}
\end{wrapfigure}

The LDA measurements were conducted using a FiberFlow LDA system, capable of measuring different velocity components at a single point in space. 2-dimensional LDA can measure two velocity components (the streamwise and wall-normal velocities), whereas a three-dimensional LDA setup can acquire all three velocity components (the streamwise, spanwise, and wall-normal velocities) from the flow. This system employs four pairs of laser beams at different wavelengths—514 nm, 488 nm, and 532 nm—from diode-pumped solid-state (DPSS) lasers. It is a class 4 continuous-wave laser system, and the beams are transmitted to the measurement probes via fibre optic cables, which focus the beams to converge to a point inside the wind tunnel, creating the measurement volume where acquisition takes place, as represented in Figure \ref{fig:LDA}.

In LDA, seeding material like atomised polyethylene glycol (PEG) particles, is introduced into the flow. These particles scatter light, causing a Doppler shift that is detected by the LDA system. This shift in Doppler frequency is processed to yield instantaneous velocity values. The raw velocity measurements, based on the LDA's frame of reference, are then converted into the global coordinate frame using a transformation matrix to enable meaningful comparisons. In the physical setup, the LDA probes were arranged on a three-axis traverse system, which had a limited range of motion especially along the span, constraining the extent of the spanwise measurement region.

\subsubsection{Multi-hole Probe}

\begin{wrapfigure}{r}{0.35\textwidth}
    \centering
    \vspace{-10pt}
    \includegraphics[width=0.4\textwidth]{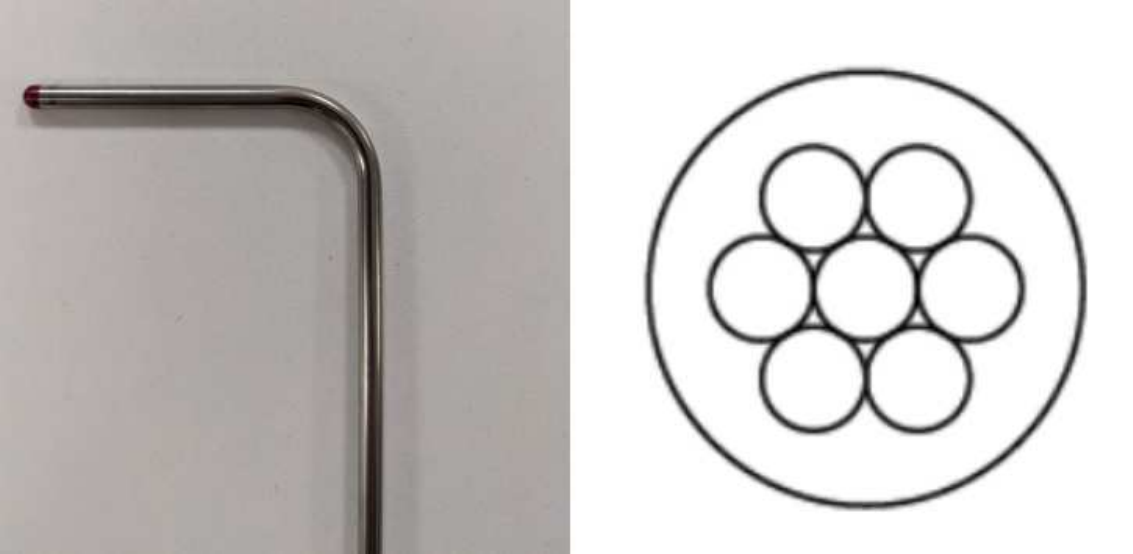}
    \caption{Seven-hole probe \cite{mhp}}
    \label{fig:MHP}
    \vspace{-10pt}
\end{wrapfigure}

The system utilises a seven-hole probe manufactured by Surrey Sensors, shown in Figure \ref{fig:MHP}. The probe is connected to the differential pressure device AllSensors' 1 INCH-D1-P4V-MINI via seven pressure tubes and a static tube, with the data acquired using a PXI. The seven differential pressure sensors were calibrated against a Furness FCO560 micromanometer, which simultaneously generates and measures differential pressures. A linear relationship between applied pressure and the voltage output of each of the differential sensors was obtained, yielding the calibration constants. 

Before each test, a no flow condition was acquired in order to account for and remove the drift in the output voltage. 

Pressure differences between the top, bottom, and side holes of the probe help determine pressure coefficients, from which velocity calibration is derived. Interpolating these coefficients from the calibration map yields pitch and yaw angles. The velocity magnitude is computed from the maximum pressure reading, and the pitch and yaw angles, velocity components are decomposed accordingly. Final velocity values after pipe corrections are obtained via linear interpolation of the calibration matrix.

Dynamic calibration involves a loudspeaker generating pink noise. The frequency response of the system is captured using the probe and a microphone. A second-order transfer function model is fitted, incorporating a correction constant \(a\), and is given by \(T_c(s) = \frac{s^2 + s \left({\omega_r}/{Q}\right) + \omega_r^2}{s^2 + s \left({\omega_r}/{a}\right) + \omega_r^2}\), where \(\omega_r\) is the resonant frequency (rad/s) and \(Q\) is the resonance quality factor \cite{mhp}. This transfer function is discretised and applied to a validation dataset to assess the effectiveness of correction and filtering. 
% A comparison of power spectral densities confirms that the filtered signals match predicted behaviour closely.

Data acquisition was set at 10 kHz, which is downsampled to 2.5 kHz for dynamic calibration, and the difference equation is applied at this frequency. For the final analysis, the signal is further subsampled to 250 Hz. This scaling implies that 1 Hz in the full-scale environment is represented by 250 Hz in the acquisition, similar to LiDAR-based systems, sometimes deployed in wind turbine hubs, ensuring accurate capture of frequency response. Further analysis comparing LDA and MHP is presented in Section \ref{LDA-MHP}.
\vspace{-0.2cm}
\subsubsection{Probe Holder Design}
\label{probedes}
\begin{wrapfigure}{l}{0.35\textwidth}
    \centering
    \vspace{-10pt}
    \includegraphics[width=0.3\textwidth]{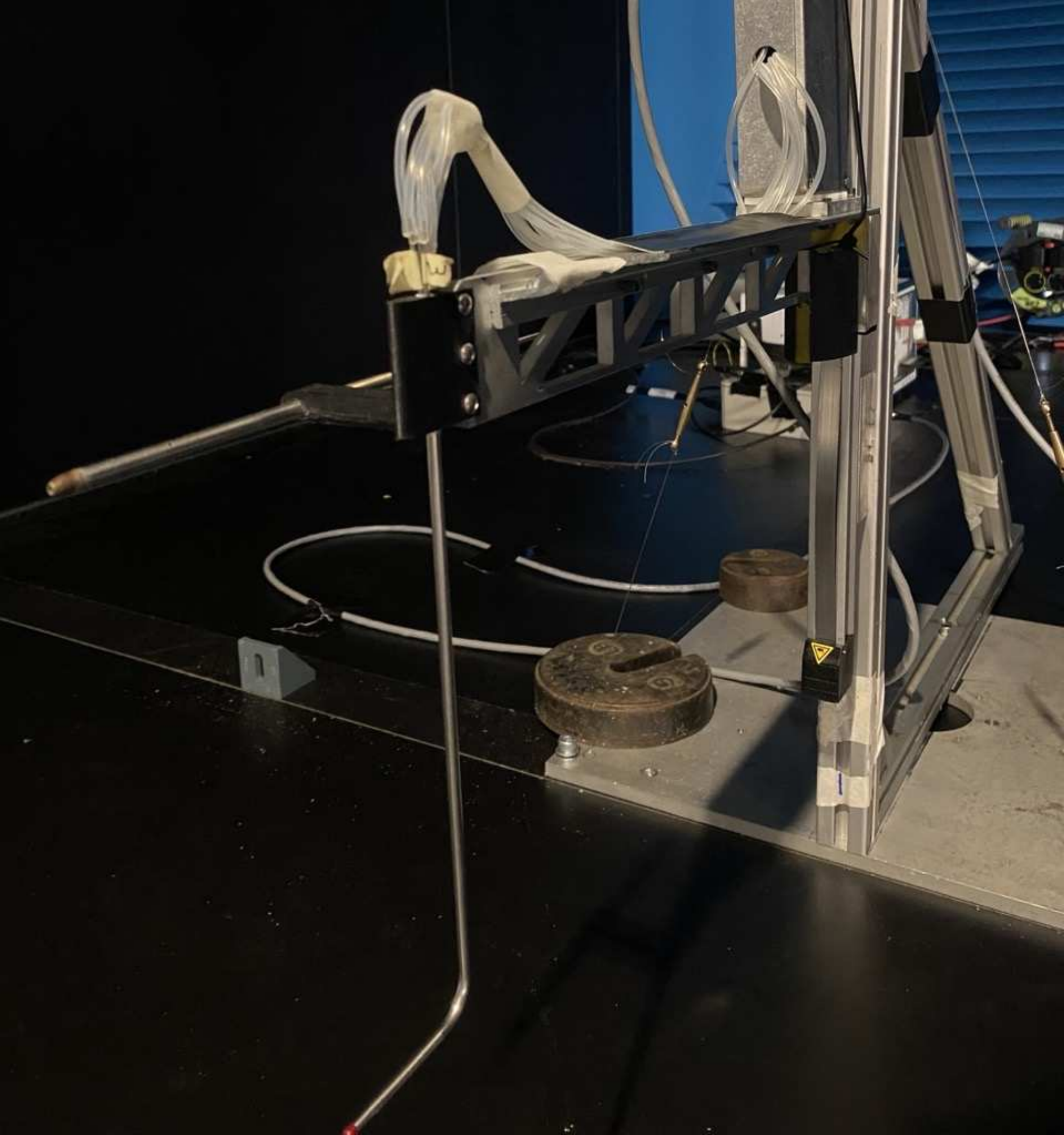}
    \caption{Modified traverse set-up}
    \label{fig:traverse}
    \vspace{-10pt}
\end{wrapfigure}

The traverse system enabled velocity measurements at various heights and spanwise positions with minimal disturbance caused to the incoming flow. Initially, a triangular Minitec aluminium frame with an encoder-controlled vertical carriage (of maximum travel height 600 mm), automated by LabVIEW, was used, by mounting the probes onto a wooden block, see Appendix \ref{oldtraverse}. However, blockage effects were observed when comparing the MHP data to LDA data as seen in Figure \ref{fig:LDA-MHP-timeseries-old}, which led to a redesign. 

The redesign aimed to reduce the frontal area, where the Minitec bars were replaced with steel cables and turnbuckles, while a repositionable aluminium base and floor-mounted rail ensured precise alignment (Appendix \ref{newtraverse}, \ref{traverse_setup}). A custom truss-style probe holder, cut from 8 mm aluminium and fitted with a 3D-printed fairing, was mounted on the traverse, illustrated by Figure \ref{fig:traverse} (Appendix \ref{cad}). The static probe and sensor box were integrated into this streamlined setup. After installation, calibrations were performed, and a cross-laser was utilised to align, to ensure accurate and repeatable probe alignment. 

\subsection{Experimental Set-up}
% \subsubsection{Wind Tunnel, LDA and Multi-hole Probe Set-up}

All the experiments were conducted at the 10'$\times$5' closed-circuit wind tunnel facility at Imperial, consisting of a lower test section of 3 m (W)$\times$1.5 m (H), and a test section length of 20 m, and an upper test section of 5.8 m (W)$\times$2.7 m (H), and a test section length of 18 m. For the purpose of conditioning the flow, before every run, spires  were carefully aligned at marked locations using measuring tape and cross-line laser, to generate the target boundary layer profiles. 

For the experiments conducted in the lower TS, the speed was set to 10 m/s at the pitot, which was placed at a height of 1 m off the ground, equivalent to 250 m in full-scale. However, in the upper test section, in order to achieve a target design velocity of 7.5 m/s at a height of 0.38 m, equivalent to a full-scale 95 m high wind turbine hub, at a 1:250 scale, the velocity was set indirectly by calibrating to the pitot reading in the lower section (approx. 25 m/s), adjusted via the MHP's system calibration and voltage mapping. 

The arrangement of `Set 1' and `Set 2' spires in the lower and the upper test sections is illustrated by Figures \ref{fig:lowerts_spires} and \ref{fig:upperts_spires}, and also in Appendix \ref{lowerspires}, \ref{upperspires}. The position of the spires along the tunnel width has been noted clearly, alongside the spanwise measurement locations that are shown by red crosses. Not all spanwise locations of the configurations shown in the Figure \ref{fig:upperts_spires} were chosen to gather measurements during the test campaign. At certain stations, only a fixed point measurement was acquired (either at 1 m height using the LDA, or at 558 mm off the ground using the MHP) to validate the spanwise uniformity, before taking full boundary layer scans. The numbering convention followed for the spanwise measurement points in each of these set-up configurations will be uniform throughout this report for easy comparison. This will be discussed further in Section \ref{results}.

\vspace{-0.3cm}
\begin{figure}[H]
    \centering    \includegraphics[width=1.1\linewidth]{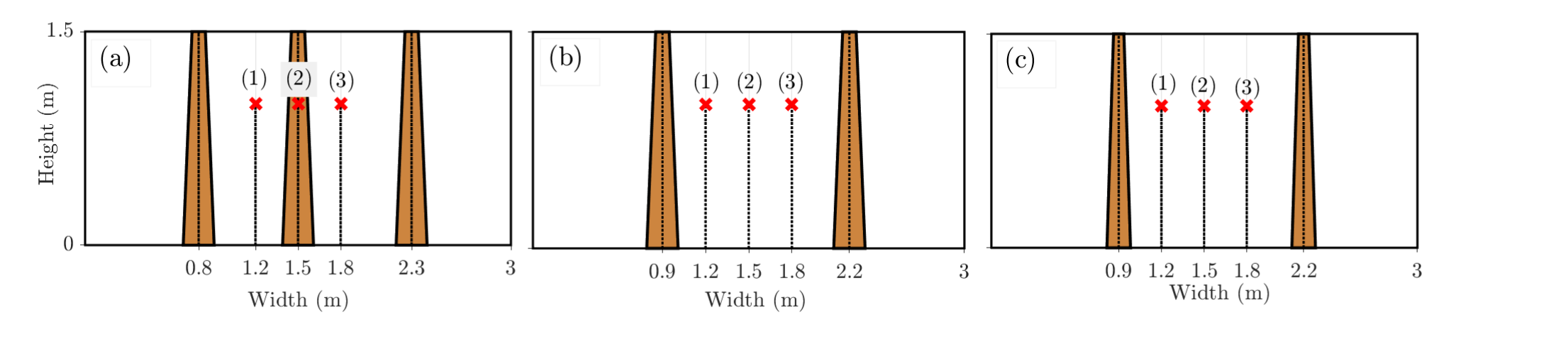}
    \vspace{-0.8cm}
    \caption{Lower TS arrangement (a) 3 spire `Set 1' (b) 2 spire `Set 1' (c) 2 spire `Set 2'}
    \label{fig:lowerts_spires}
\end{figure}

\vspace{-0.5cm}

\begin{figure}[H]
    \centering    \includegraphics[width=1\linewidth]{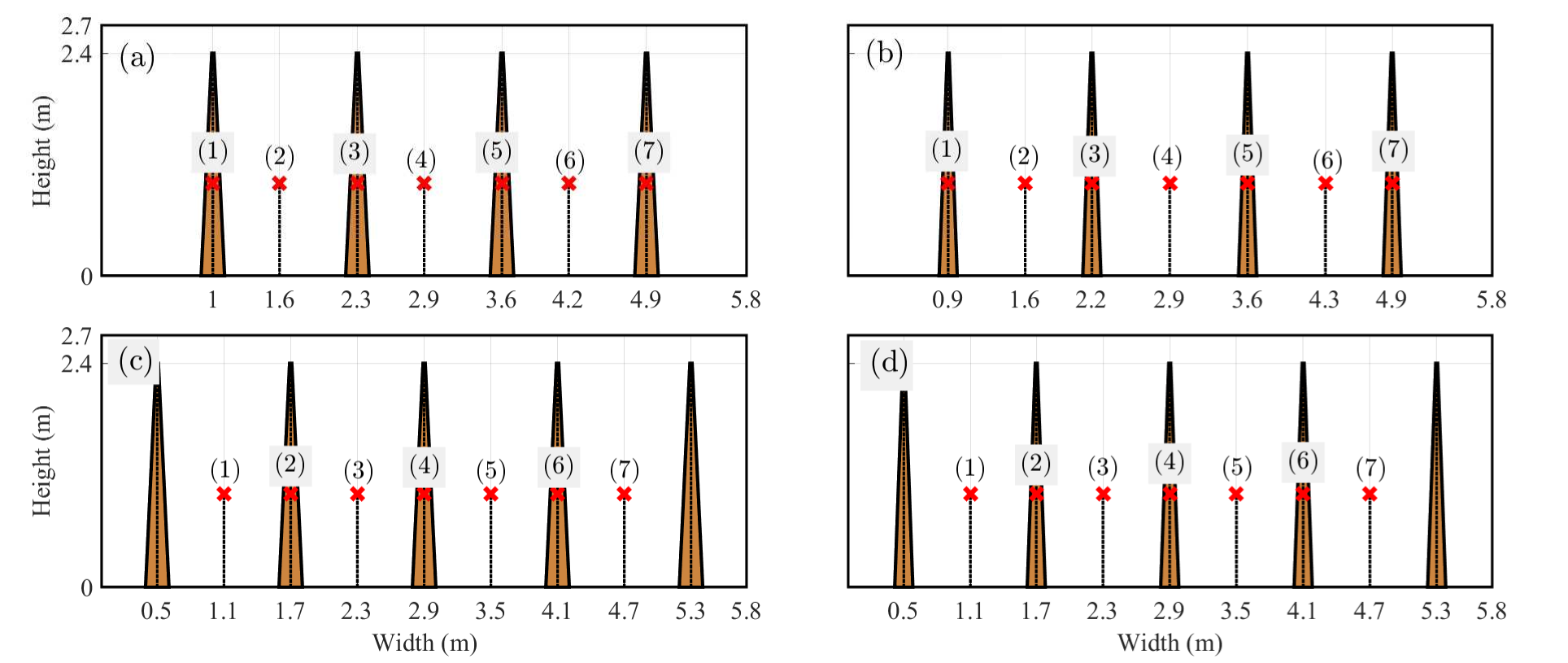}
    \caption{Upper TS arrangement: 4 spires (a) `Set 1' (b) `Set 2'; 5 spires (c) `Set 1' (d) `Set 2'}
    \label{fig:upperts_spires}
\end{figure}
\vspace{-0.6cm}

For experiments carried out in the lower test section, once the sufficient seeding was established, the LDA probe's datum position in the wind tunnel was recorded, serving as the reference point for all the subsequent tests. The lowest measurable height that could be reached using LDA was 72 mm above the tunnel floor, due to limitations introduced by beam obstruction. Scans were performed via vertical traverse and spanwise optical adjustments (Appendix \ref{ldascan}).

Since LDA could not be used in the upper test section due to optical and geometric limitations, the seven-hole probe was chosen as the primary measurement tool for that section.  A pre-calibrated multi-hole probe was initially aligned with a cross-line laser for accurate orientation, following which it was mounted on a traverse system laterally movable using a pre-aligned Minitec rail, shown in Appendix \ref{traverse_setup}. Pressure data from the MHP was recorded using LabVIEW, triggered either by the LDA (lower TS) or the traverse (upper TS) system. A no-flow condition was acquired before each run.

Lastly, before moving to the upper TS, a second phase of the validation campaign was conducted in the lower TS utilising the redesigned traverse/ multi-hole probe holder system, using both 2D LDA and MHP, with the LDA set sufficiently upstream of the probe. This aided in comparing the two measurement techniques and thereby verifying the reliability of the multi-hole probe, whose time-resolved velocities were analysed. A validation test conducted after the traverse redesign confirmed the reliability of the probe for use in the upper section, with results presented in Sections \ref{LDA-MHP-time} and \ref{LDA-MHP}.

\newpage
\section{Results}
\label{results}
\subsection{Lower Test-Section Analysis}

\subsubsection{3 Spires `Set 1'}

`Set 1' design was implemented to replicate `Profile 1'('s) mean velocity and turbulence intensity, as outlined in Section \ref{unify_standards}. The aim of `Profile 1' is to establish a commonality between the various wind standards, thereby providing a reference framework with a velocity of 25 m/s at a height of 10 m.

3 spires of `Set 1' were installed in the lower test section of the wind tunnel. The measurement points, indicated by red crosses in Figure \ref{fig:lowerts_spires}, follow the same spanwise scan location numbering as used in the plots below, to aid in straightforward interpretation. Measurements were taken using 2D LDA, whose spanwise extent was limited by the reach of the traverse system. Here, $U$ refers to the streamwise velocity component, and $W$ denotes the wall-normal velocity.

\begin{figure}[H]
    \centering
    \begin{minipage}{0.47\linewidth}
        \centering
        \includegraphics[width=1.15\linewidth]{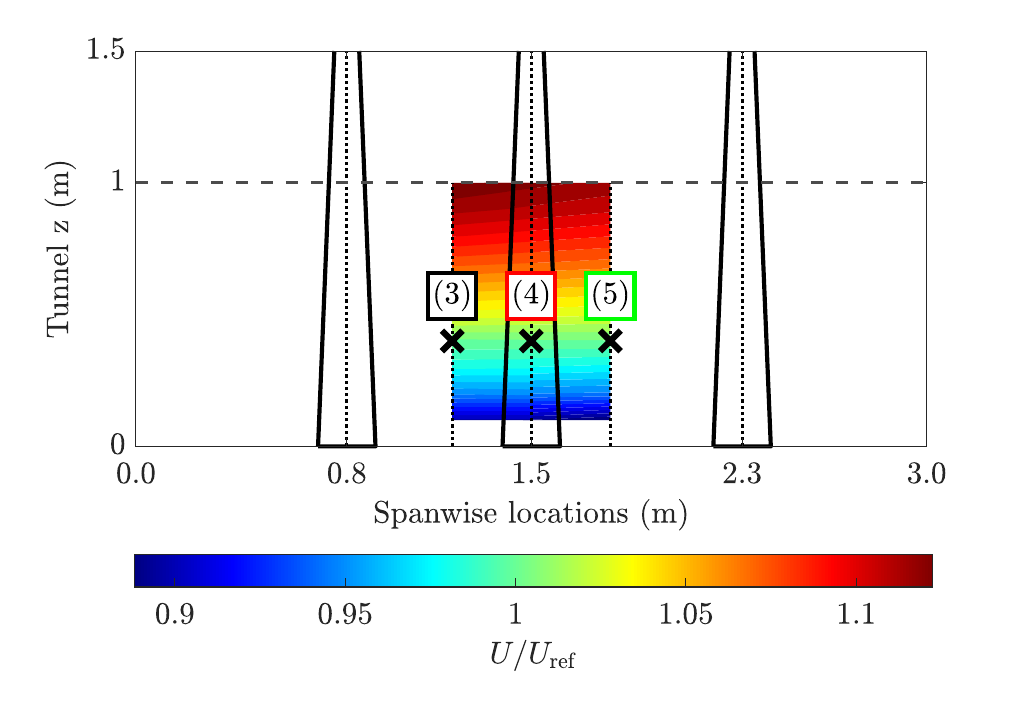}
        \caption{Contour plot of $U/U_{ref}$, using `Set 1' 3 spires, in the lower TS}
        \label{fig:L-set1-3spires-normVel-contour-full}
    \end{minipage}\hfill
    \begin{minipage}{0.47\linewidth}
        \centering
        \includegraphics[width=0.94\linewidth]{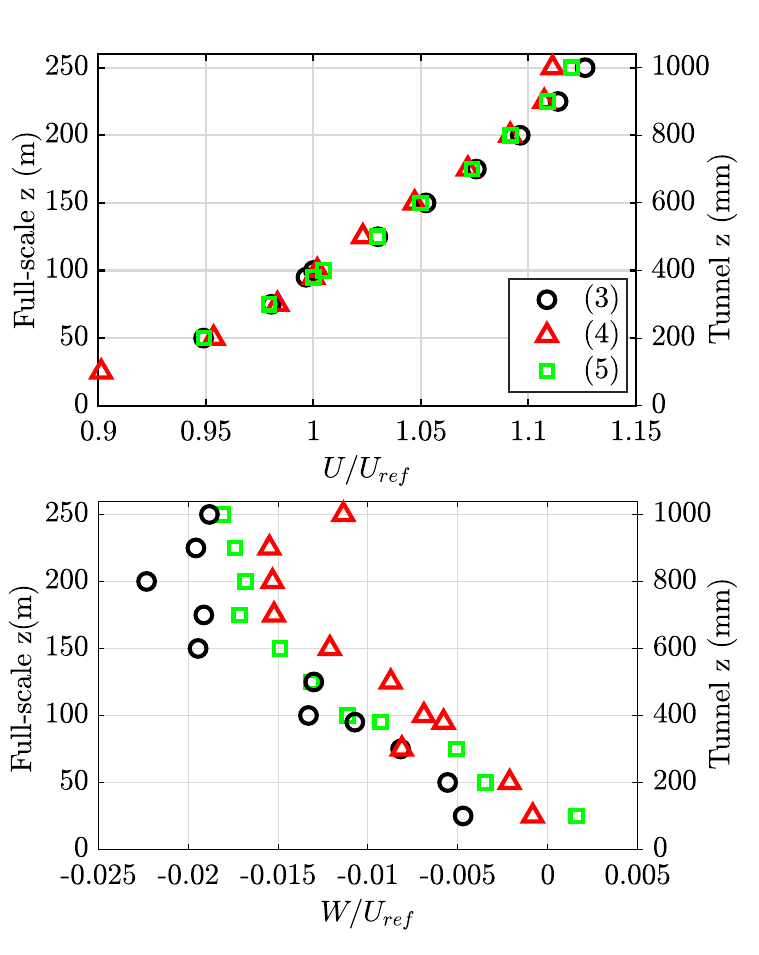}
        \caption{2D plots of $U/U_{ref}$ and $W/U_{ref}$, using `Set 1' 3 spires, in the lower TS}
        \label{fig:L-set1-3spires-normVel-2d}
    \end{minipage}
\end{figure}

Initially, spanwise uniformity in the flow was verified by taking preliminary velocity measurements at three fixed heights along the span. Following this, a complete scan was carried out at the three span positions, where data was acquired up to a height of 1 m, as shown in Figure \ref{fig:L-set1-3spires-normVel-contour-full}. 

The $U$ velocities have been normalised by the reference velocity $U_{ref}$ at the representative hub height for wind turbines off the coast of the North Sea and the English Channel. Hence, for all the test cases to follow, $U_{ref}$ is to be interpreted as $ U_{hub} = U$ at 95 m in full-scale (corresponding to a height of 380 mm off the tunnel floor, in a test-scale of 1:250). For precise comparison, a 2D plot of normalised $U$ and $W$ measurements at the three spanwise locations is demonstrated in Figure~\ref{fig:L-set1-3spires-normVel-2d}.
\vspace{0.7cm}

\begin{figure}[H]
    \centering
    \begin{minipage}{0.48\linewidth}
        \centering
        \vspace{0.2cm}
        \includegraphics[width=1\linewidth]{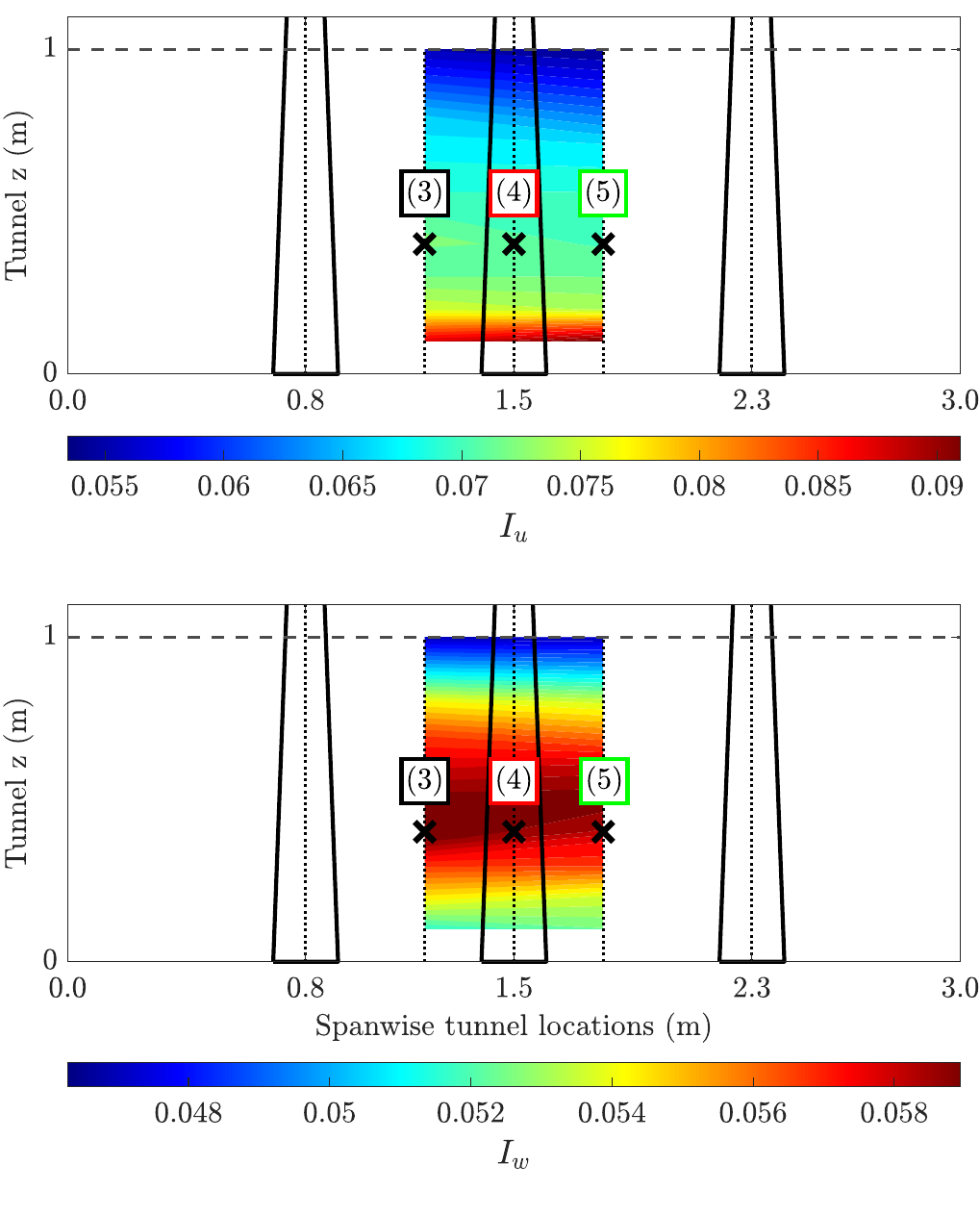}
        \caption{Contour plots of $I_u$ and $I_w$, using `Set 1' 3 spires, in the lower TS}
        \label{fig:L-set1-3spires-turbint-contour}
    \end{minipage}\hfill
    \begin{minipage}{0.46\linewidth}
        \centering
        \includegraphics[width=0.9\linewidth]{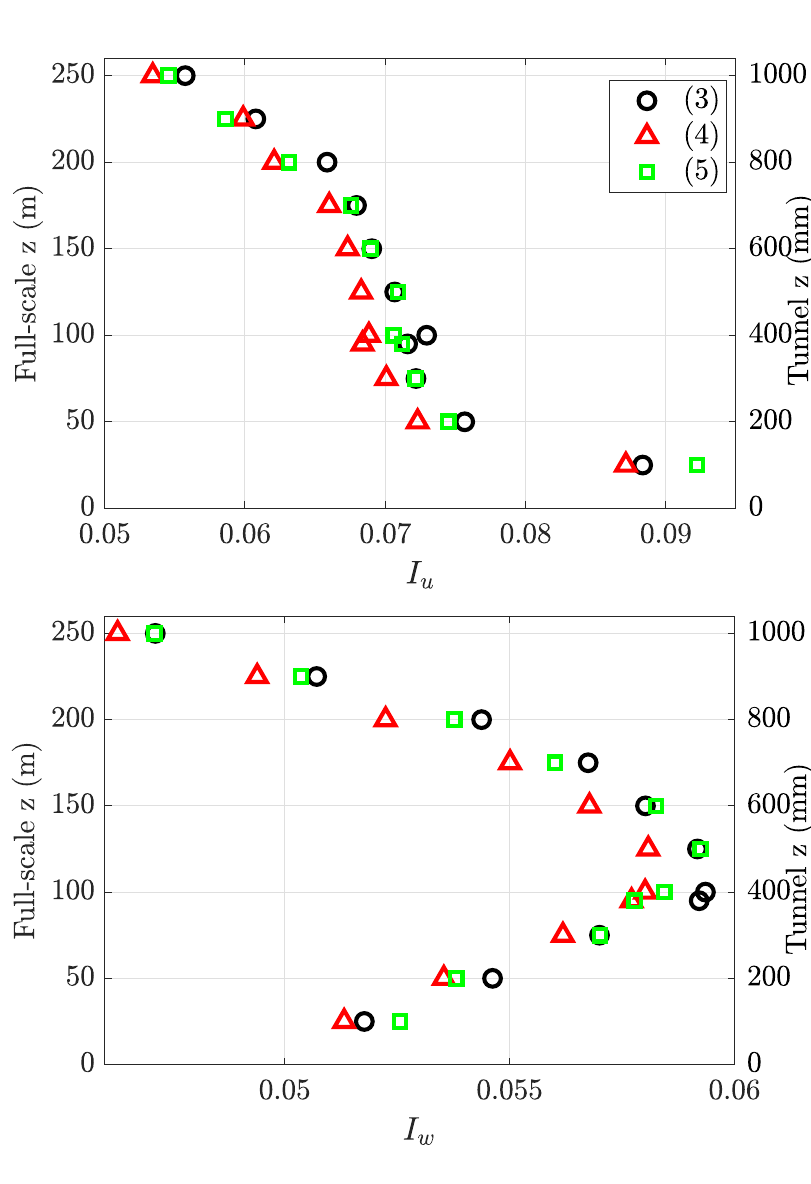}
        \caption{2D plots of $I_u$ and $I_w$, using `Set 1' 3 spires, in the lower TS}
        \label{fig:L-set1-3spires-turbint-2d}
    \end{minipage}
\end{figure}

\newpage
Focus is then narrowed down to representing only the measured 1 m height region of the tunnel test section for convenient analyses, as in Figure~\ref{fig:L-set1-3spires-turbint-contour}, showing the variation of turbulence intensities in $U$ and $W$, i.e., $I_u$ and $I_w$ respectively, computed using Equation \ref{IU}. Similarly, a 2D plot as in Figure~\ref{fig:L-set1-3spires-turbint-2d} also represents the variation in $I_u$ and $I_w$ across the three spanwise stations.
\vspace{0.7cm}

\vspace{0.35cm}
\subsubsection{2 Spires `Set 1'}

Following from the previous test case, two `Set 1' spires were installed in the lower test section. Figure \ref{fig:lowerts_spires} shows their positions and the LDA measurement points, with a matching numbering system used in the plots below.

Spanwise uniformity was assessed using 2D LDA measurements, supplemented by a multi-hole probe test case at the centre for comparing the two techniques, and to validate the use of multi-hole probe without the LDA. Initial results showed unexpected discrepencies due to the blockage effect caused by the MHP traverse (set-up shown in Appendix \ref{oldtraverse}, \ref{refoldtraverse}), which also affected the LDA measurements taken at the tunnel centre location, since the LDA's measurement point was too close to the MHP. The resulting measurements obtained from this test is shown in Appendix \ref{badlda}. However, these blockage effects were tackled and eliminated during the validation stage, by redesigning the probe holder and the traverse set-up. The comparison results verifying that the MHP is capable of taking measurements as close to LDA as possible is discussed in the upcoming Sections \ref{LDA-MHP-time} and \ref{LDA-MHP}.  

\begin{figure}[H]
    \centering
    \begin{minipage}{0.48\linewidth}
        \centering
        \includegraphics[width=0.97\linewidth]{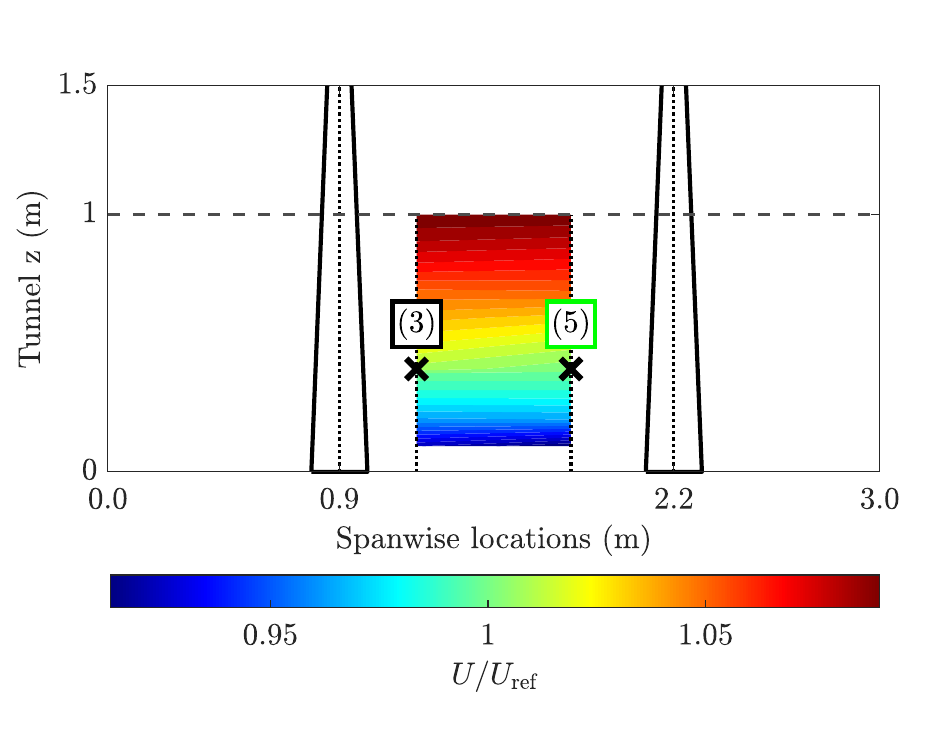}
        \caption{Contour plot of $U/U_{ref}$ using `Set 1' 2 spires, in the lower TS}
        \label{fig:L-set1-2spires-normVel-contour-full}
    \end{minipage}\hfill
    \begin{minipage}{0.48\linewidth}
        \centering
        \includegraphics[width=0.87\linewidth]{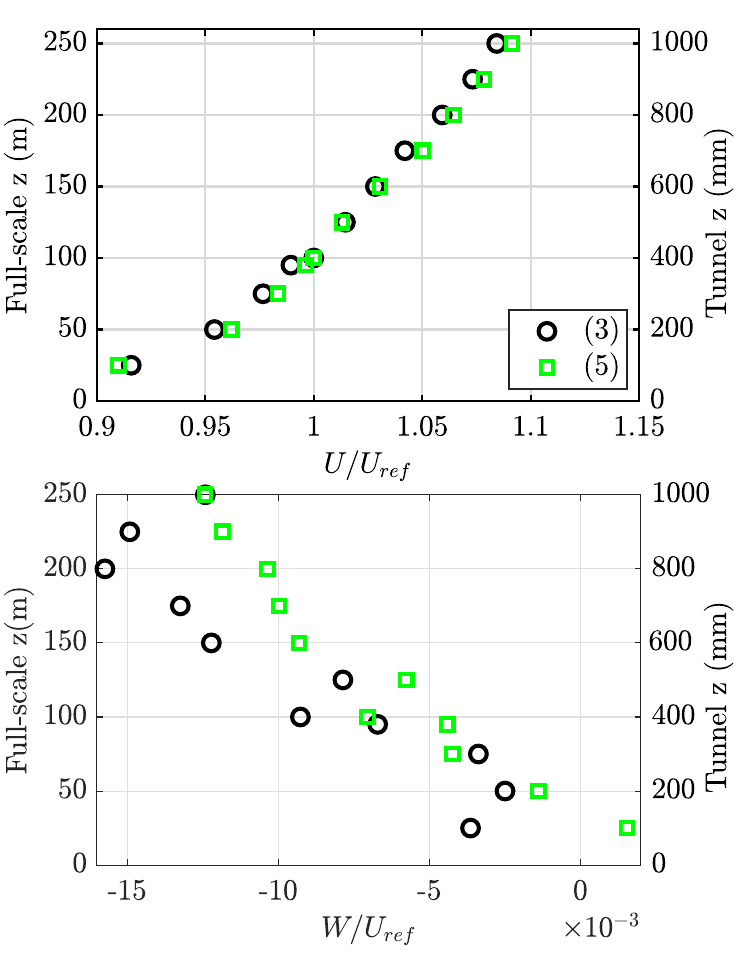}
        \caption{2D plots of $U/U_{ref}$ and $W/U_{ref}$ using `Set 1' 2 spires, in the lower TS}
        \label{fig:L-set1-2spires-normVel-2d}
    \end{minipage}
\end{figure}

The LDA measurements, with the aforementioned traverse removed (and hence free from the unexpected data), for the normalised velocities are depicted in Figures \ref{fig:L-set1-2spires-normVel-contour-full}, and \ref{fig:L-set1-2spires-normVel-2d}, via the contour, and the 2D plots respectively. 

\begin{figure}[H]
    \centering
    \begin{minipage}{0.48\linewidth}
        \centering
        \includegraphics[width=0.95\linewidth]{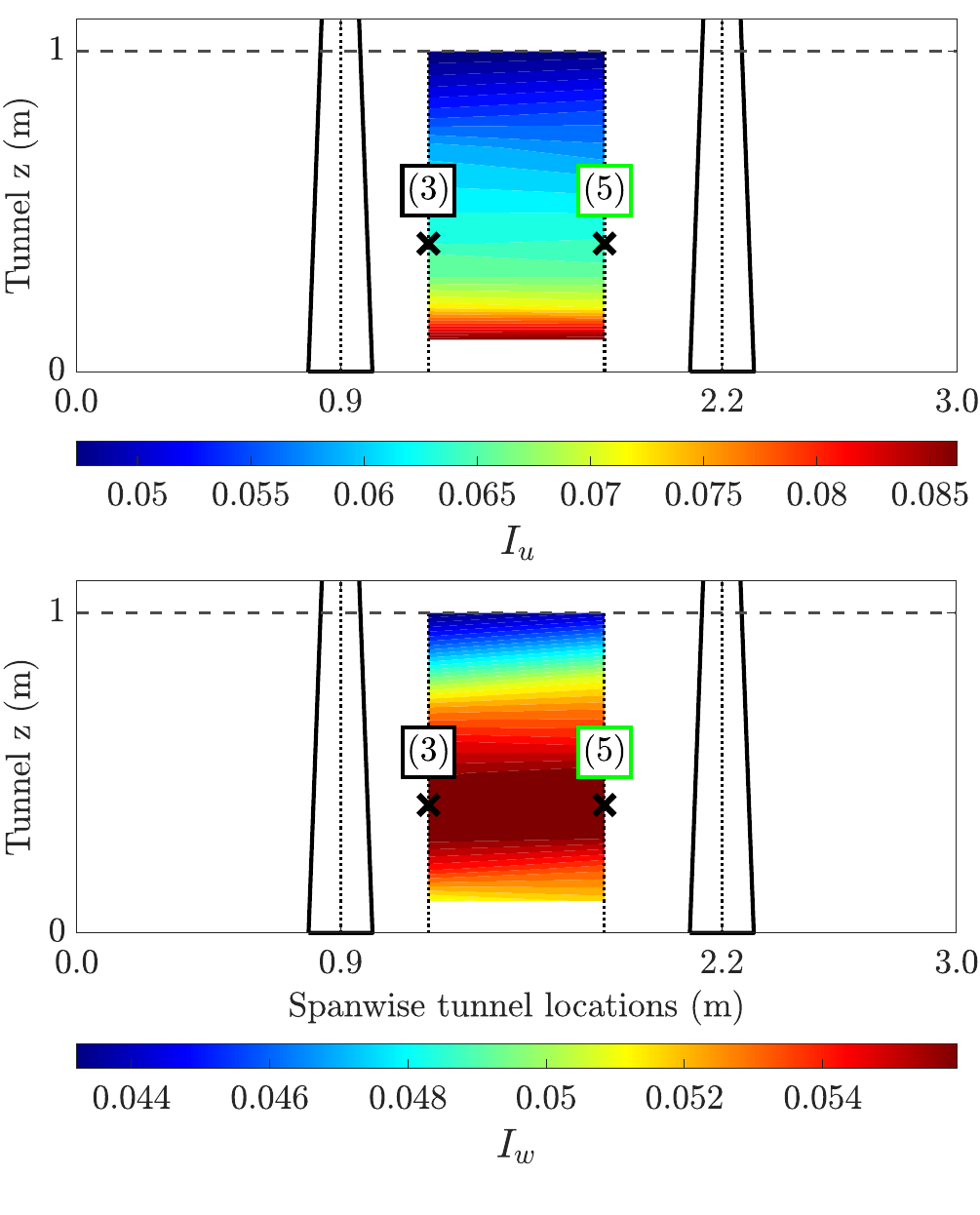}
        \caption{Contour plots of $I_u$ and $I_w$ using `Set 1' 2 spires, in the lower TS}
        \label{fig:L-set1-2spires-turbint-contour}
    \end{minipage}\hfill
    \begin{minipage}{0.46\linewidth}
        \centering
        \includegraphics[width=0.92\linewidth]{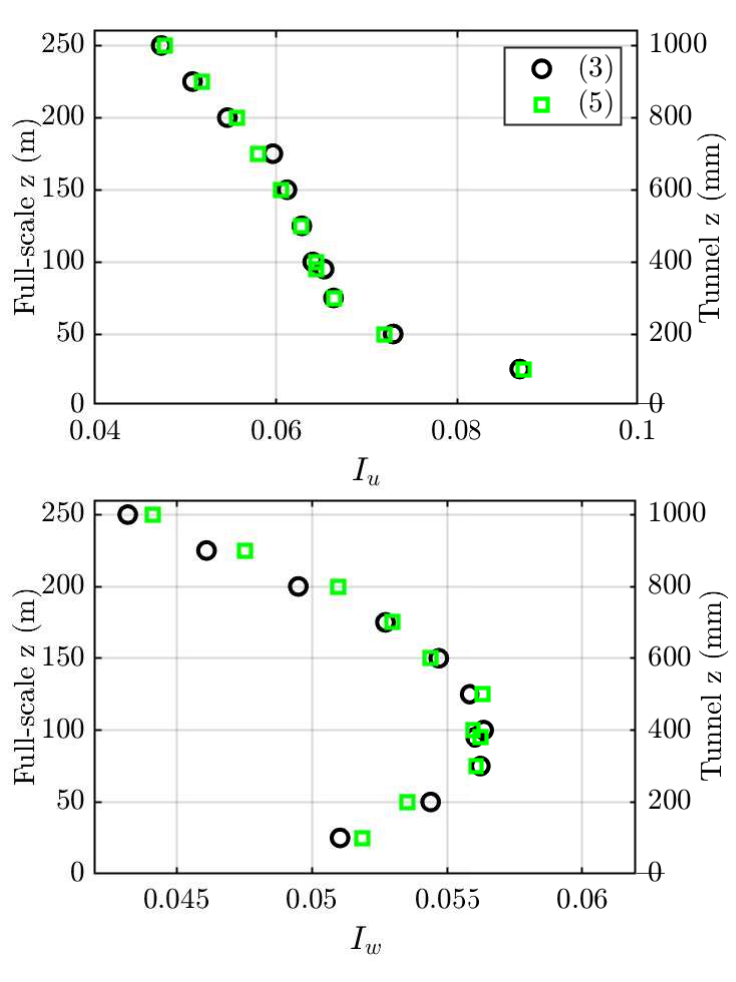}
        \caption{2D plots of $I_u$ and $I_w$ using `Set 1' 2 spires, in the lower TS}
        \label{fig:L-set1-2spires-turbint-2d}
    \end{minipage}
\end{figure}

Similarly, the flow uniformity in the $I_u$, and the $I_w$ values in the two (unaffected) spanwise readings, can be seen in Figures \ref{fig:L-set1-2spires-turbint-contour}, and \ref{fig:L-set1-2spires-turbint-2d}.

\newpage
\subsubsection{2 Spires `Set 1' with Roughness}

\begin{figure}[H]
    \centering
    \begin{minipage}{.47\linewidth}
        \centering
        \includegraphics[width=1.05\linewidth]{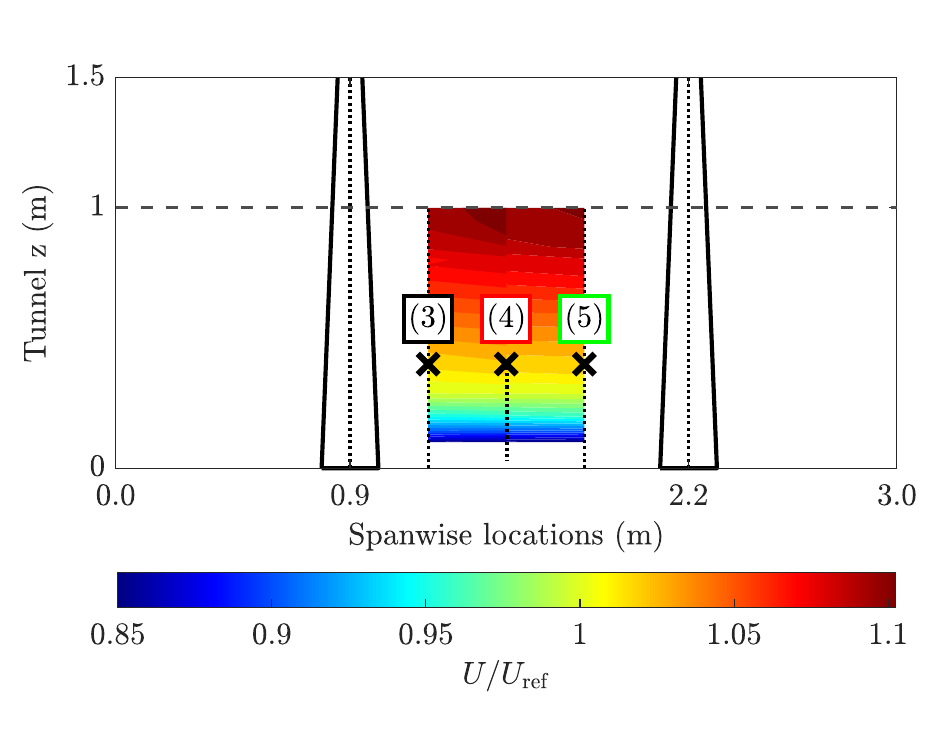}
        \caption{Contour plot of $U/U_{ref}$, using `Set 1' 2 spires+green mesh, in the lower TS}
        \label{fig:L-set1-2spiresR-normVel-contour}
    \end{minipage}\hfill
    \begin{minipage}{.47\linewidth}
        \centering
        \includegraphics[width=0.9\linewidth]{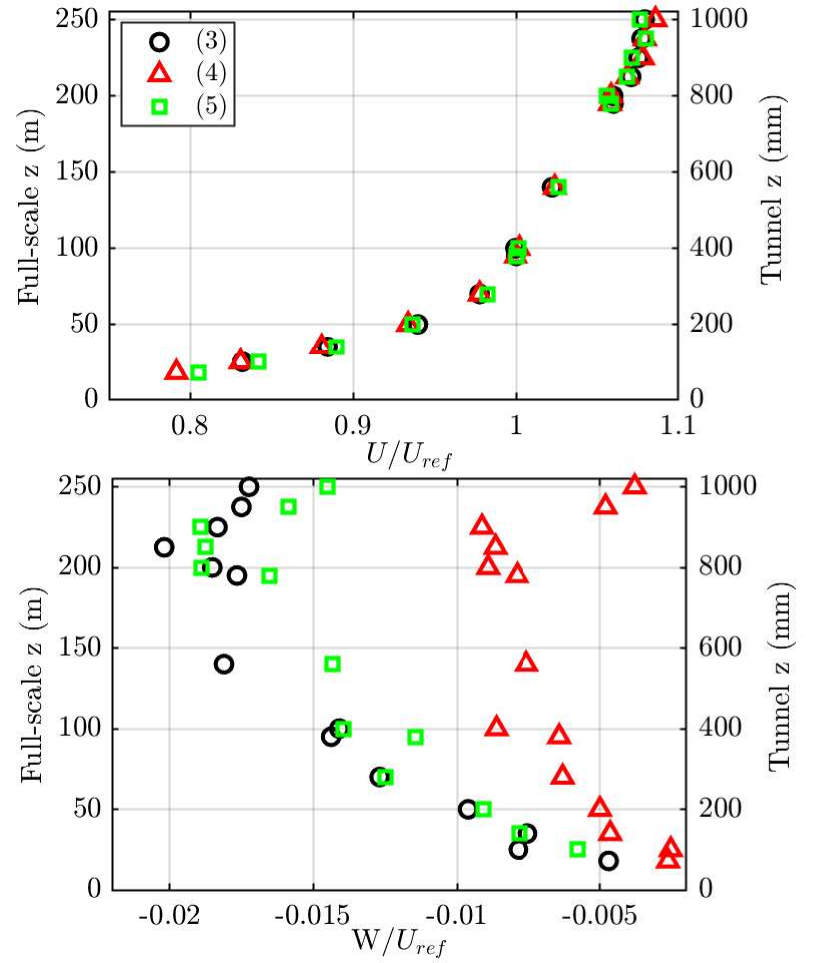}
        \caption{2D plots of $U/U_{ref}$, and $W/U_{ref}$, using `Set 1' 2 spires+green mesh, in the lower TS}
        \label{fig:L-set1-2spiresR-normVel-2d}
    \end{minipage}
\end{figure}

\begin{figure}[H]
    \centering
    \begin{minipage}{0.48\linewidth}
        \centering
        \vspace{0.4cm}
        \includegraphics[width=1\linewidth]{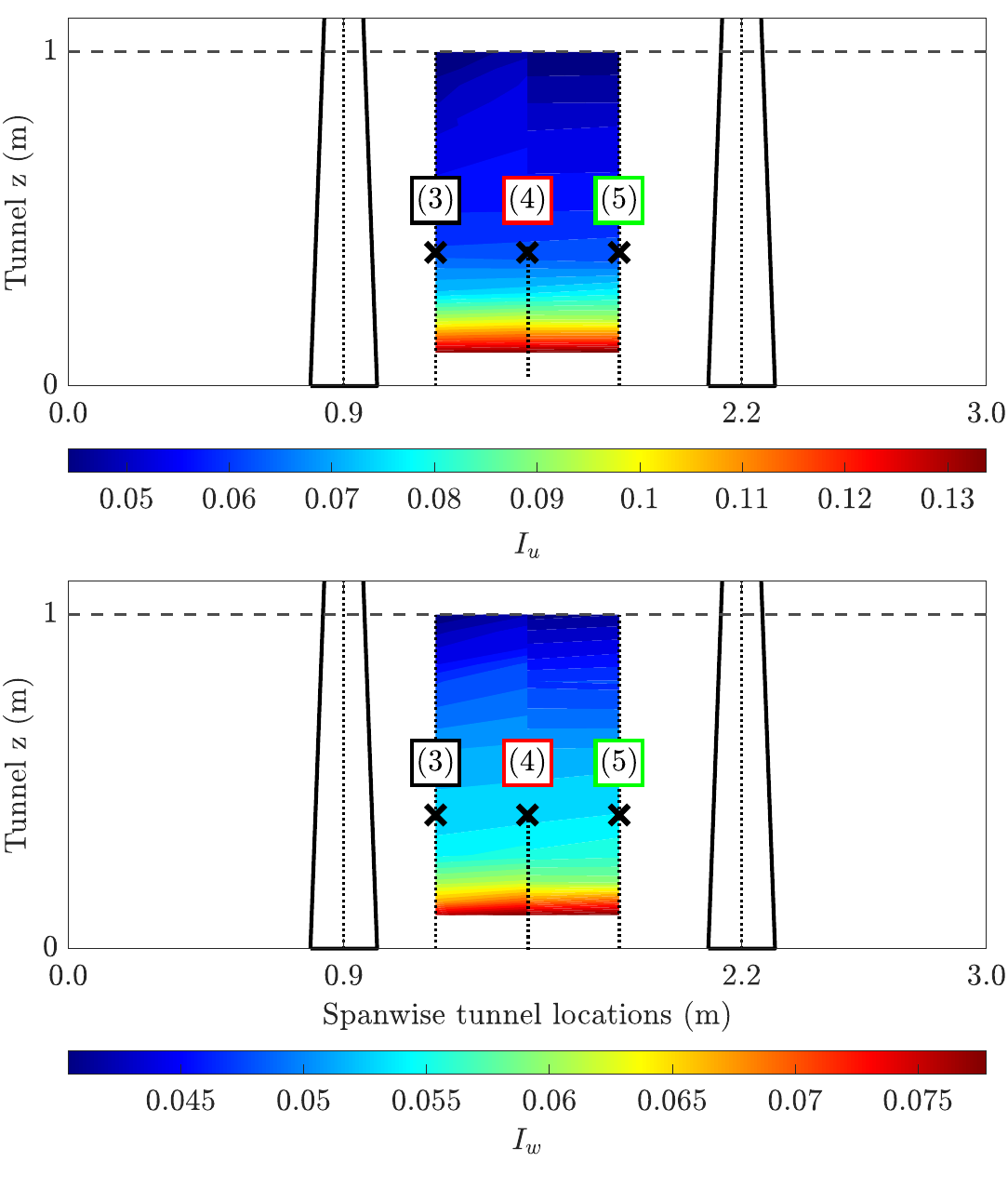}
        \caption{Contour plots of $I_u$ and $I_w$ using `Set 1' 2 spires+green mesh, in the lower TS}
        \label{fig:L-set1-2spiresR-turbint-contour}
    \end{minipage}\hfill
    \begin{minipage}{0.46\linewidth}
        \centering
        \includegraphics[width=0.92\linewidth]{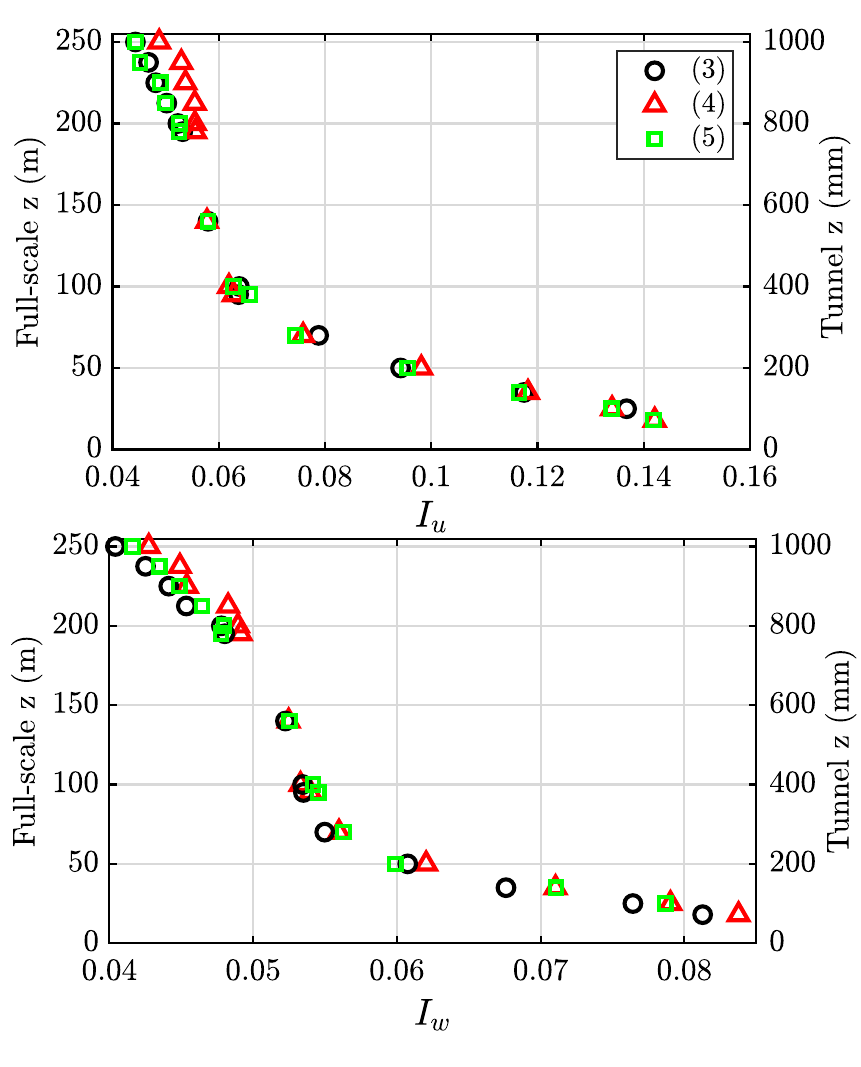}
        \caption{2D plots of $I_u$ and $I_w$ using `Set 1' 2 spires+green mesh, in the lower TS}
        \label{fig:L-set1-2spiresR-turbint-2d}
    \end{minipage}
\end{figure}

A modified setup was implemented, based on the earlier 2-spire configuration, with the same spacing as before, but with added roughness elements (green mesh). This modification was aimed to obtain a better match with the target turbulence intensity profile, discussed in Section \ref{R}. Flow uniformity in the wind tunnel was establishged from the plots of normalised $U$, and $W$, depicted in Figures \ref{fig:L-set1-2spiresR-normVel-contour}, and \ref{fig:L-set1-2spiresR-normVel-2d}. Additionally, turbulence intensity profiles for the $U$ and $W$ components are also visualised via similar plots, as seen in Figures \ref{fig:L-set1-2spiresR-turbint-contour}, and \ref{fig:L-set1-2spiresR-turbint-2d}. 

\vspace{-0.5cm}

\vspace{-1cm}

\subsubsection{2 Spires `Set 2'}

The `Set 2' spires, slightly narrower than `Set 1', were developed specifically to produce `Profile WF', a velocity profile that is typical of incident on a wind farm, for example, in the English Channel. 

\begin{figure}[H]
    \centering
    \begin{minipage}{0.49\linewidth}
        \centering
        \includegraphics[width=0.9\linewidth]{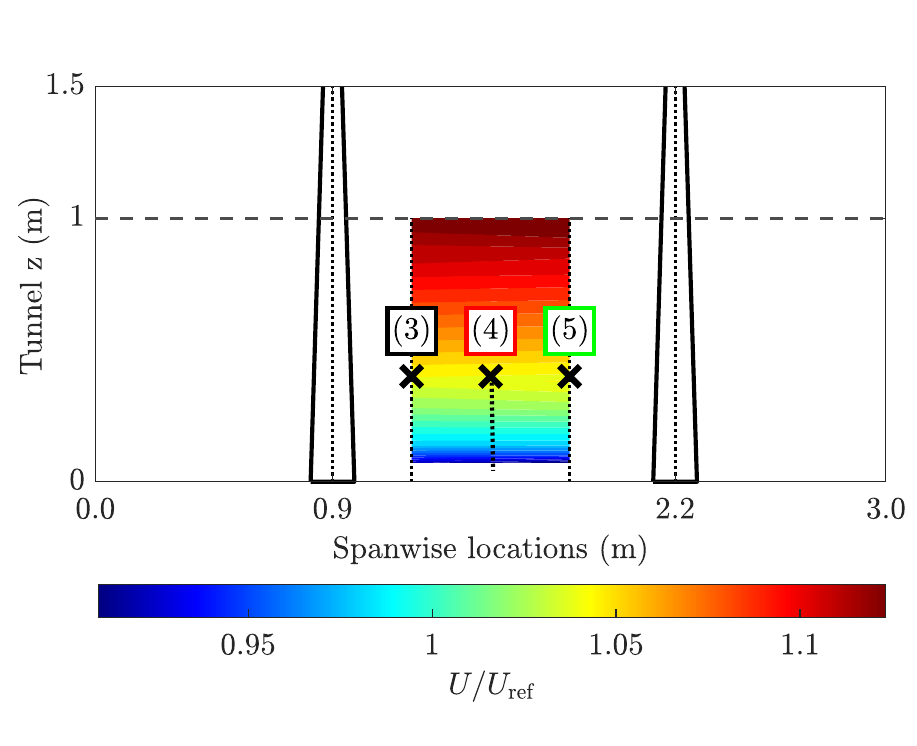}
        \vspace{-0.6cm}
        \caption{Contour plot of $U/U_{ref}$ using `Set 2' 2 spires, in the lower TS}
        \label{fig:L-set2-2spires-normVel-contour}
    \end{minipage}
    \hfill
    \begin{minipage}{0.49\linewidth}
        \centering
        \includegraphics[width=0.795\linewidth]{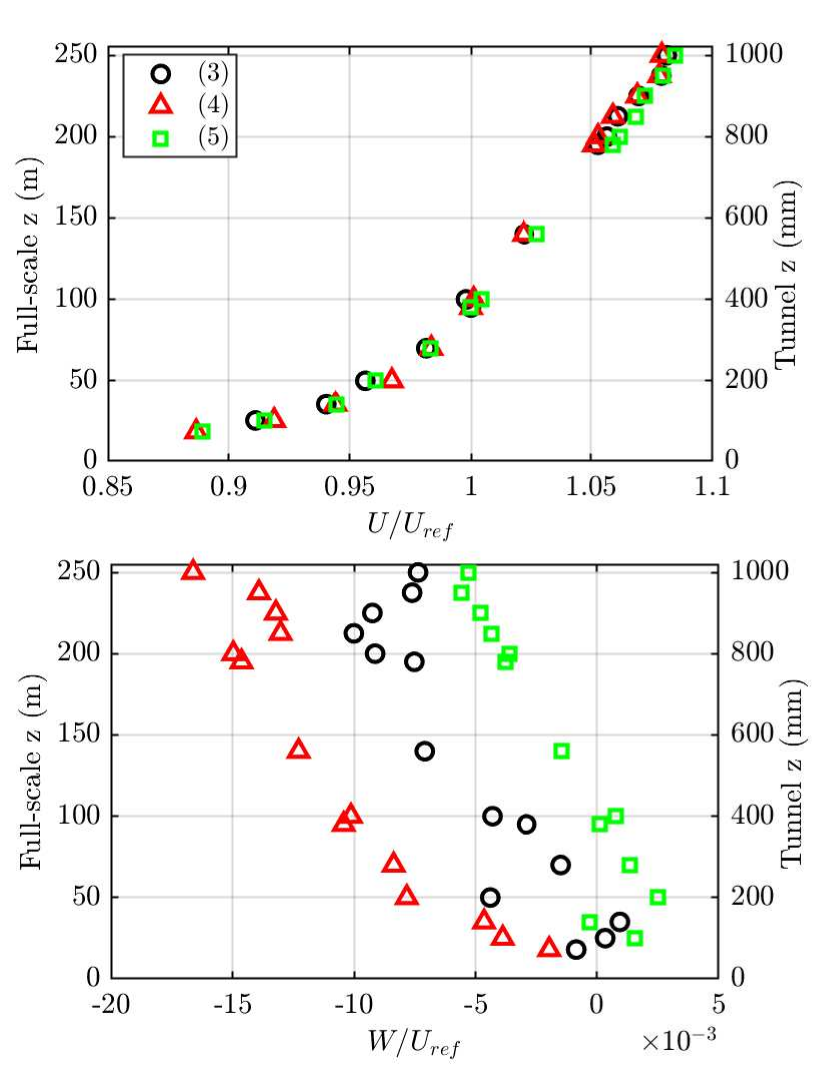}
        \caption{2D plots of $U/U_{ref}$ and $W/U_{ref}$ using `Set 2' 2 spires, in the lower TS}
        \label{fig:L-set2-2spires-normVel-2d}
    \end{minipage}
\end{figure}

\begin{figure}[H]
    \centering
    \begin{minipage}{.48\linewidth}
        \centering
        \includegraphics[width=0.87\linewidth]{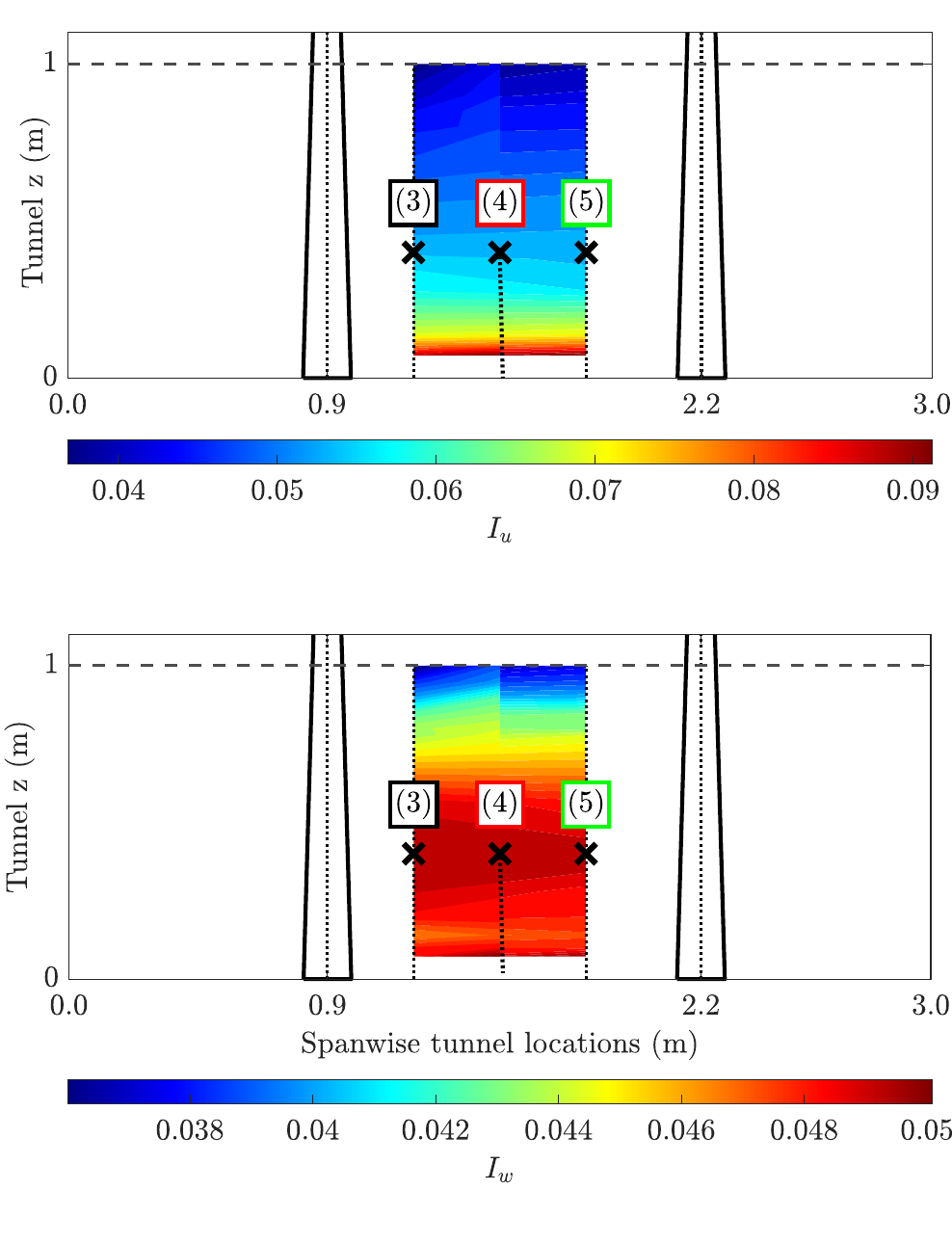}
        \vspace{-0.5cm}
        \caption{Contour plots of $I_u$, and $I_w$, using `Set 2' 2 spires, in the lower TS}
        \label{fig:L-set2-2spires-turbint-contours}
    \end{minipage}\hfill
    \begin{minipage}{.46\linewidth}
        \centering
        \includegraphics[width=0.87\linewidth]{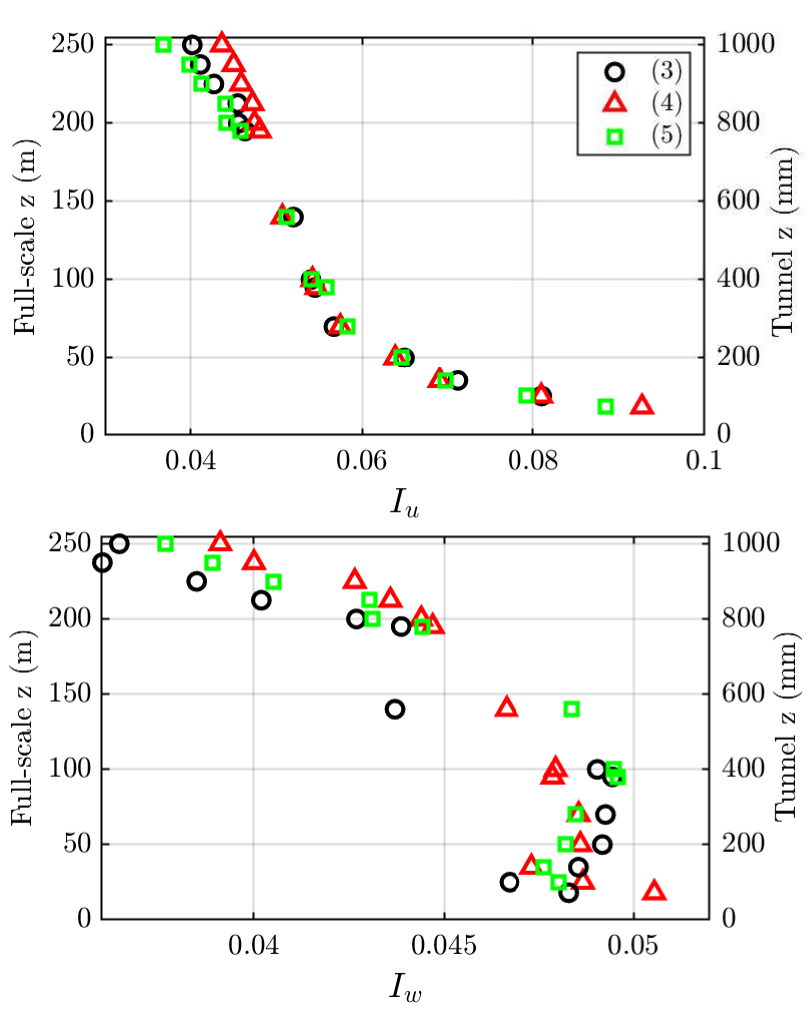}
        \caption{2D plots of $I_u$, and $I_w$, using `Set 2' 2 spires, in the lower TS}
        \label{fig:L-set2-2spires-turbint-2d}
    \end{minipage}
\end{figure}

The spires' placement, and the LDA measurement points in the lower test section are shown in Figure \ref{fig:lowerts_spires}, following the same numbering system as the plots shown below. The normalised profile of $U$ is presented with respect to the full tunnel test section via Figure~\ref{fig:L-set2-2spires-normVel-contour}, as well as 2D plots in Figure~\ref{fig:L-set2-2spires-normVel-2d} to showcase the extent of flow uniformity at the LDA sampling locations. Furthermore, turbulence intensities in both $U$ and $W$ are visualised by means of Figures~\ref{fig:L-set2-2spires-turbint-contours} and~\ref{fig:L-set2-2spires-turbint-2d}.

\vspace{-0.5cm}

\subsubsection{LDA vs Multi-hole Probe}
\label{LDA-MHP-time}
\begin{figure}[H]
    \centering
    \begin{minipage}{0.47\linewidth}
        \centering
        \includegraphics[width=1\linewidth]{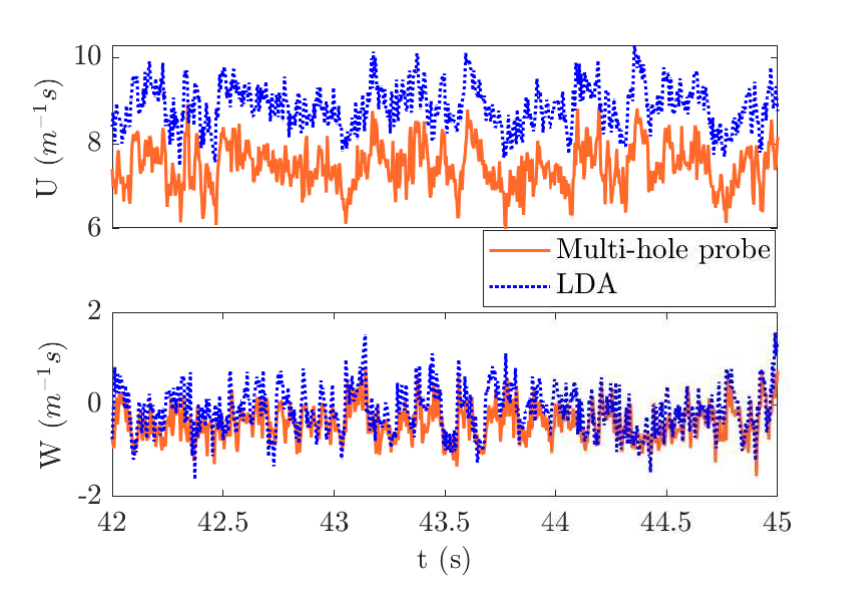}
        \caption{Time series of LDA vs MHP, using the original traverse system}
        \label{fig:LDA-MHP-timeseries-old}
    \end{minipage}\hfill
    \begin{minipage}{0.47\linewidth}
        \centering
        \includegraphics[width=1.03\linewidth]{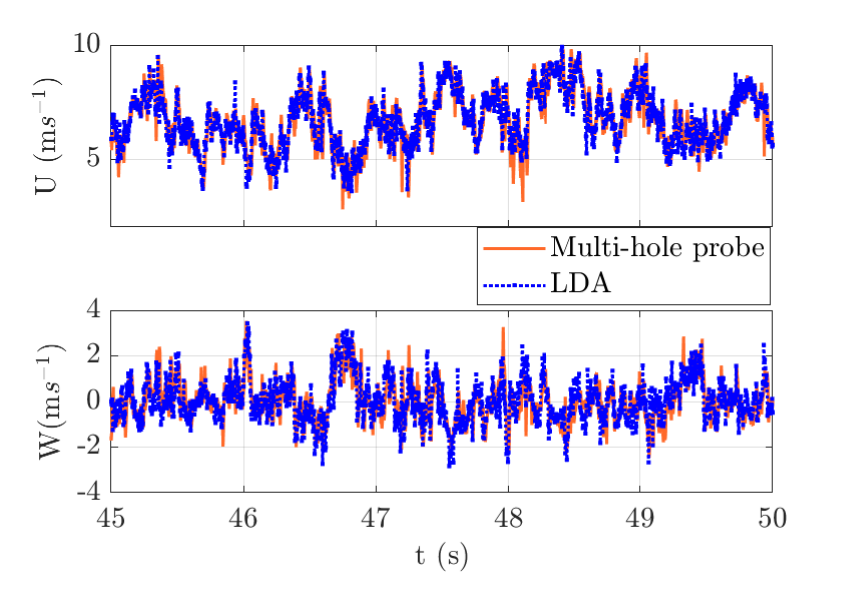}
        \caption{Time series of LDA vs MHP, using the modified traverse system}
        \label{fig:LDA-MHP-timeseries-new}
    \end{minipage}
\end{figure}

Validating the multi-hole probe against the LDA in the lower test section was pivotal to ensure reliable measurements in the upper test section, where only the multi-hole probe was available. A preliminary test was performed with the 2 spires of `Set 1', involving both LDA and multi-hole probe, but proper validation was hindered due to blockage effects from the traverse, leading to a redesign of the traverse system. This has been detailed in Section \ref{probedes}. Comparison of the time series data recorded before and after the redesign, has been presented in Figures \ref{fig:LDA-MHP-timeseries-old} and \ref{fig:LDA-MHP-timeseries-new}, respectively. In the time series shown in Figure \ref{fig:LDA-MHP-timeseries-old}, the mean value recorded by the multihole probe appears to be shifted lower, which is attributed to a blockage effect caused by the static probe being positioned too close to the traverse in the original setup. Whereas, Figure \ref{fig:LDA-MHP-timeseries-new} demonstrates the multi-hole probe's effectiveness in tracking the $U$, and $W$ velocities, portraying good agreement with the LDA measurements (in the modified traverse set-up). Further analysis of the same is presented in Section \ref{LDA-MHP}. 

% Furthermore, spectral analysis confirms that all the critical turbulence energy components are captured, which when compared with the LDA measurements verify the efficacy of the calibration procedure and the data processing methods followed. 

\subsection{Upper Test-Section Analysis}

Similar experiments were implemented in the upper test-section, reflecting those done in the lower test-section using the `Set 1' and `Set 2' spires, following successful validation of the multi-hole probe via comparison with the LDA in the lower test-section. 

As with the previous setup, the `Set 1' spires are indented to generate `Profile 1' (framework marine profile to establish commonality across all the reviewed standards), while the `Set 2' spires are designed to replicate the `Profile WF' (i.e., Profile Wind Farm, which is suitable for modelling wind farms in English Channel and the North Sea). Further design details of the spires implemented in the upper test-section can be found in Section \ref{spire_design}.

\subsubsection{5 Spires `Set 1'}

In the upper test-section, a `Set 1' 5 spire configuration was implemented, where all the measurements were taken using the multi-hole probe. The spire arangement and the spanwise measurement locations are illustrated in Figure \ref{fig:upperts_spires}, with the same numbering convention being followed in the plots below. All the labelled spanwise locations were measured in this test configuration. Here, $U$ represents the streamwise velocity, $V$ is the spanwise velocity, and $W$ is the wall-normal velocity.

\begin{figure}[H]
    \centering    \includegraphics[width=0.75\linewidth]{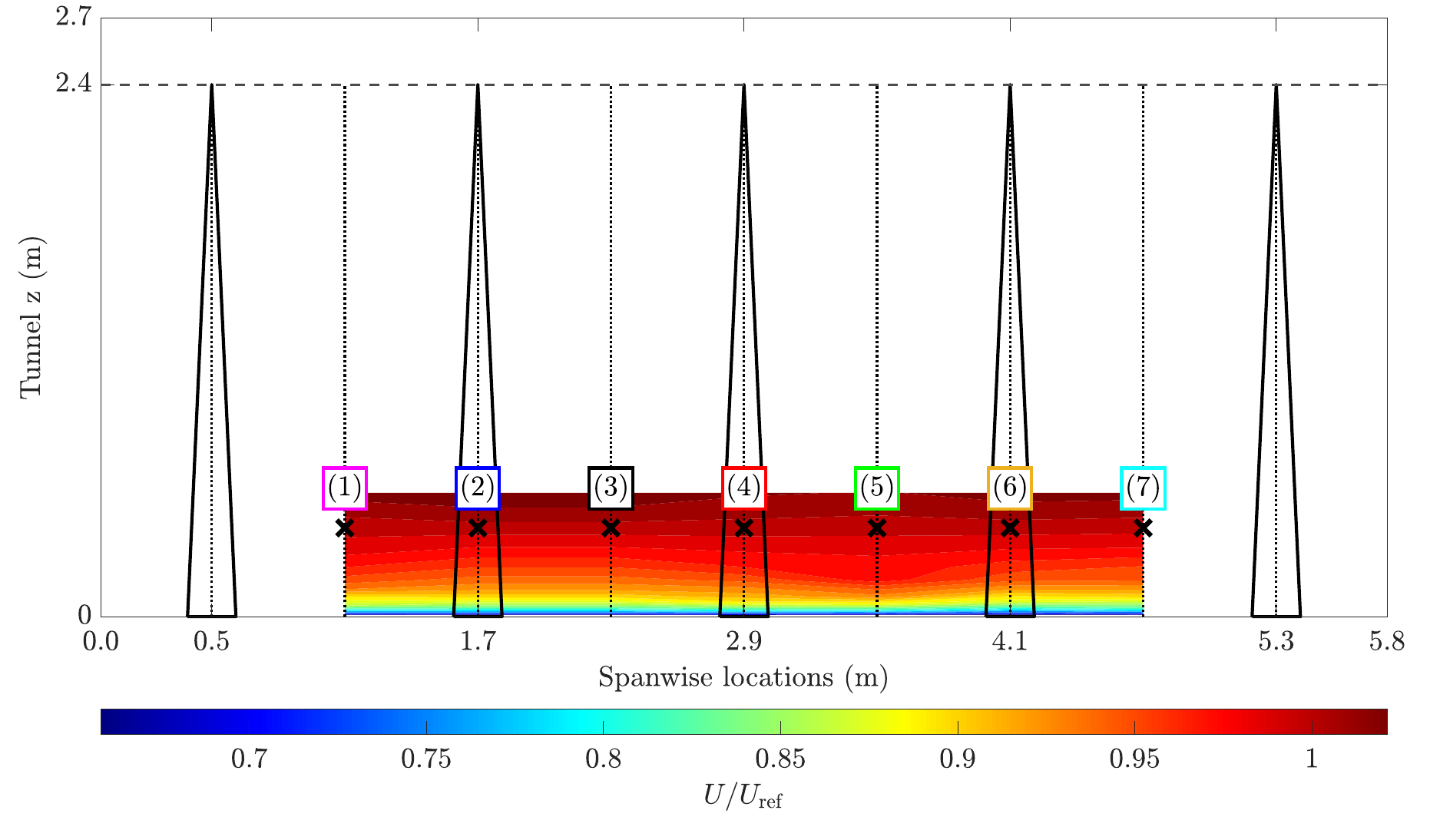}
    \caption{Contour plot of $U/U_{ref}$, using `Set 1' 5 spires, in the upper TS, relative to the test section dimensions}
    \label{fig:U-set1-5spires-normVel-contour-full}
\end{figure}

\begin{figure}[H]
    \centering    \includegraphics[width=0.8\linewidth]{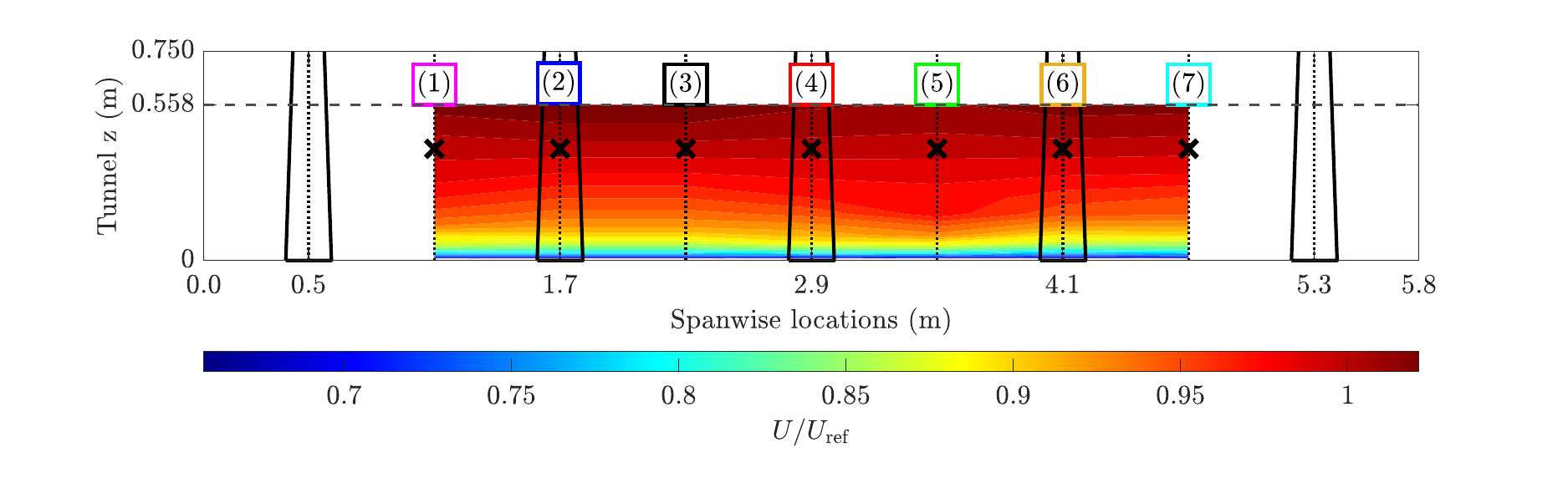}
    \vspace{-0.5cm}
    \caption{Contour plot of $U/U_{ref}$, using `Set 1' 5 spires, in the upper TS}
    \label{fig:U-set1-5spires-normVel-contour-part}
\end{figure}

\begin{figure}[H]
\hspace{5cm}
    \centering    
\includegraphics[width=1.1\linewidth]{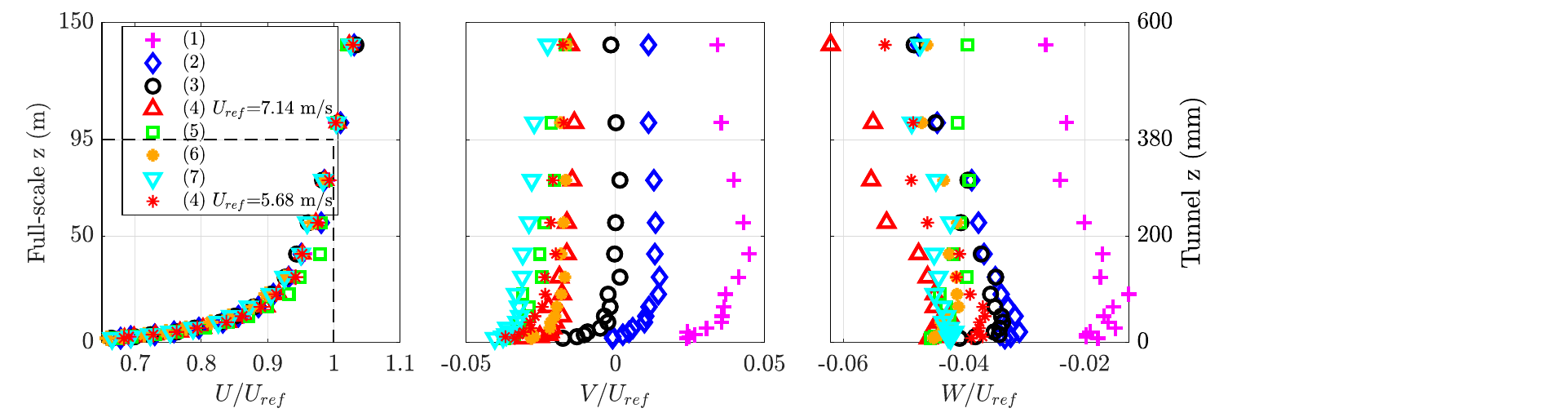}
    \caption{2D plot of $U/U_{ref}$,  $V/U_{ref}$, and  $W/U_{ref}$, using `Set 1' 5 spires, in the upper TS}
    \label{fig:U-set1-5spires-normVel-2d}
\end{figure}

\vspace{-0.5cm}
The measurement range of the multi-hole probe, relative to the wind tunnel section 2.7 m high by 5.8 m wide, along with the spanwise uniformity in the normalised $U$ velocity ($U/U_{ref}=U/U_{hub}$) is shown in Figure \ref{fig:U-set1-5spires-normVel-contour-full}. A zoomed in version of Figure \ref{fig:U-set1-5spires-normVel-contour-full}, focusing on the vertical measurement from the floor until a height of 558 mm is shown in Figure \ref{fig:U-set1-5spires-normVel-contour-part}. 2D plots have been employed to observe similar information, as seen in Figure \ref{fig:U-set1-5spires-normVel-2d}. It should also be noted that the tests at spanwise location (4) were performed at two different test speeds, corresponding to the $U$ at hub height (95 m in full-scale, representative of the wind turbine farms in the British Channel), i.e., $U_{ref}=7.14$ and $5.68$ m/s.

\begin{figure}[H]
    \centering    \includegraphics[width=0.8\linewidth]{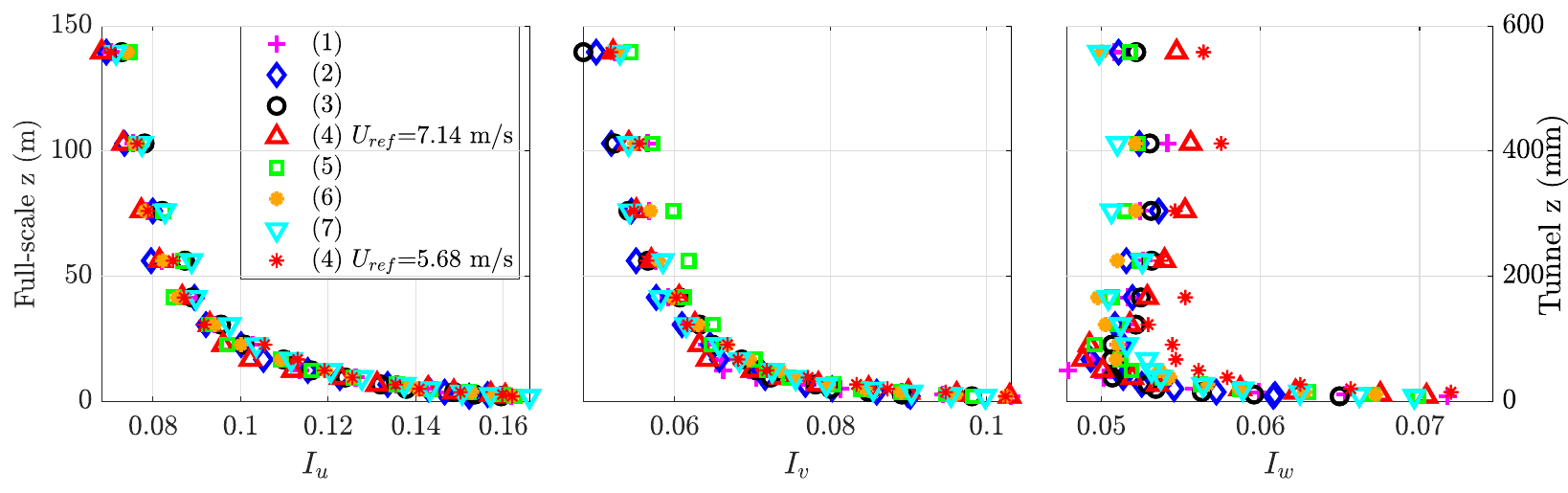}
    \caption{2D plots of $I_u$, $I_v$, and $I_w$, using `Set 1' 5 spires, in the upper TS}
    \label{fig:U-set1-5spires-turbint-2d}
\end{figure}

\begin{figure}[H]
    \centering    \includegraphics[width=0.77\linewidth]{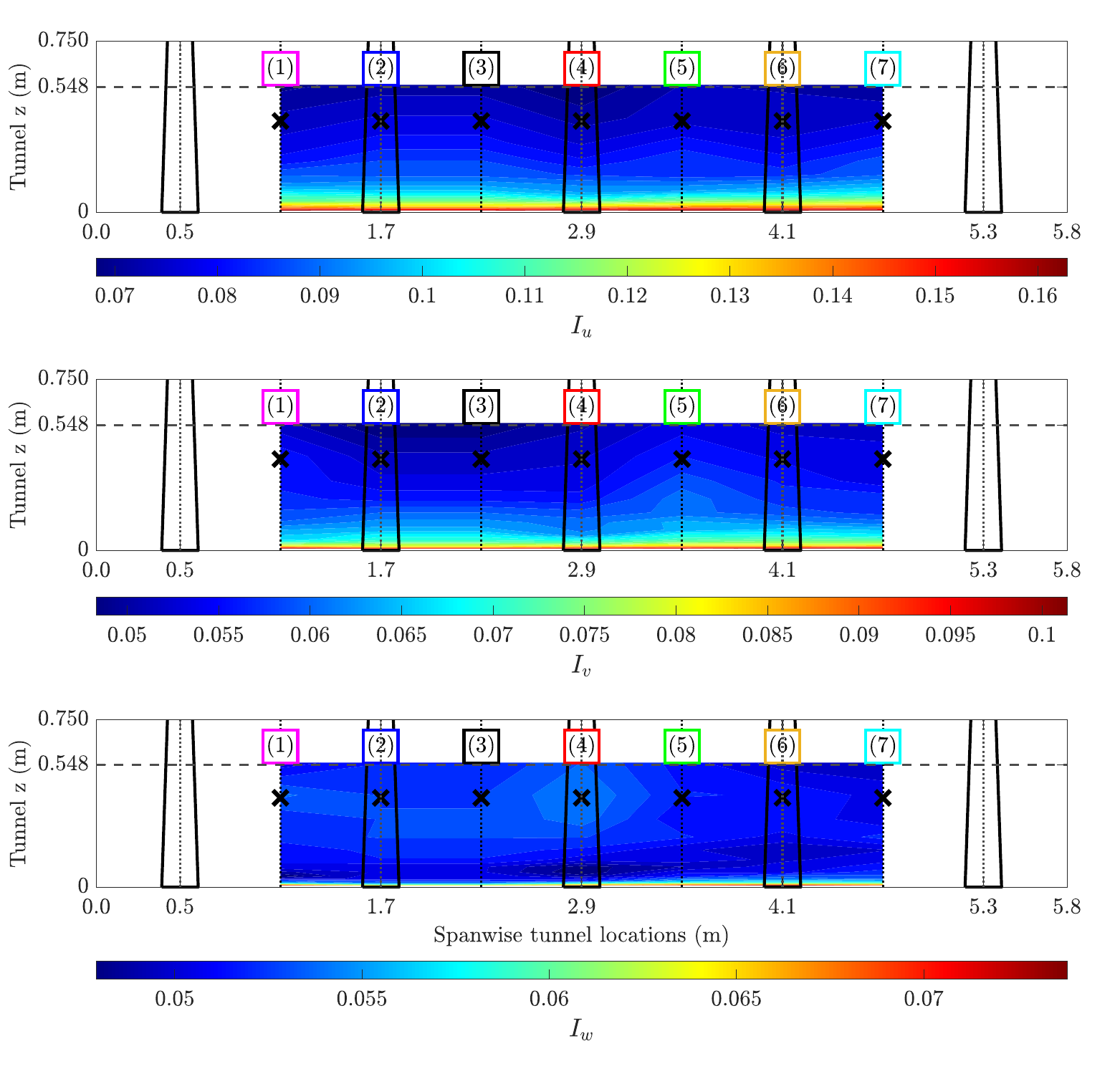}
    \vspace{-0.5cm}
    \caption{Contour plots of $I_u$, $I_v$, and $I_w$, using `Set 1' 5 spires, in the upper TS}
    \label{fig:U-set1-5spires-turbint-contours}
\end{figure}

\vspace{-0.5cm}
\newpage
Corresponding plots of turbulence intensities, $I_u$, $I_v$, and $I_w$ are also presented in Figures \ref{fig:U-set1-5spires-turbint-2d}, and \ref{fig:U-set1-5spires-turbint-contours}, with the (4) test case performed at two different test speeds as mentioned before. 

\subsubsection{5 Spires `Set 2'}

Based on the spanwise uniformity results from the 5 spires `Set 1' configuration, the number of spanwise location for this test case was reduced. Measurements focused on spanwise location around the centre of the tunnel and at the outermost span, omitting the intermediate measurement points. 

\begin{figure}[H]
    \centering    \includegraphics[width=0.75\linewidth]{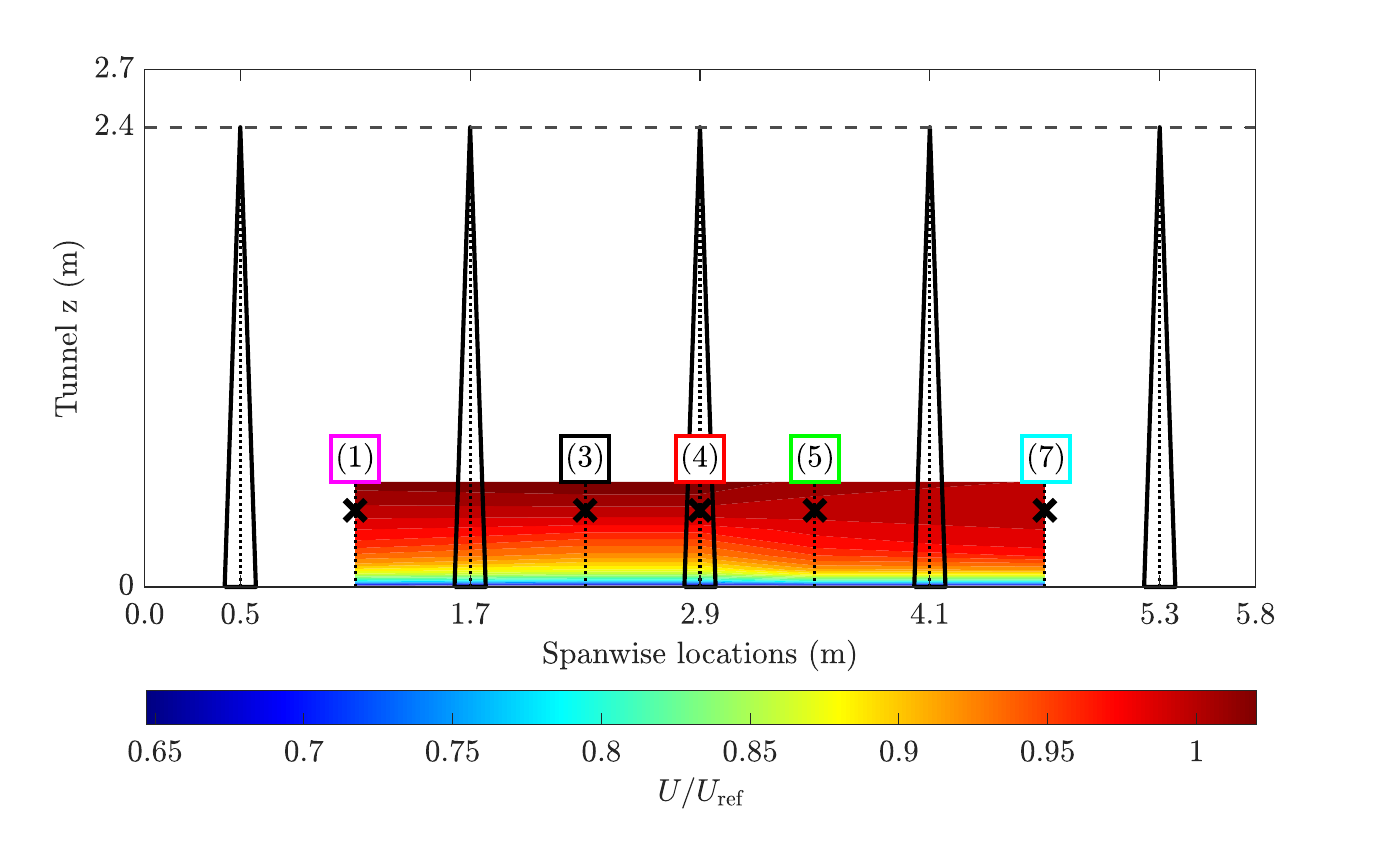}
    \caption{Contour plot of $U/U_{ref}$, using `Set 2' 5 spires, in the upper TS, relative to the test section dimensions}
    \label{fig:U-set2-5spires-normVel-contour-full}
\end{figure}

\begin{figure}[H]
    \centering    \includegraphics[width=0.8\linewidth]{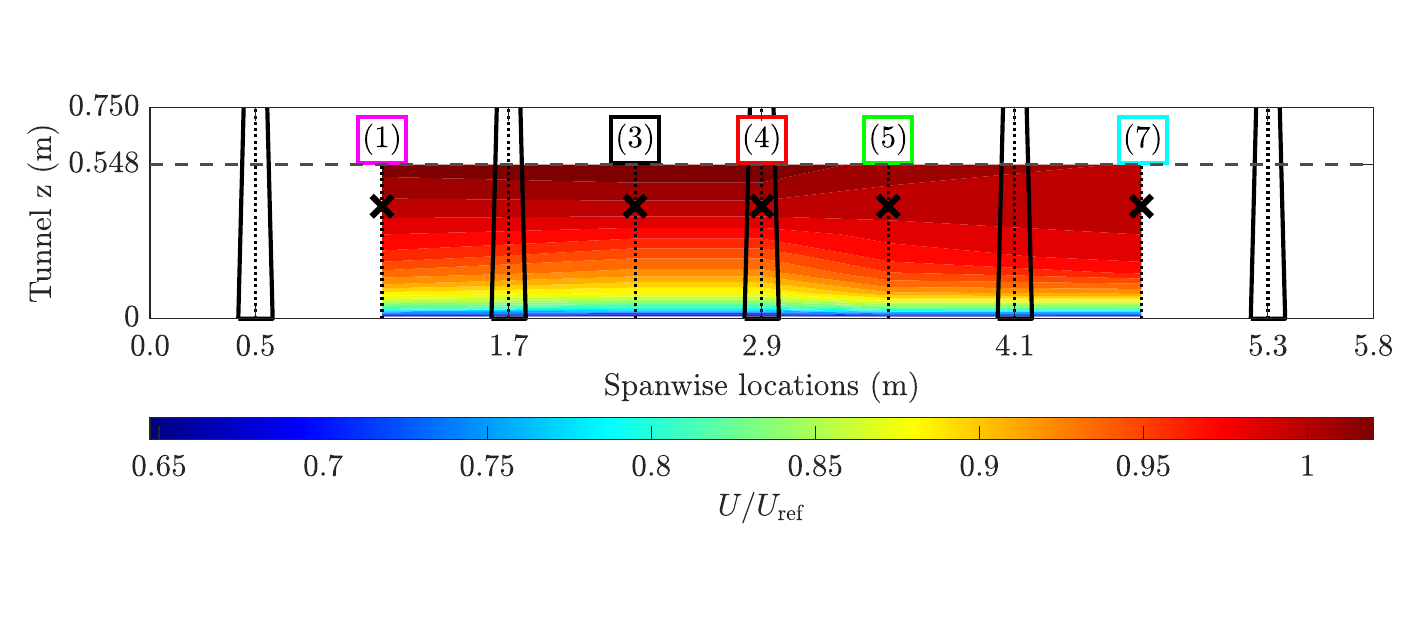}
    \vspace{-1cm}
    \caption{Contour plot of $U/U_{ref}$, using `Set 2' 5 spires, in the upper TS}
    \label{fig:U-set2-5spires-normVel-contour-part}
\end{figure}

\begin{figure}[H]
    \centering    \includegraphics[width=0.9\linewidth]{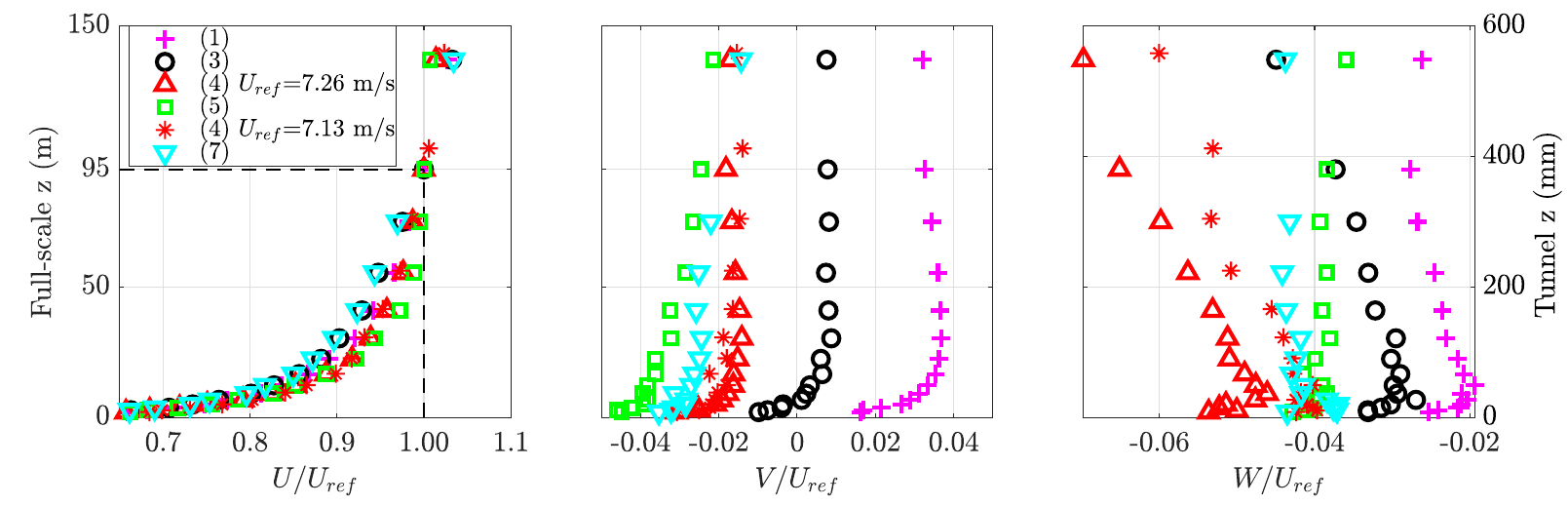}
    \caption{2D plot of $U/U_{ref}$,  $V/U_{ref}$, and  $W/U_{ref}$, using `Set 2' 5 spires, in the upper TS}
    \label{fig:U-set2-5spires-normVel-2d}
\end{figure}

Contour plots of $U/U_{ref}$ across the full section and a version focused on just the boundary layer measured are shown in Figures \ref{fig:U-set2-5spires-normVel-contour-full} and \ref{fig:U-set2-5spires-normVel-contour-part}, alongside 2D plots of all the normalised velocity components in Figure \ref{fig:U-set2-5spires-normVel-2d}.  
\vspace{-0.7cm}

\vspace{-0.5cm}

\begin{figure}[H]
    \centering    \includegraphics[width=0.8\linewidth]{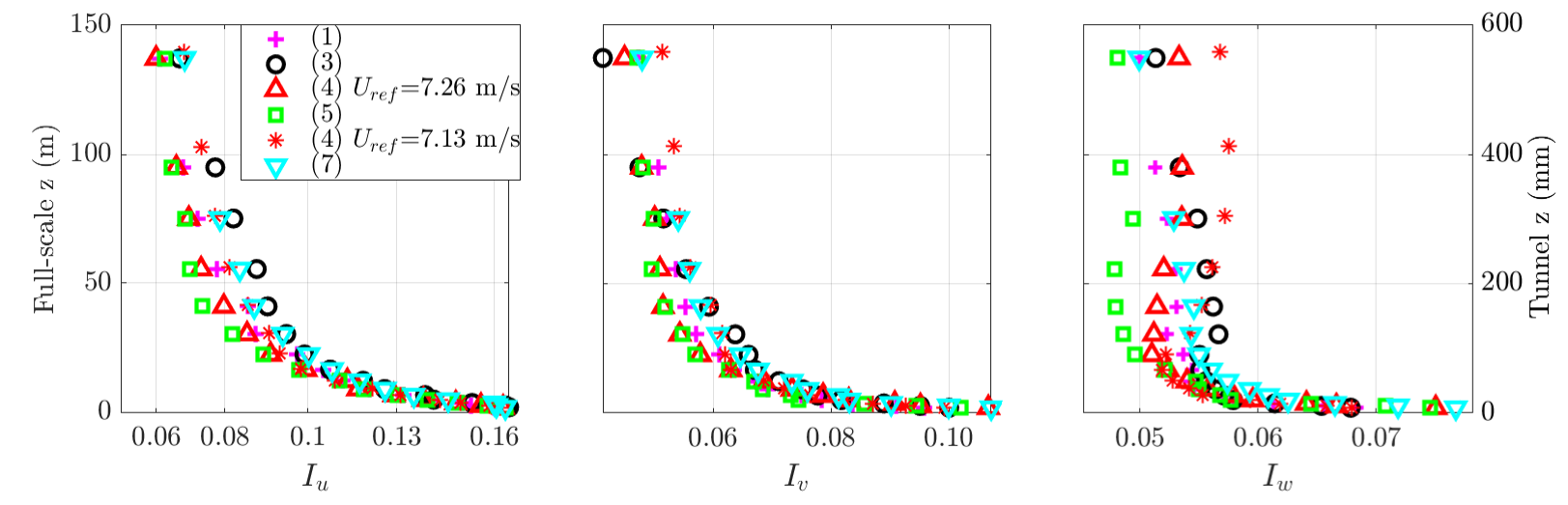}
    \caption{2D plots of $I_u$, $I_v$, and $I_w$, using `Set 2' 5 spires, in the upper TS}
    \label{fig:U-set2-5spires-turbint-2d}
\end{figure}

\begin{figure}[H]
    \centering    \includegraphics[width=0.73\linewidth]{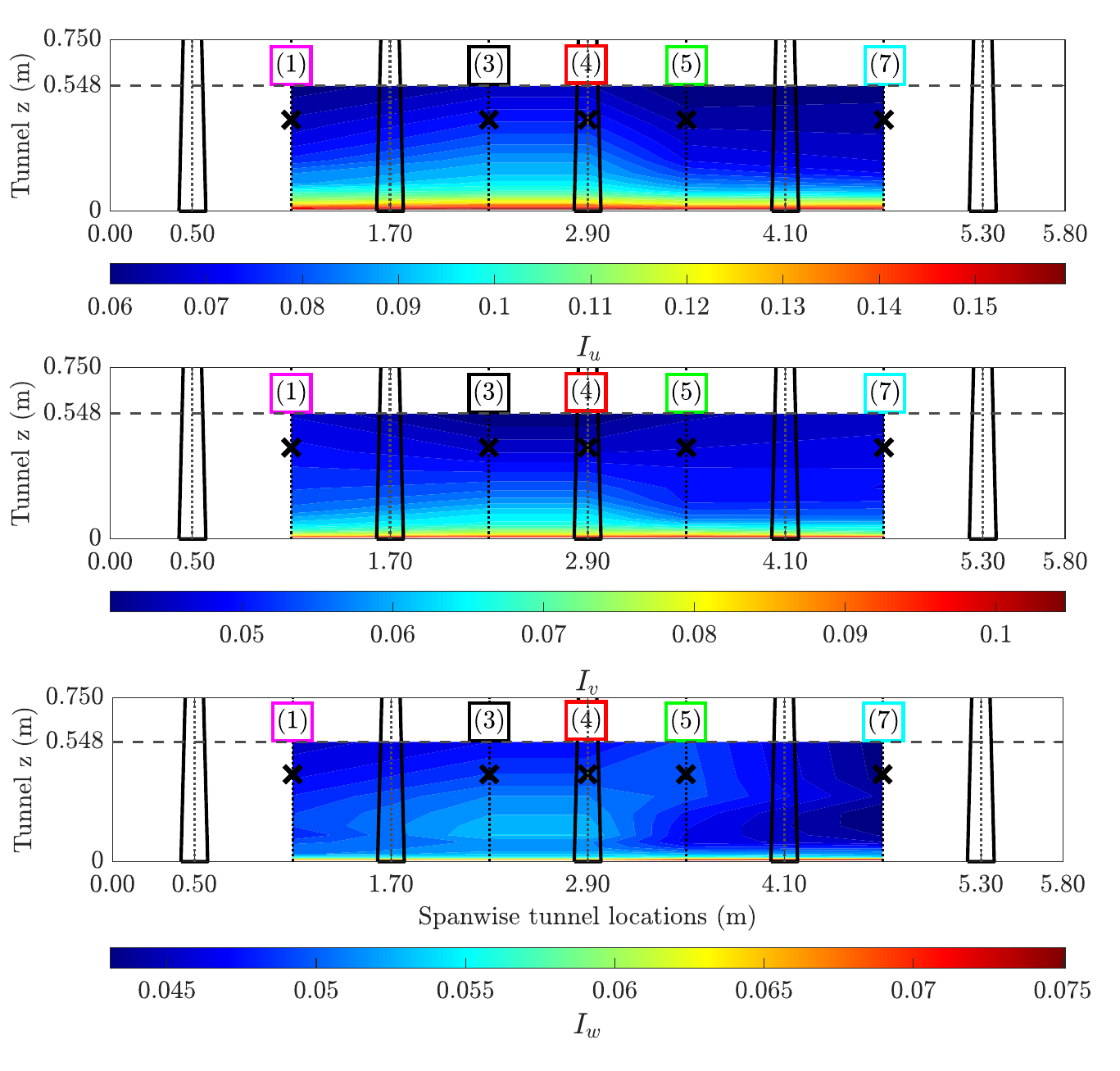}
    \vspace{-0.5cm}
    \caption{Contour plots of $I_u$, $I_v$, and $I_w$, using `Set 2' 5 spires, in the upper TS}
    \label{fig:U-set2-5spires-turbint-contours}
\end{figure}

\newpage
The 2D and contour plots for turbulence intensity in $U$, $V$, and $W$ are also provided in Figures\ref{fig:U-set2-5spires-turbint-contours} and \ref{fig:U-set2-5spires-turbint-2d}, respectively

\subsubsection{4 Spires `Set 2'}

\label{4spireset2}
Measurements were recorded near the tunnel centre and the outer edges to get an idea of the maximum spanwise extent of the flow across the tunnel width. This was done, since the focus was on developing the desired profile using the `Set 2' 4 spires for the `Profile WF (Wind Farm)', i.e. the profile which is suitable for the inflow of wind farms in the English Channel and North Sea. 

\begin{figure}[H]
    \centering    \includegraphics[width=0.8\linewidth]{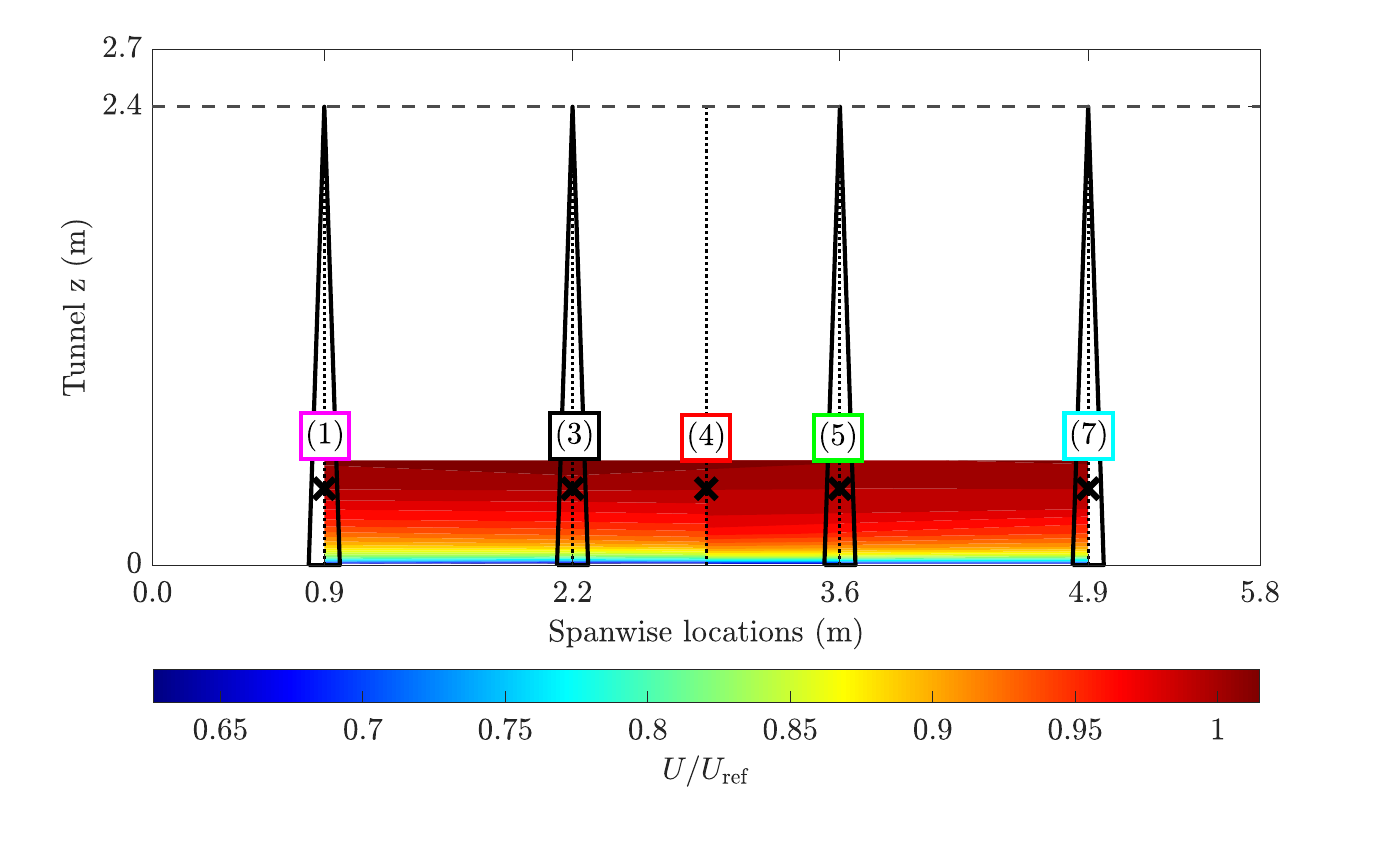}
    \caption{Contour plot of $U/U_{ref}$, using `Set 2' 4 spires, in the upper TS, relative to the test section dimensions}
    \label{fig:U-set2-4spires-normVel-contour-full}
\end{figure}

\begin{figure}[H]
    \centering   \vspace{-1cm} \includegraphics[width=0.75\linewidth]{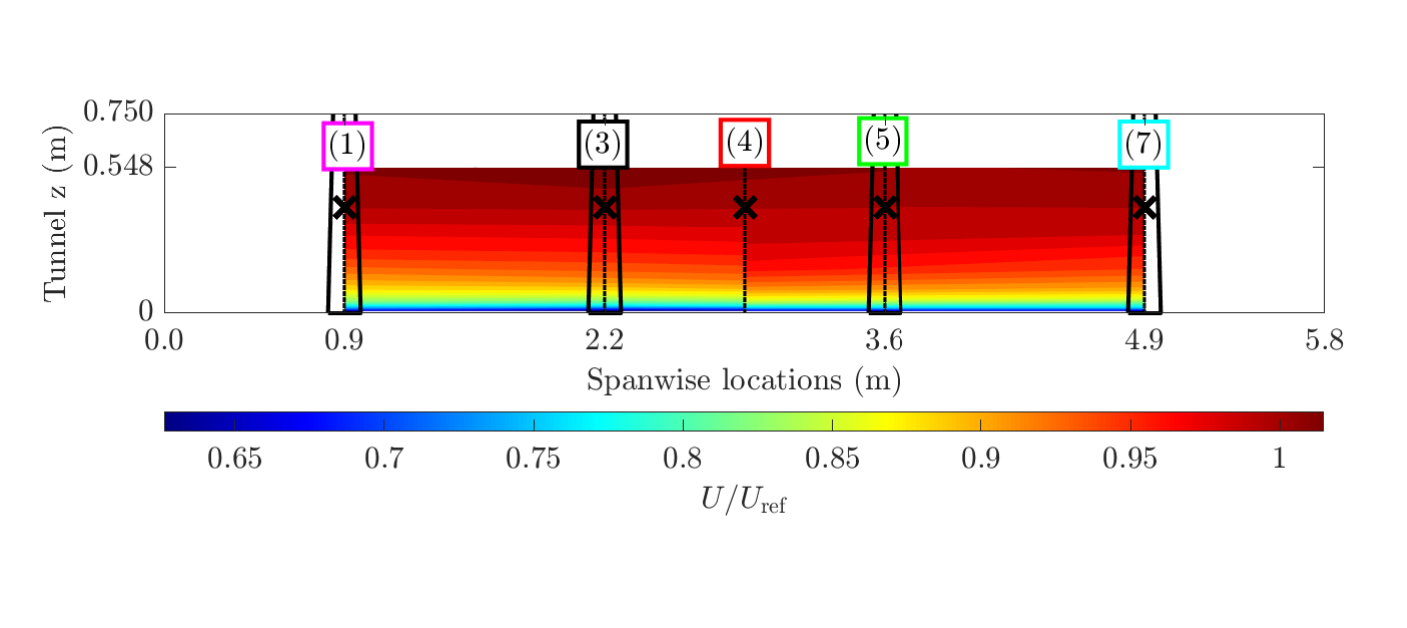}
    \vspace{-1cm}
    \caption{Contour plot of $U/U_{ref}$, using `Set 2' 4 spires, in the upper TS}
    \label{fig:U-set2-4spires-normVel-contour-part}
\end{figure}

\begin{figure}[H]
    \centering    \includegraphics[width=0.85\linewidth]{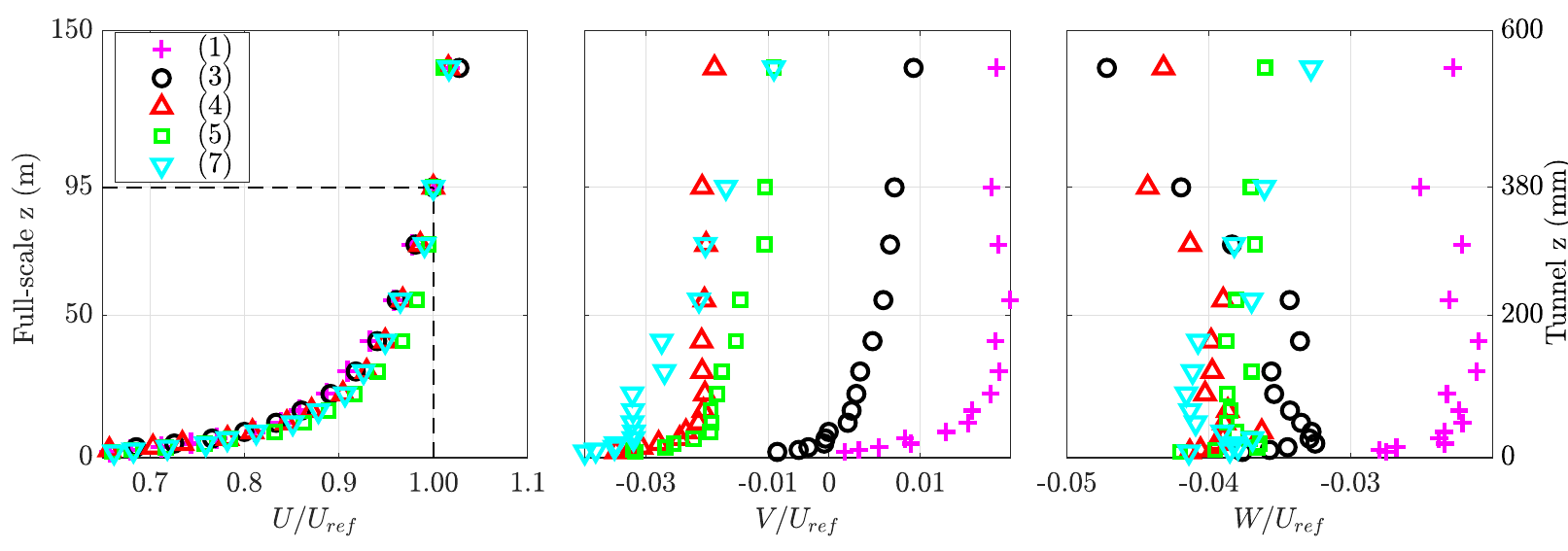}
    \caption{2D plot of $U/U_{ref}$,  $V/U_{ref}$, and  $W/U_{ref}$, using `Set 2' 4 spires, in the upper test-section}
    \label{fig:U-set2-4spires-normVel-2d}
\end{figure}

\begin{figure}[H]
    \centering    \includegraphics[width=0.9\linewidth]{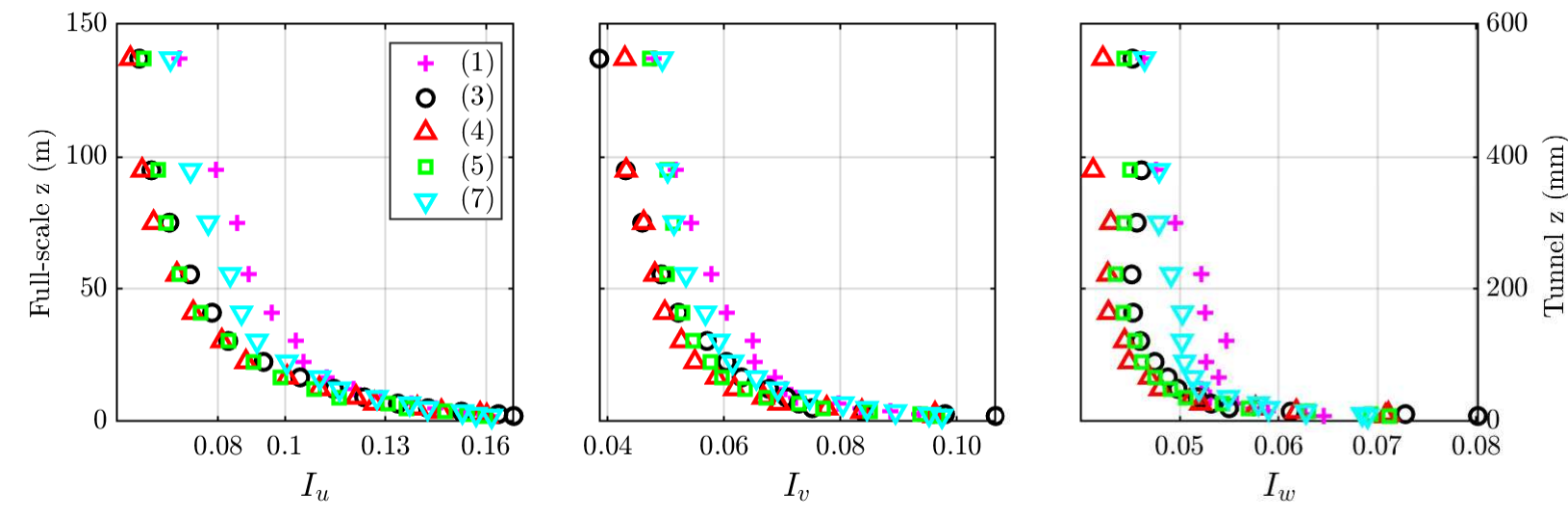}
    \caption{2D plots of$I_u$, $I_v$, and $I_w$, using `Set 2' 4 spires, in the upper TS}
    \label{fig:U-set2-4spires-turbint-2d}
\end{figure}

\begin{figure}[H]
    \centering    \includegraphics[width=0.67\linewidth]{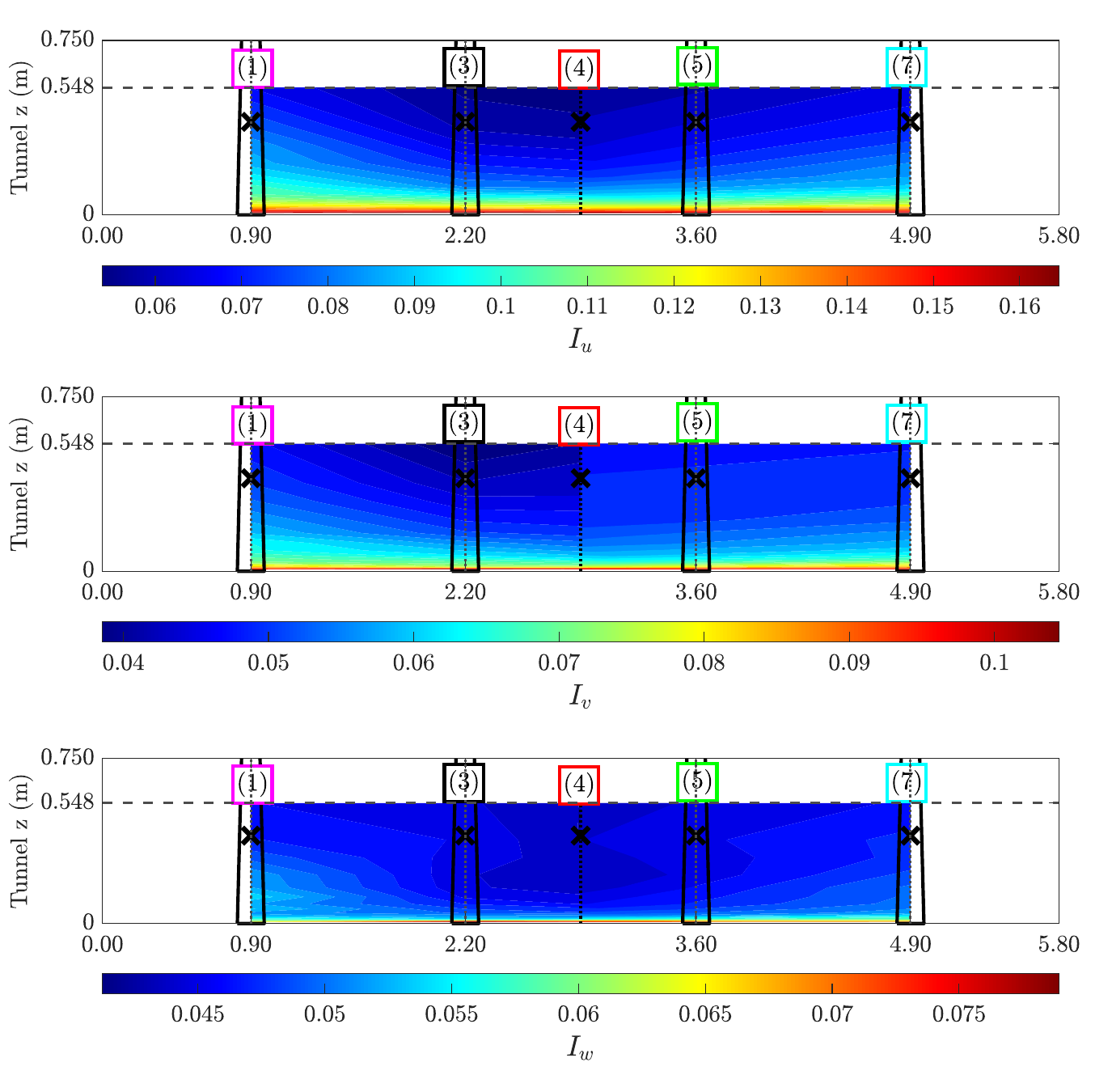}
    \caption{Contour plots of $I_u$, $I_v$, and $I_w$, using `Set 2' 4 spires, in the upper TS}
    \label{fig:U-set2-4spires-turbint-contours}
\end{figure}

Contour plots of $U/U_{ref}$, with $U_{ref}=U_{hub}$, where the hub height is 95 m for typical wind farm turbines in the aforementioned territory, are shown for both the full tunnel section and a zoomed-in view in Figures \ref{fig:U-set2-4spires-normVel-contour-full} and \ref{fig:U-set2-4spires-normVel-contour-part}, respectively. Respective 2D plots can be seen in Figure \ref{fig:U-set2-4spires-normVel-2d}. Furthermore, 2D and contour plots of turbulence intensities in $U$, and $W$, is demonstrated via Figures \ref{fig:U-set2-4spires-turbint-2d} and \ref{fig:U-set2-4spires-turbint-contours}. 

\subsubsection{4 spires `Set 1'}

% \paragraph{4 Spires}

% \begin{figure}[H]
%     \centering
%     \begin{minipage}{.47\linewidth}
%         \centering
%         \includegraphics[width=1.15\linewidth]{Images/Upper_TS/Set1/4Spires/meanU_contour_full.pdf}
%         \caption{2 spires - contour}
%         \label{fig:ds_3spire_set1}
%     \end{minipage}\hfill
%     \begin{minipage}{.47\linewidth}
%         \centering
%         \includegraphics[width=1.05\linewidth]{Images/Upper_TS/Set1/4Spires/meanU_contour_part.pdf}
%         \caption{3 spires - contours}
%         \label{fig:ds_3spire_set2}
%     \end{minipage}
% \end{figure}

For the 4 spire `Set 1' testing, fewer spanwise locations were used to save time, as key flow behaviour was already observed in the outermost spanwise locations, as described in Section \ref{4spireset2}. The spire arrangement is illustrated in Figure \ref{fig:upperts_spires}. 

\begin{figure}[H]
    \centering    \includegraphics[width=0.75\linewidth]{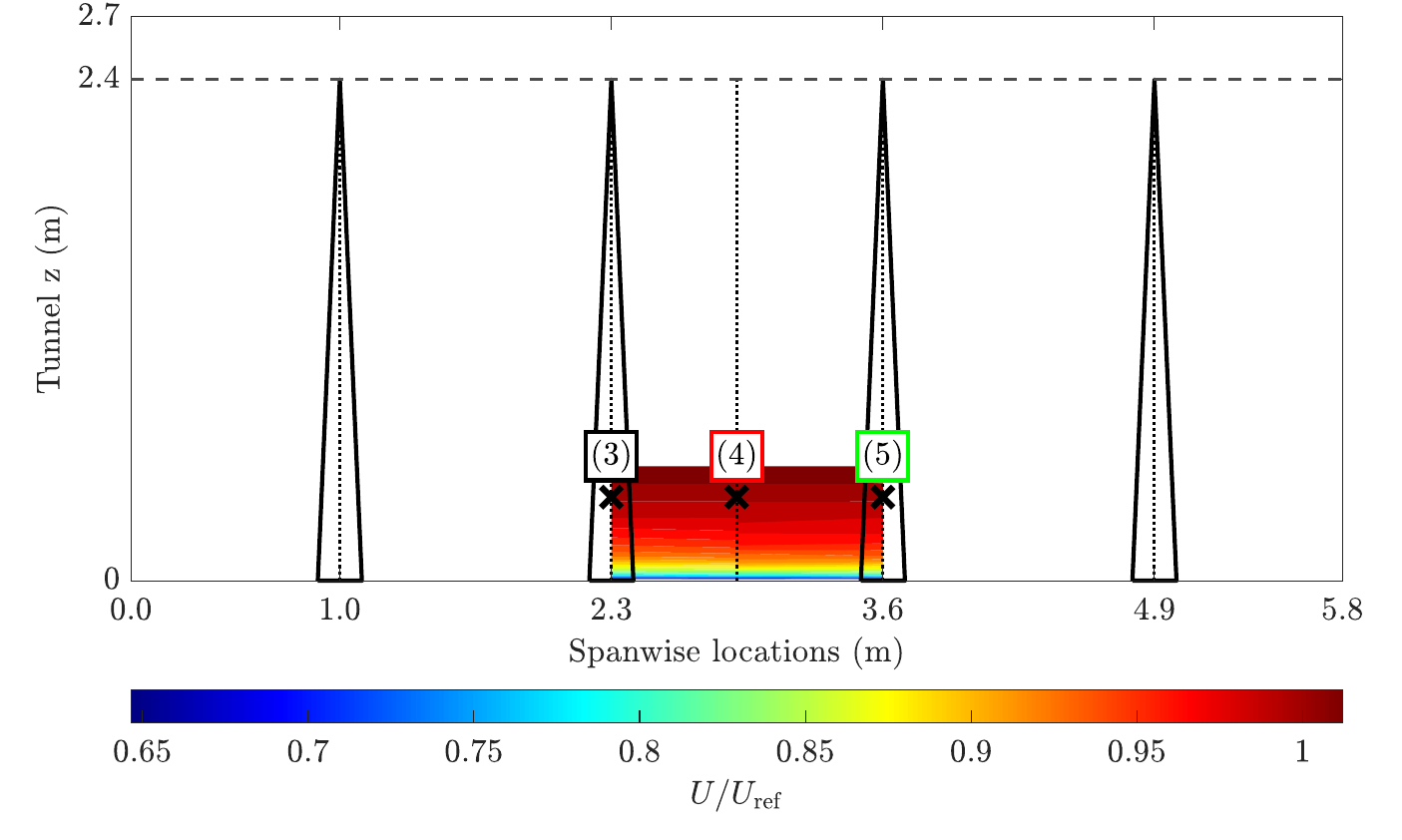}
    \caption{Contour plot of $U/U_{ref}$, using `Set 1' 4 spires, in the upper TS, relative to the test section dimensions}
    \label{fig:U-set1-4spires-normVel-contour-full}
\end{figure}

\begin{figure}[H]
    \centering    \includegraphics[width=0.8\linewidth]{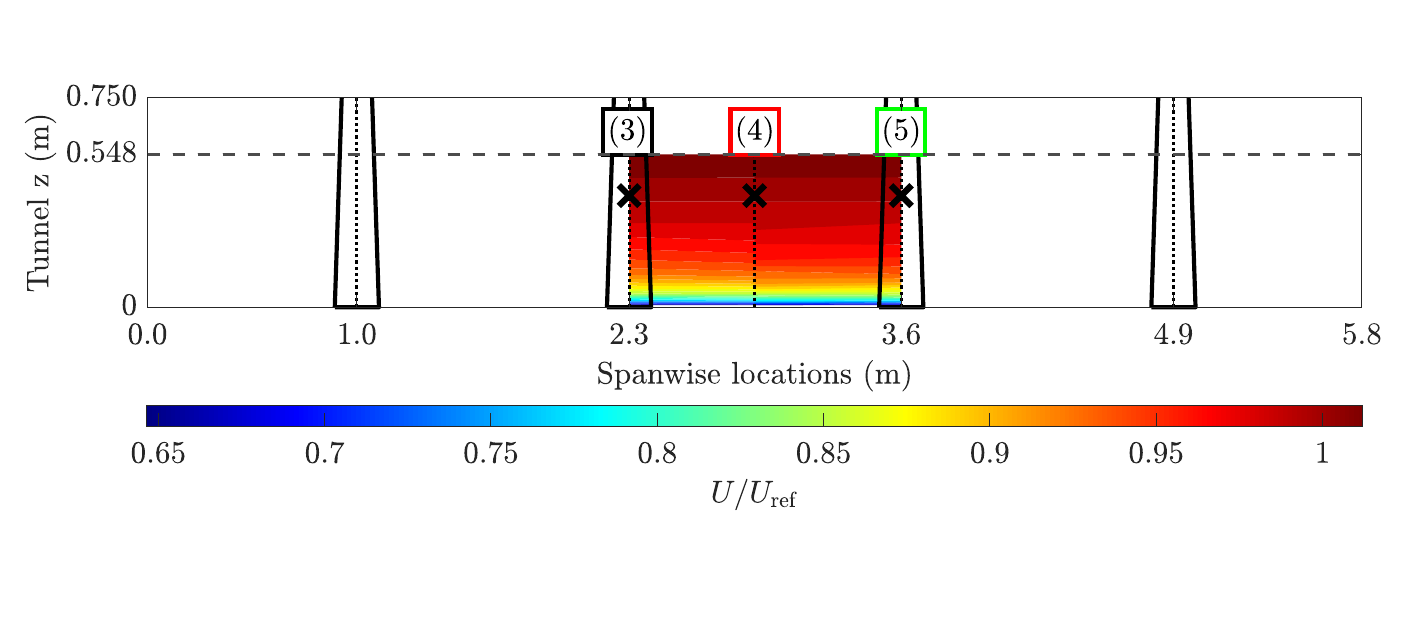}
    \vspace{-1cm}
    \caption{Contour plot of $U/U_{ref}$, using `Set 1' 4 spires, in the upper TS}
    \label{fig:U-set1-4spires-normVel-contour-part}
\end{figure}

Contour plots of normalised $U$ are presented relative to the full tunnel dimensions in Figure \ref{fig:U-set1-4spires-normVel-contour-full}, with Figure \ref{fig:U-set1-4spires-normVel-contour-part} focusing only on the specific measurement region. As before, $U_{ref}=U_{hub}=U$ at 95 m hub height. 

\begin{figure}[H]
    \centering    \includegraphics[width=1\linewidth]{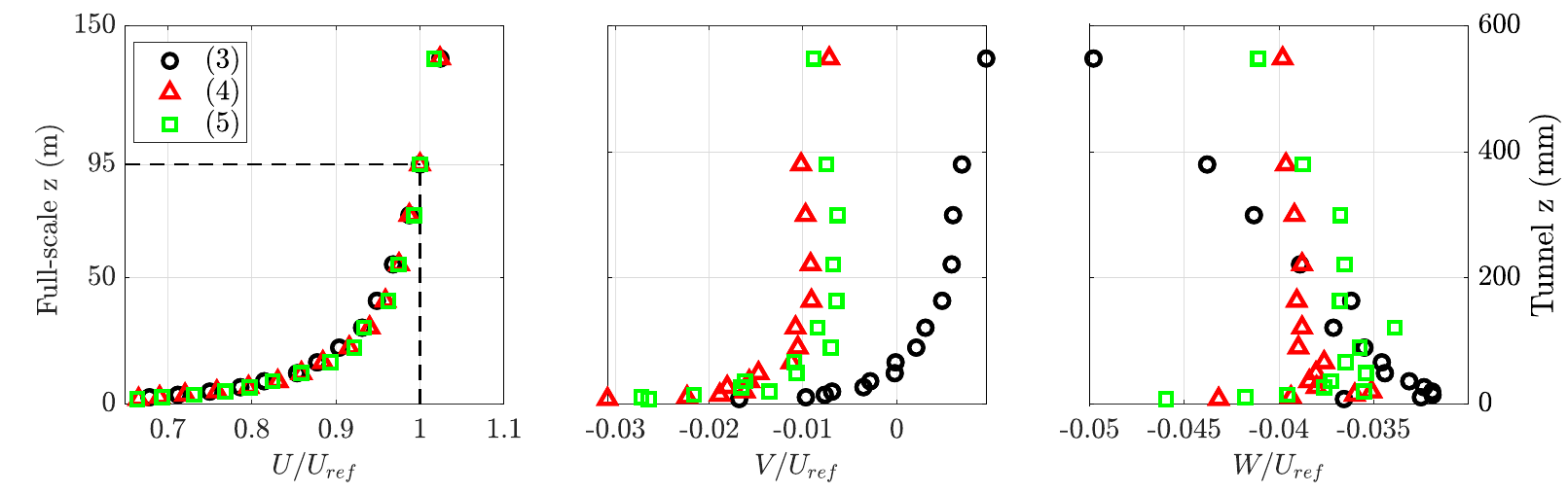}
    \caption{2D plot of $U/U_{ref}$,  $V/U_{ref}$, and  $W/U_{ref}$, using `Set 1' 4 spires, in the upper TS}
    \label{fig:U-set1-4spires-normVel-2d}
\end{figure}

To complement with these, the 2D plot represented in Figure \ref{fig:U-set1-4spires-normVel-2d} also showcases the uniformity in normalised $U$, $V$, and $W$ velocities. 

\vspace{-0.5cm}

\begin{figure}[H]
    \centering    \includegraphics[width=1\linewidth]{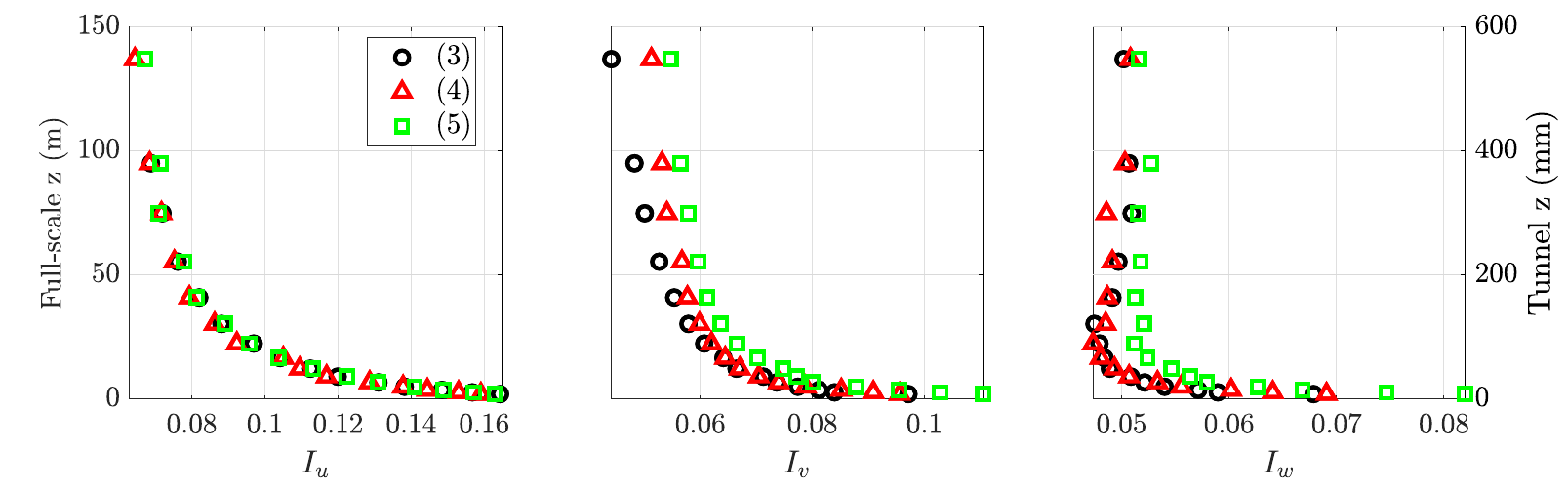}
    \caption{2D plots of$I_u$, $I_v$, and $I_w$, using `Set 1' 4 spires, in the upper TS}
    \label{fig:U-set1-4spires-turbint-2d}
\end{figure}

\begin{figure}[H]
    \centering    \includegraphics[width=0.72\linewidth]{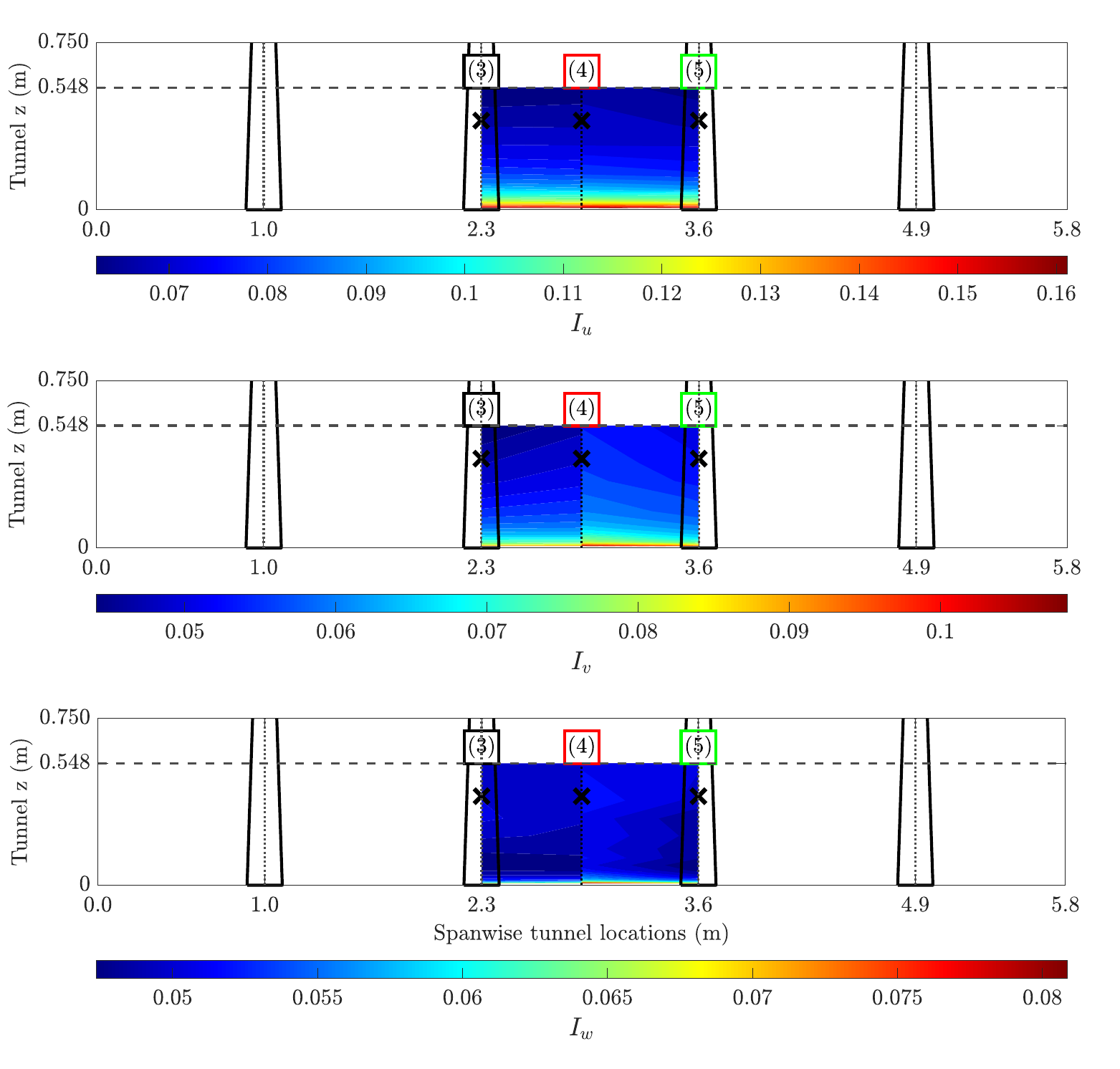}
    \vspace{-0.5cm}
    \caption{Contour plots of $I_u$, $I_v$, and $I_w$, using `Set 1' 4 spires, in the upper TS}
    \label{fig:U-set1-4spires-turbint-contours}
\end{figure}

\newpage
Similarly, the plots showing 2D and contour plots depicting the distribution of turbulence intensities $I_u$, $I_v$, and $I_w$, across the three spanwise locations are also shown by Figures \ref{fig:U-set1-4spires-turbint-contours} and \ref{fig:U-set1-4spires-turbint-2d}.

\vspace{-0.5cm}
\newpage
\section{Evaluation and Discussion}
\subsection{Lower Test Section}

\subsubsection{General Overview}

\textbf{`Set 1' 3 Spires} configuration exhibits a strong spanwise uniformity in the normalised $U$ velocity, and also in the turbulence intensities, shown in Figures \ref{fig:L-set1-3spires-normVel-contour-full}, \ref{fig:L-set1-3spires-turbint-contour}, and \ref{fig:L-set1-3spires-turbint-2d}. The $U_{mean}$ recorded for the fixed reference height of 1 m, at the spanwise locations in Figure \ref{fig:lowerts_spires} (a), showed a maximum deviation of about 3.2\%, which is within the experimental uncertainty of LDA measurements. Bazan et al. \cite{lda_accuracy} has reported a possible uncertainty of 5\% in the LDA measurements acquired, but this would vary considerably depending on the experimental setup, alignment of the lasers, precise calibration, etc. In Figure \ref{fig:L-set1-3spires-normVel-2d}, a small variation in the measurements of $W$ at the spanwise locations can be seen, without a clear symmetric pattern. These discrepancies are likely attributed to the minor misalignments in the 2D LDA setup, where the LDA's measurement axes are not perfectly aligned with the tunnel's axes. This leads to errors in the $W$ readings due to the cross contamination of the $V$, and $W$, from the $U$ values measured. Additionally, LDA measurements are very sensitive to beam alignment and optical distortions, including the light scattering capabilities of the seeding particles, and the slight spanwise inconsistencies in the seeding density that can lead to biased statistics in $W$, whilst $U$ remains better resolved. These errors in the normalised $W$ could also be amplified, since $W$ is usually much smaller and more fluctuating than $U$. 

\textbf{`Set 1' 2 Spires'} results reveal the consistency in the spanwise uniformity in normalised $U$, and the turbulence intensity profiles, illustrated by Figures \ref{fig:L-set1-2spires-normVel-contour-full} to \ref{fig:L-set1-2spires-turbint-2d}. The maximum error in the spanwise measurement of $U_{mean}$ at 1 m height, was about 2.8\%. There are several reasons for the error amplification in $W$, in Figure \ref{fig:L-set1-2spires-normVel-2d}, including the sensitivity of the LDA to axis misalignment, and the greater sensitivity of the vertical velocity to local geometric imperfections like spire misplacement, non-uniform seeding, etc. The mid-section scan for this configuration is omitted from the results, as it was taken during the preliminary LDA-MHP comparison test, when the MHP traverse caused a significant flow blockage, causing flow interference (results shown in Appendix \ref{badlda}). A redesign of the probe holder and the traverse system is detailed in Section \ref{probedes}, with the results in Sections \ref{LDA-MHP-time} and \ref{LDA-MHP}, which then facilitated a reliable comparison and validation. 

\textbf{`Set 1' 2 Spires with Roughness} test case was introduced after assessing the above results against the target `Profile 1', in Figures \ref{fig:lowerts_standards_U}, and \ref{fig:lowerts_standards_iU}. Hence, it was concluded that an additional roughness (R) would be necessary in order to match the target. As a preliminary test, green mesh was used, but not the roughness elements and their distribution as stated by Irwin's method in Table \ref{tab:geometry_table}. A notable irregularity in the trend of normalised $W$ is observed in the tunnel mid-section at location (4) as in Figure \ref{fig:L-set1-2spiresR-normVel-2d}, which could be indicative of more vertical mixing in the middle, due to the presence of roughness (since this behaviour was not observed in any of the other test cases). The maximum deviation of (4) from the values at locations (3) and (5) is relatively small, only about 2.5\% of $U_{ref}$, occurring near the measurement point at the height of 1 m. Nevertheless, the spanwise uniformity across other measured parameters remained high. 

\textbf{`Set 2' 2 Spires} show consistent results with the previous observations. Good spanwise uniformity portrayed in $U/U_{ref}$, $I_u$, and $I_w$, but the $W/W_{ref}$ still showing small variations (less than approx. 1.5\% of $U_{ref}$), which is quite insignificant and consistent with the sources of errors previously described. 

\subsubsection{LDA vs MHP}
\label{LDA-MHP}

\begin{wrapfigure}{r}{0.47\textwidth}
    \centering
    \vspace{-0.5cm}
    \includegraphics[width=0.53\textwidth]{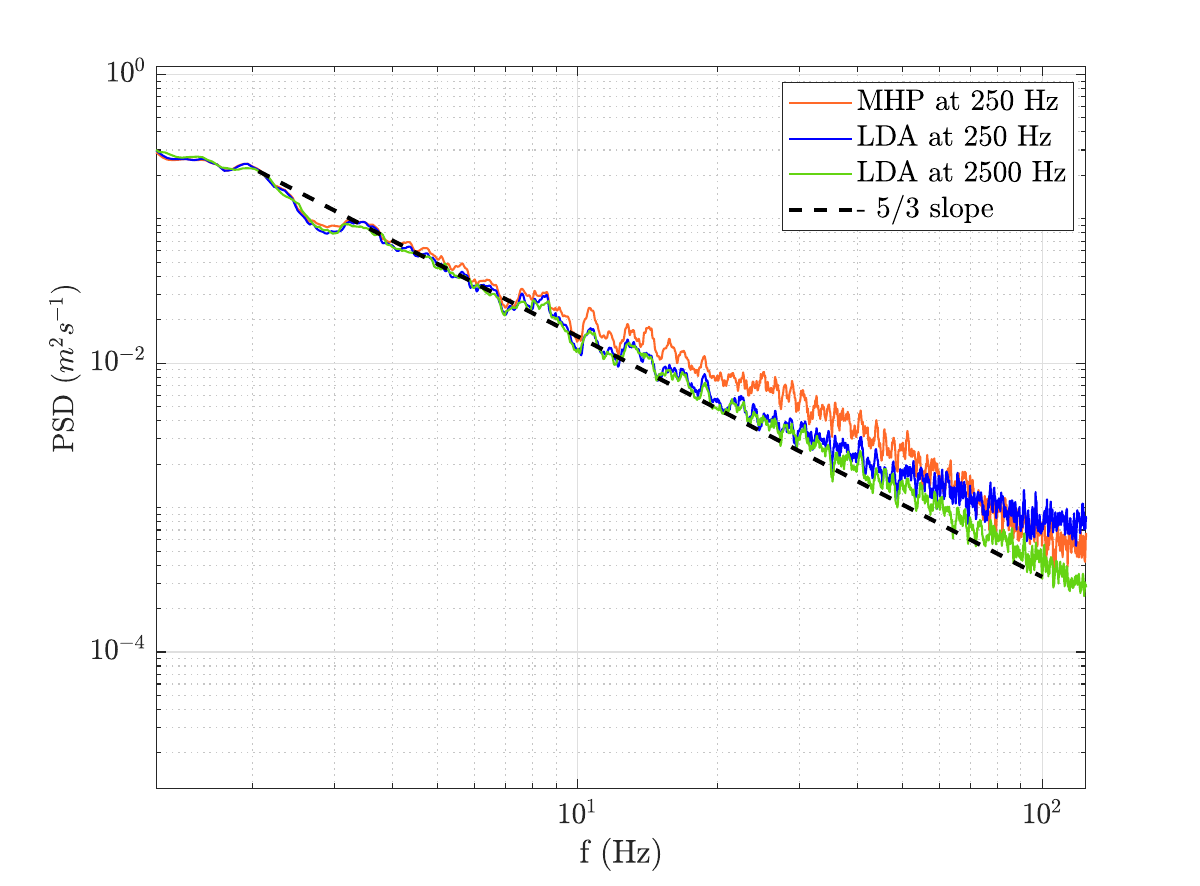}
    \centering
    \caption{Comparing the spectra of LDA and MHP, at different sampling rates}
    \label{fig:spectra_MHP_LDA}
\end{wrapfigure}

Comparing the PSD reveals a strong agreement between the LDA and the multi-hole probe (MHP), as shown in Figure~\ref{fig:spectra_MHP_LDA}. The MHP data were sampled at 250 Hz, in line with the experimental scale considerations, just the same as all the MHP data. The original LDA sampling rates were at several kHz depending on the dataset, so to ensure comparability, LDA data was sampled uniformly, and then resampled to 2500 Hz, and 250 Hz for this particular analysis. The resulting spectra thereby confirm that the resampling did not significantly affect the fidelity of capturing the turbulence information. The spectra from both measurement techniques follow a similar inertial subrange trend, which is as expected, showcasing behavior consistent with Kolmogorov's $-5/3$ power law (shown by the -5/3 slope). This agreement validates the ability of both systems to accurately resolve turbulence over the range of relevant frequencies, indicating reasonably precise calibration and measurement fidelity.

To support the spectra findings, a time series comparison was also performed earlier in Figures \ref{fig:LDA-MHP-timeseries-old} and \ref{fig:LDA-MHP-timeseries-new}, where the data series acquired utilising the modified probe holder/ traverse design shows excellent agreement between the LDA and the MHP measurements, with the MHP accurately capturing both the mean flow, and the unsteady fluctuations. Quantitative error analysis between the original LDA and MHP data showcased discrepancies of less than 1\% for the streamwise velocity $U$, under 2\% for the wall-normal component $W$, a magnitude error of just 1.7\%, and standard deviations of different velocities to all be within 5\%, indicating consistent measurements. 

Integral length scales from both LDA and MHP data show good agreement, confirming the capability of the MHP to resolve the low-frequency components that dictate these scales. A comparison of the integral length scales from various methods is presented in Table \ref{tab:Lux_LDAMHP}. Across the three tests, a maximum error of approximately 12.5\% (between the MHP at 250 Hz, and the LDA at 2500 Hz) was observed when using the von K\'arm\'an method. This further reinforces the validity of the MHP measurements against the reference LDA.  From the spectral analysis, it can be deciphered that the key energy-carrying turbulence is captured by both techniques, underscoring the validity of the calibration and data processing methods adopted in this project.

\begin{table}[H]
\centering
\begin{tabular}{c|c|c|c}
\textbf{Method}              & \textbf{MHP (250 Hz)} & \textbf{LDA (250 Hz)} & \textbf{LDA (2500 Hz)} \\ \hline \hline
von K\'arm\'an fit (m)           & 0.45                  & 0.42                  & 0.40                   \\
Spectra peak (m)             & 0.52                  & 0.51                  & 0.55                   \\
y-intercept (m)              & 0.49                  & 0.46                  & 0.44                   \\
Autocorrelation (m)          & 0.60                  & 0.54                  & 0.54                   \\
\end{tabular}
\caption{Comparison of $L_{u,x}$, for LDA and MHP}
\label{tab:Lux_LDAMHP}
\end{table}
\vspace{-0.4cm}
\subsubsection{Comparison with `Profile 1'}
\label{R_result}
(For the comparison plots of the wind tunnel measurements with the target profiles, only the centre tunnel test data, i.e. position (2) according to Figure \ref{fig:lowerts_spires}, have been considered, since spanwise uniformity has been proven to be sufficient, via the aforementioned results)

The baseline clean flow in the lower test section of the wind tunnel exhibited very low turbulence intensity, just as expected for an empty section. Therefore, to simulate the framework target profile established via the commonalities between various wind engineering standards (25 m/s at 10 m height), passive devices like spires and roughness elements are essential to produce the required turbulence. 

\begin{figure}[H]
    \centering
    \begin{minipage}{.47\linewidth}
        \centering
        \includegraphics[width=1.07\linewidth]{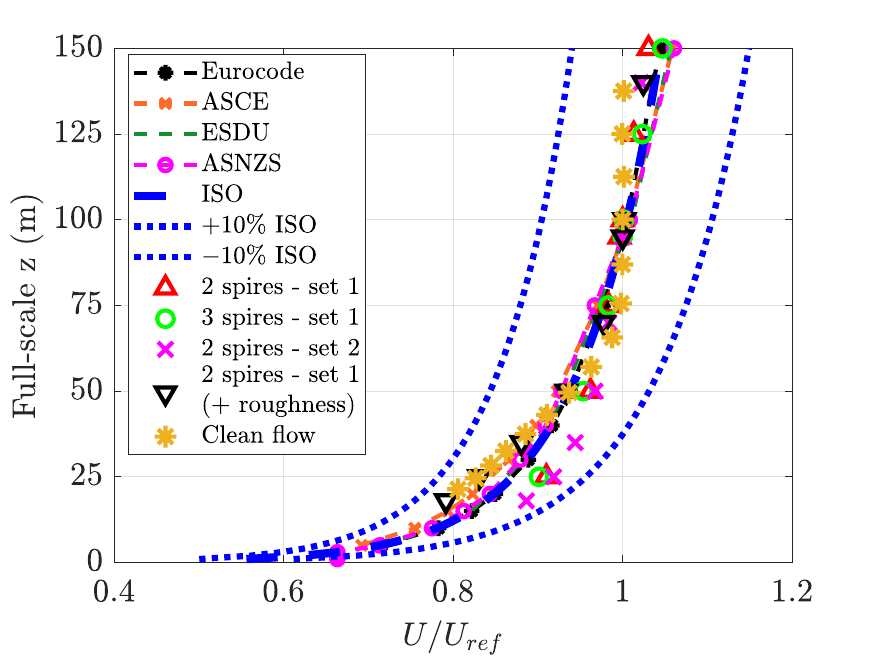}
        \caption{Lower TS configurations vs `Profile 1', via normalised velocity in $U$}
        \label{fig:lowerts_standards_U}
    \end{minipage}\hfill
    \begin{minipage}{.47\linewidth}
        \centering
        \includegraphics[width=1.07\linewidth]{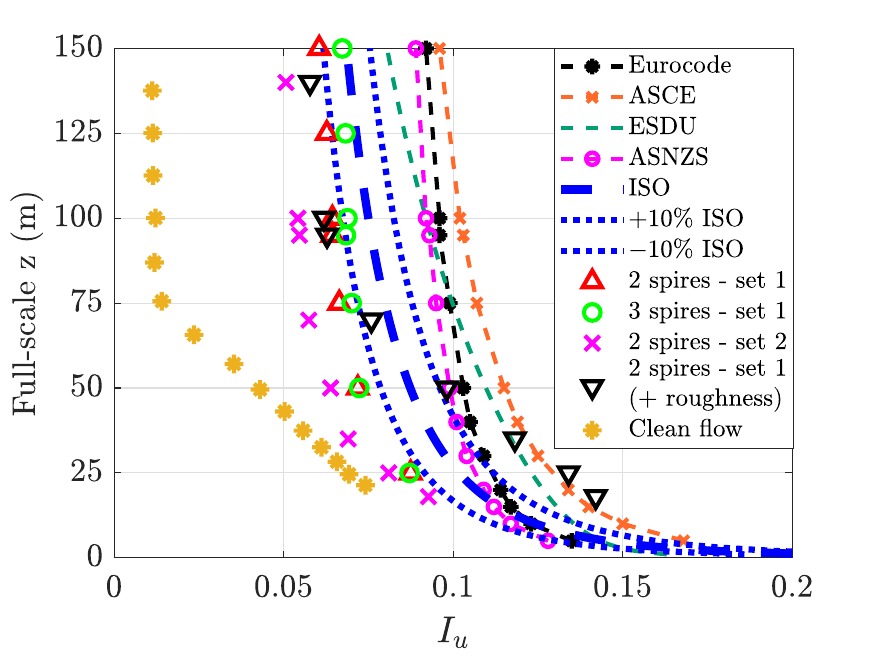}
        \caption{Lower TS configurations vs `Profile 1', via the turbulence intensity in $U$}
        \label{fig:lowerts_standards_iU}
    \end{minipage}
\end{figure}

The `Set 1' spires (2 spires, 3 spires, 2 spires with roughness) were designed to match `Profile 1' having an $\alpha=0.11$, as detailed in Section \ref{spire_design}. These configurations outperformed the `Set 2' 2 spires case (designed to replicate `Profile WF', $\alpha=0.07$), as expected. `Set 1' 3 spires shows the closest match to the required `Profile 1', as seen in Figures \ref{fig:lowerts_standards_U} and \ref{fig:lowerts_standards_iU}. However, the resulting turbulence intensity from the `Set 1' 3 spires case was lower than desired, prompting the addition of roughness (green mesh in this project, not the roughness recommended by Irwin). Further details about this can be found in Section \ref{R}, with the test section arrangement shown in Appendix \ref{greenmesh}. Test performed with the green mesh reveals that it provided higher roughness than required, as excessive turbulence intensity was recorded. Hence, it shows that only minimal surface roughness would be required to adequately match the turbulence characteristics required near the floor. Figure \ref{fig:lowerts_standards_U} presents that all the `Set 1' configurations showed an excellent agreement in the $U/U_{ref}$ profiles, which collapsed across all heights, portraying the consistency of the velocity field induced by the spire arrangements. 
\vspace{-0.35cm}

\vspace{-0.7cm}
\subsubsection{Comparison with `Profile WF'}

\begin{figure}[H]
    \centering
    \begin{minipage}{.47\linewidth}
        \centering
        \includegraphics[width=1.07\linewidth]{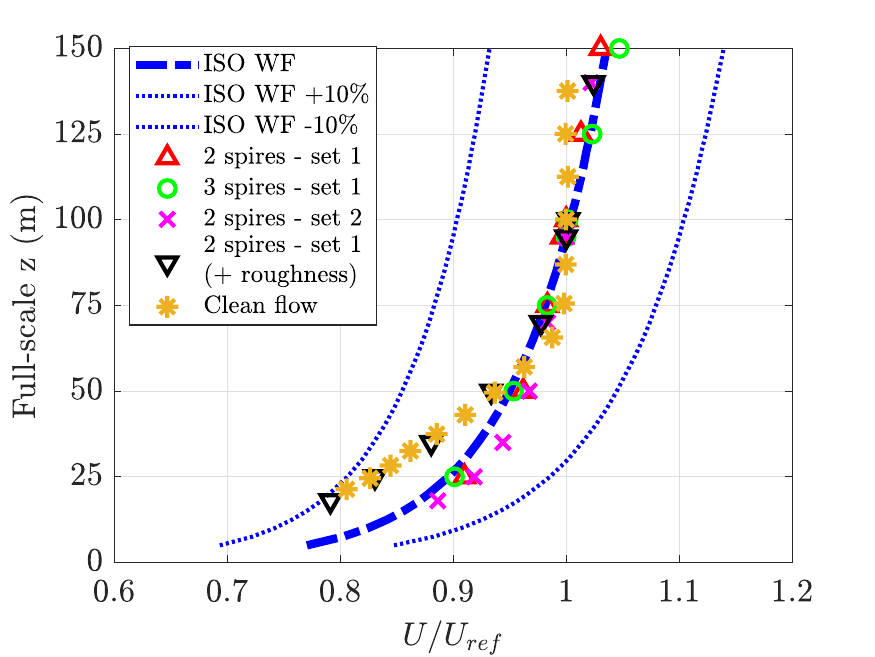}
        \caption{Lower TS configurations vs `Profile WF', via normalised velocity in $U$}
        \label{fig:lowerts_iconic_U}
    \end{minipage}\hfill
    \begin{minipage}{.47\linewidth}
        \centering
        \includegraphics[width=1.07\linewidth]{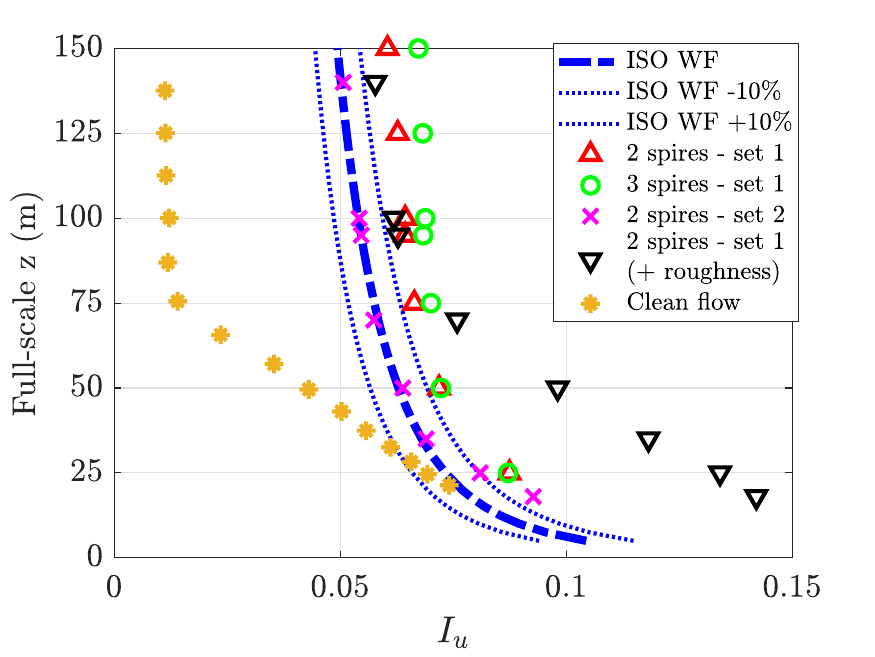}
        \caption{Lower TS configurations vs `Profile WF', via turbulence intensity in $U$}
        \label{fig:lowerts_iconic_iU}
    \end{minipage}
\end{figure}

Results confirm that the `Set 2' 2 spires configuration, which was specifically designed to replicate `Profile WF (Wind Farm)' (most suitable for modelling wind farms in the English Channel and North Sea), provides the closest match to the target profiles as shown in Figures \ref{fig:lowerts_iconic_U} and \ref{fig:lowerts_iconic_iU}. This expected behaviour validates the effectiveness of the designed spires. As speculated, the `Set 1' spires, which are originally designed for `Profile 1', produced a higher turbulence intensity than required, due to their greater width compared to the refined `Set 2' spires, resulting in stronger production of turbulence. The normalised profile from the `Set 2' spires closely aligns with the `Profile WF' target, whilst the `Set 1' deviates, which is consistent with the design it is intended for and its performance expectations.  

\subsection{Upper Test Section}

\subsubsection{General Overview}

\textbf{`Set 1' 5 Spires} excellent spanwise uniformity in the streamwise velocity component $U$, but not in the spanwise $V$, and in the wall-normal $W$ components, with an indication of the presence of minor anomalies, demonstrated in Figures \ref{fig:U-set1-5spires-normVel-contour-full} to \ref{fig:U-set1-5spires-normVel-2d}.  

The slight misalignment of the probe relative to the flow direction, leading to small errors in pitch and yaw, could be a reason for the deviation in the $V$ and $W$ readings from zero, in the form of systematic offsets. This leads to a portion of the $U$, being misinterpreted as $V$, and $W$. Error propagation analysis was performed for $U$, $V$, and $W$, to quantify this, assuming that the uncertainty introduced by the cross-line laser, which was used for aligning the probe is $\pm4\degree$ \cite{crossline_laser}. For the 5 spire `Set 1' test case, this analysis demonstrated that the maximum deviations in V and W (for test cases (4) and (5)) were obtained to be $-0.2997\pm0.3092$ m/s, and $-0.2868\pm0.3117$ m/s, respectively. Therefore, it can be concluded that the fluctuations observed in the spanwise direction are within the stipulated uncertainty limits, hence supporting the validity of the measurements using the MHP. Furthermore, calibration issues for the pitch and yaw can also cause $V$ and $W$ to appear shifted systematically, since calibration relates pressure differences in the probe holes to flow angles, which causes any small error to shift the $V$, and $W$ baseline. Any contamination or blockage in the pressure probe hole can also cause the pressure difference readings to be skewed, causing a bias in $V$ and $W$. 

While $I_u$, and $I_v$ show consistent spanwise uniformity, $I_w$ presents some variation at the higher heights. Nevertheless, the overall profiles of $I_u$, $I_v$, and $I_w$, align well across all spanwise locations, as shown by Figures \ref{fig:U-set1-5spires-turbint-2d} and \ref{fig:U-set1-5spires-turbint-contours}. The data from different velocity test cases also collapse onto each other, which demonstrates reliable flow consistency in the measurements acquired. 

\textbf{`Set 1' 4 Spires} good spanwise uniformity in normalised $U$, and also in $I_u$, $I_v$, and $I_w$, with similar discrepencies notes in $V$, and $W$ as in the `Set 1' 5 spires case, represented in Figures \ref{fig:U-set1-4spires-normVel-contour-full} to \ref{fig:U-set1-4spires-turbint-contours}. $V$ component appears to be more strongly influenced by $U$ in this case, owing from the misalignment of the probe. $I_v$ profile shows symmetry between positions 3 and 5, which are on either sides of the tunnel centre-line location 4. 

\textbf{`Set 2' 5 Spires} Though a good consistency between the various spanwise positions are observed in the velocity and turbulence intensity distributions, in Figures \ref{fig:U-set2-5spires-normVel-contour-full} to \ref{fig:U-set2-5spires-turbint-contours}, there is a certain amount of spanwise non-uniformity observed in spanwise locations (3) and (7). This could be due to local flow behaviour, since a similar trend has not been observed in other test cases.

\textbf{`Set 2' 4 Spires} Positions close to the wall (runs (1) and (7)), show a higher turbulence intensity, likely caused by interactions between the turbulence induced by the spire and the boundary layer developing along the tunnel walls, demonstrated by Figure \ref{fig:U-set2-4spires-normVel-2d} and \ref{fig:U-set2-4spires-turbint-contours}. Due to the close proximity of these locations to the wall, recirculation zones may also exist. Aside from these edge positions, the remaining spanwise stations show good uniformity in the normalised velocity and turbulence intensity profiles as seen in Figures \ref{fig:U-set2-4spires-normVel-contour-full} to \ref{fig:U-set2-4spires-turbint-contours}. 

\subsubsection{Comparison with `Profile 1'}

(For all the comparison plots of the wind tunnel measurements with the target profiles, only the centre tunnel test data, i.e. position (4) according to Figure \ref{fig:upperts_spires}, have been considered, since the spanwise uniformity has been proven to be sufficient according to the previously discussed results)

\begin{figure}[H]
    \centering
    \begin{minipage}{.47\linewidth}
        \centering
        \includegraphics[width=1.1\linewidth]{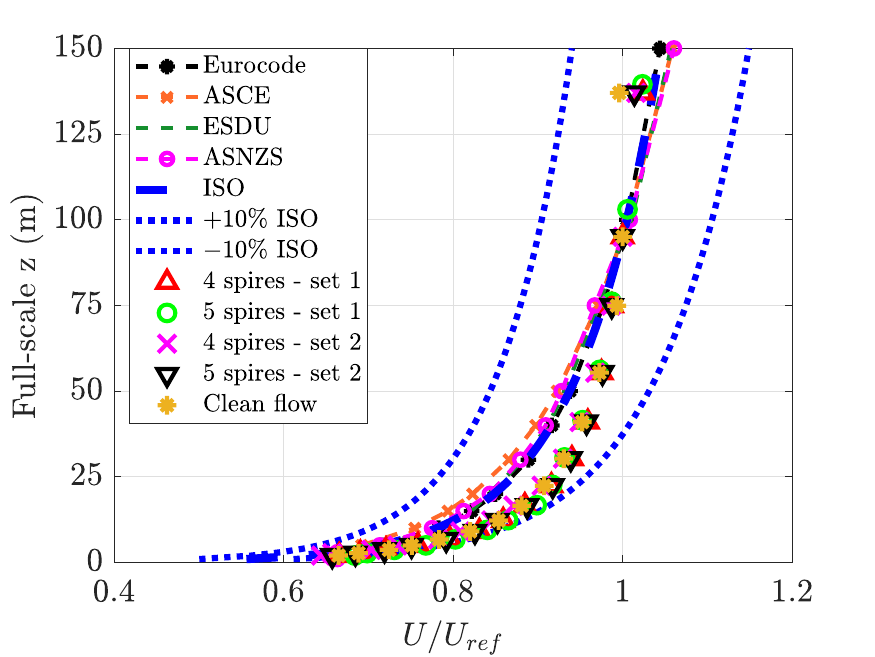}
        \caption{Upper TS configurations vs `Profile 1', via normalised velocity in $U$}
        \label{fig:upperts_standards_U}
    \end{minipage}\hfill
    \begin{minipage}{.47\linewidth}
        \centering
        \includegraphics[width=1.1\linewidth]{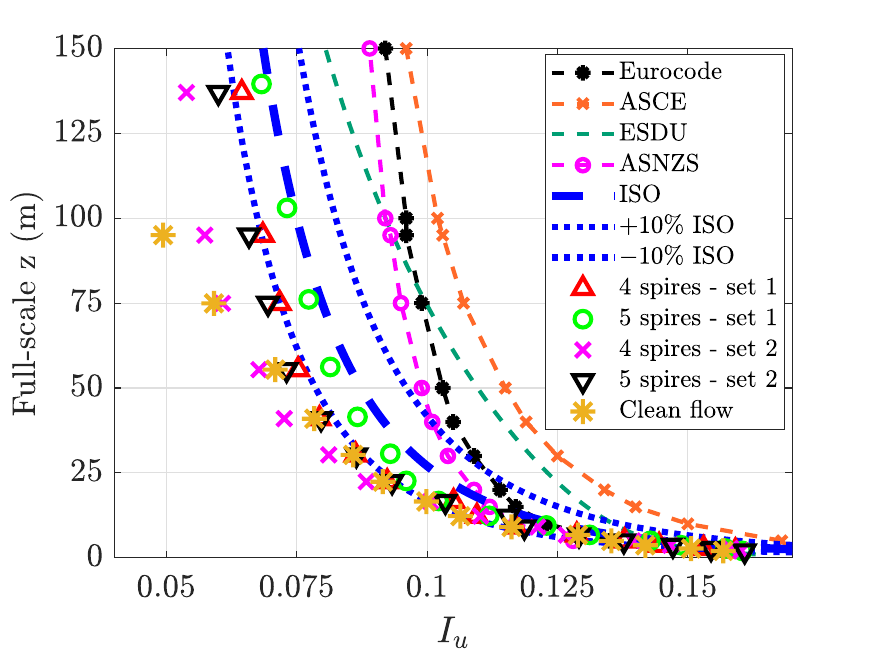}
        \caption{Upper TS configurations vs `Profile 1', via turbulence intensity in $U$}
        \label{fig:upperts_standards_iU}
    \end{minipage}
\end{figure}

The clean flow with no spires exhibit inherently high turbulence intensity in the upper test section, which is a known feature of the tunnel. Therefore, this limits how low of a turbulence intensity can be achieved regardless of the spire configuration used. For example, the $I_u$ is 9\% in the upper section and 7\% in the lower section at a height of about 25 m, whereas at about 75 m, $I_u$ is approximately 6\% in the upper section, whereas only 1\% in the lower section.`Set 1' closely resembled the ISO `Profile 1', which indicates that the spires are functioning as designed. In particular, 4 spire `set 1' shows the closest match to within $\pm$10\%, as in Figure \ref{fig:upperts_standards_iU}. The velocity profile produced by all the spire configurations collapse onto a common trend line, to within $\pm$10\% showing the consistency and reliability in the flow development proven by the profiles in Figure \ref{fig:upperts_standards_U}. 

\subsubsection{Comparison with `Profile WF'}

\begin{figure}[H]
    \centering
    \begin{minipage}{.47\linewidth}
        \centering
        \includegraphics[width=1.08\linewidth]{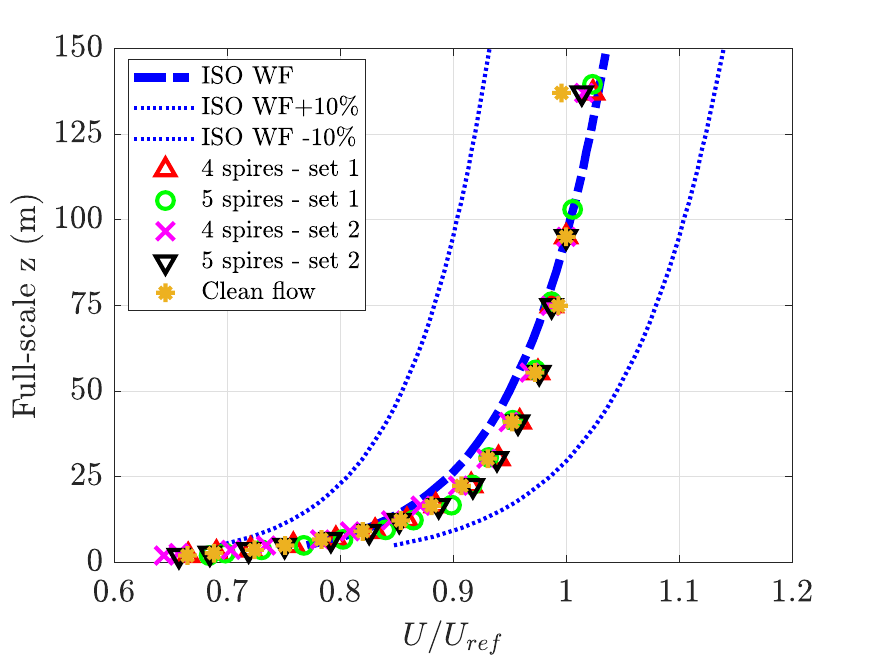}
        \caption{Upper TS configurations vs `Profile WF', via normalised velocity in $U$}
        \label{fig:upperts_iconic_U}
    \end{minipage}\hfill
    \begin{minipage}{.47\linewidth}
        \centering
        \includegraphics[width=1.1\linewidth]{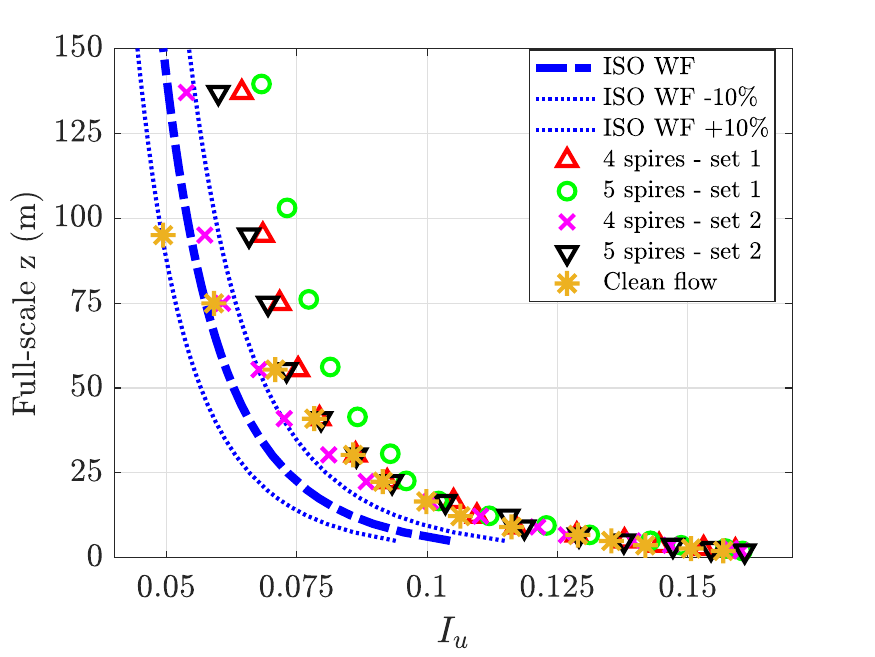}
        \caption{Upper TS configurations vs `Profile WF', via turbulence intensity in $U$}
        \label{fig:upperts_iconic_iU}
    \end{minipage}
\end{figure}

`Set 2' spires were designed to effectively match the target `Profile WF', for modelling the inflow of wind farms in the English Channel and North Sea. The clean flow is shown to have a higher baseline turbulence intensity in the upper TS. This poses constraints on the profile, which can be developed in the wind tunnel to match the `Profile WF' that requires reduced $I_u$ at lower heights. Nevertheless, the `Set 2' spires still show a very strong match overall.  At the hub height, which is a critical point in wind turbines, the 4 spire `set 2' shows a reasonable agreement with the target, with a difference in $I_u$ by about 4\%, presented in Figure \ref{fig:upperts_iconic_iU}. For the normalised velocity profile, the 4 spire `set 2' shows the closes match as in Figure \ref{fig:upperts_iconic_U}. Furthermore, there is an indication of consistent and stable mean flow development, since the $U/U_{ref}$ corresponding to all the setups collapse pretty well. 

\newpage
\subsection{Power Spectral Density}

\begin{figure}[H]
    \centering
    \begin{minipage}{.48\linewidth}
        \centering \vspace{0.8cm}
        \includegraphics[width=\linewidth, height=6.8cm]{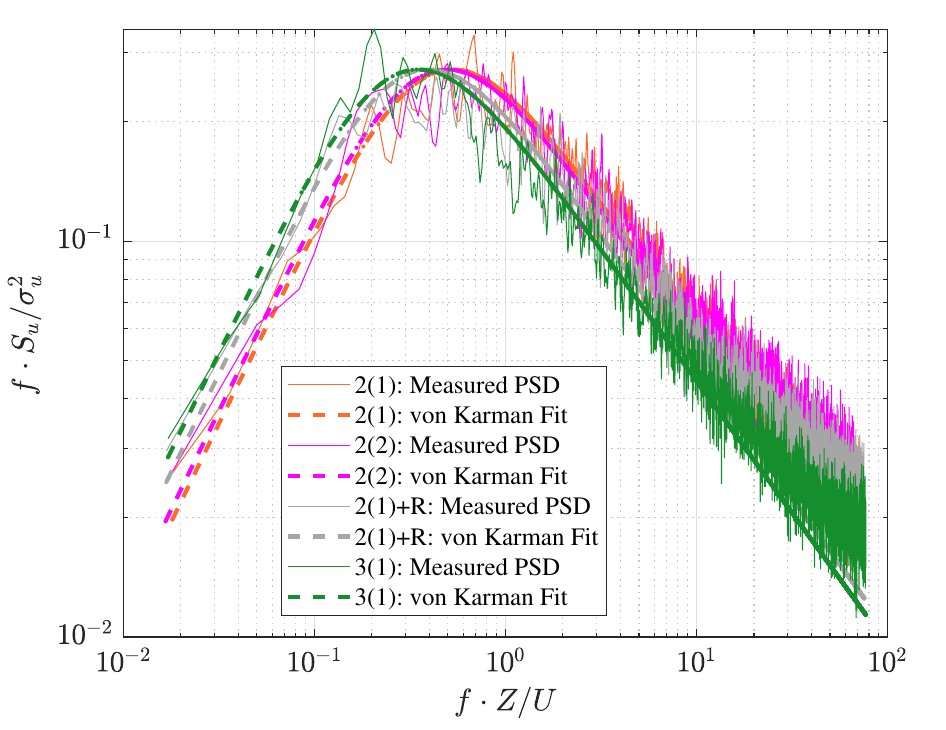}
        \caption{PSD of the test cases in the lower TS at hub height, using LDA subsampled to 2500 Hz; e.g., 2(1)+R = 2 spire set 1 + roughness (green mesh).}
        \label{fig:psd_lower}
    \end{minipage}\hfill
    \begin{minipage}{.48\linewidth}
        \centering
        \includegraphics[width=\linewidth, height=7cm]{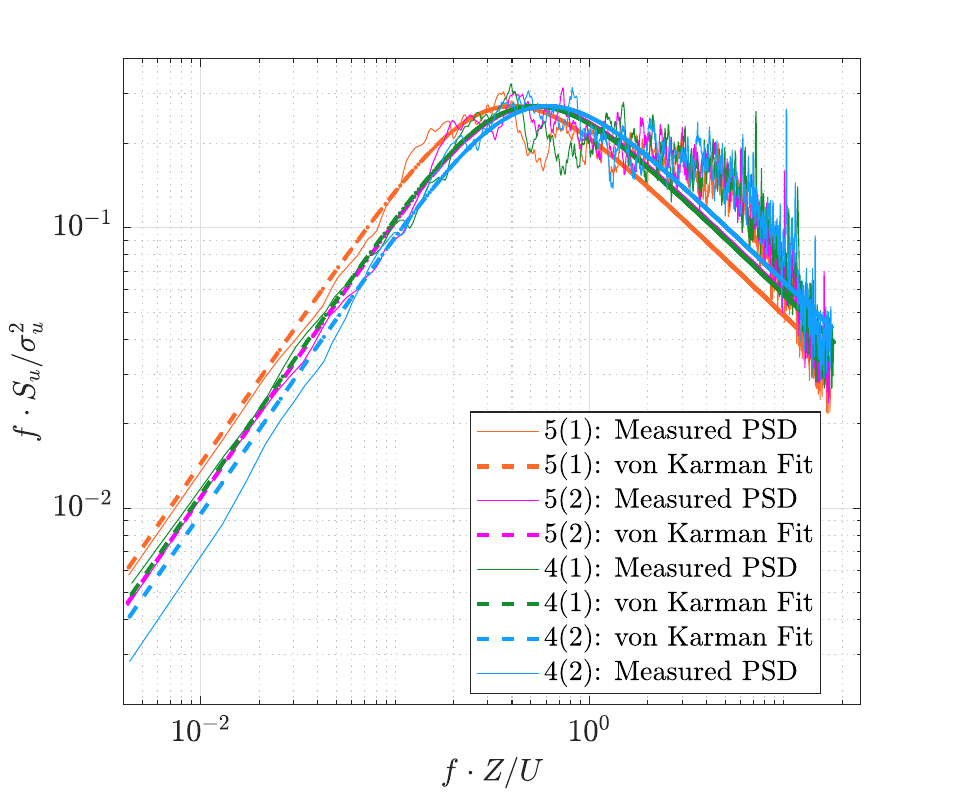}
        \caption{PSD of the test cases in the upper TS at hub height, using MHP subsampled to 250 Hz; e.g., 5(1) = 5 spires from set 1.}
        \label{fig:psd_upper}
    \end{minipage}
\end{figure}

The Power Spectral Density (PSD) of various test cases from the lower and upper test sections, at the 95 m hub height (representative value of the hub height of wind farms situated in the British Channel and the North Sea) are shown in Figures \ref{fig:psd_lower}, and \ref{fig:psd_upper}, respectively. The lower TS measurements were all acquired using LDA, equally sampled to 2.5 kHz, and the PSD was calculated using a built-in function in MATLAB, pwelch. Whereas, the upper TS measurements were obtained using MHP, subsampled to 250 Hz.

The PSD curves obtained from the experimental data, are also compared against the theoretical von K\'arm\'an spectrum, which models the turbulence energy distribution across various frequencies in homogeneous, isotropic turbulence. All the configurations exhibit the -5/3rd slope in the inertial sub range as expected, consistent with the power law decay of turbulent energy. Comparing the `Set 1' 2 spire, and the `Set 1' 2 spire with roughness (R) shows that the former with higher $I_u$, as seen in Figure \ref{fig:lowerts_standards_iU}, also gives a higher area under the PSD curve (i.e. higher the square of the r.m.s of the velocity fluctuations) by approx. 9.6\%, at an arbitrary height of 100 m.

Looking at the results from configurations from different number of spires, it is evident that number of spires is related to greater blockage, hence higher energy levels and more uniform turbulence generation; for example, PSD of the 5 spire `Set 1' shows 25\% more energy than 4 spire `Set 1' (taking the area after the $S_u$ vs $f$), at the hub height. Similarly, wider spires generate larger and stronger vortices that concentrates energy at lower frequencies. This is evident from the dominance of the 5(1) configuration, in Figure \ref{fig:psd_upper}. Similarly, the energy given by 4 spire `set 2' is the least, as it has the least $I_u$, whereas the area under the 5 spire `Set 1' is the maximum (highest $I_u$ in Figure \ref{fig:upperts_iconic_iU}).

On the other hand, the narrower spires `Set 2' tend to produce more energy at higher frequencies. The von K\'arm\'an fit aligns well with all the configurations, except for 4 spire `set 2' case, which shows a small shift to the right. Overall, it can be concluded that the trends seen in PSD are reflected in the changes in $I_u$, thereby the effect of change in geometry, and the number of spires on the $I_u$ has been revealed. 

\newpage
\subsection{Length Scales}
\label{lux_compare}
The 3 spire `Set 1', and 5 spire `Set 1' configurations were chosen for length scale comparisons with ESDU and Eurocode standards, due to their closest match to the standard profiles, as in Figures \ref{fig:lowerts_iconic_U}, \ref{fig:lowerts_iconic_iU} and \ref{fig:upperts_iconic_U}, \ref{fig:upperts_iconic_iU}, respectively. von K\'arm\'an fit and autocorrelation methods were used to compute the length scales, since the other methods in Section \ref{lxmethods} stem from the theoretical von Ka\'rma\'n fit. 

\begin{figure}[H]
\centering
\begin{minipage}[c]{0.55\textwidth}
    \centering
    \includegraphics[width=\linewidth]{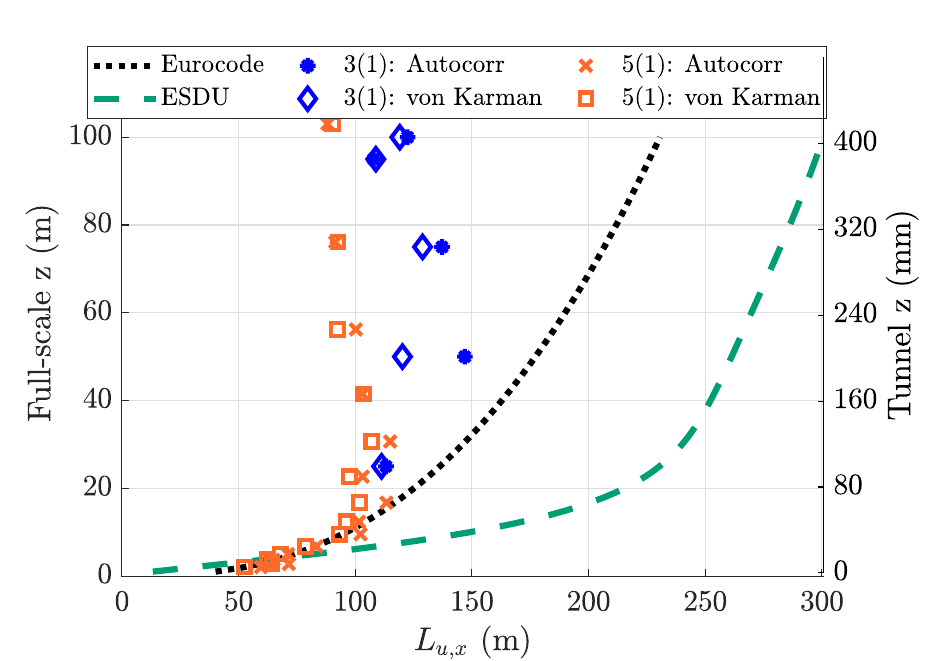}
    \caption{$L_{u,x}$ from standards vs 5 spire `Set 1' and 3 spires `Set 1', at 95 m (hub height)}
    \label{fig:Lux_standards}
\end{minipage}
\hfill
\begin{minipage}[c]{0.4\textwidth} 
    \centering
    \begin{tabular}{c|c}
    \textbf{Method}              & \textbf{Value (m)} \\ \hline \hline
    von K\'arm\'an fit               & 0.30               \\
    Spectra peak                 & 0.29               \\
    y-intercept                  & 0.35               \\
    Autocorrelation              & 0.31               \\
    \end{tabular}
    \captionof{table}{Comparison of $L_{u,x}$ for the 2 Spires `Set 2', at 95 m (hub height), using LDA subsampled to 2500 Hz}
    \label{tab:Lux_hubht}
\end{minipage}
\end{figure}

Results from these two methods showed good agreement across both the chosen test cases, as demonstrated in Figure \ref{fig:Lux_standards}. On comparing the wind tunnel results with the standard profiles, it can be observed that length scales increase with height initially and then plateau off to a nearly constant value, whereas the standard profiles show that the length scales increase with height in reality, from Figure \ref{fig:Lux_standards}. In essence, a considerable difference can be seen between the length scales from experimental data and those from standards. This is an expected trend, as pointed out by Counihan \cite{coun06}, and Kozmar \cite{kozmar2011wind}. The standards assume idealised ABL conditions, which usually differ from the wind tunnel environment. Furthermore, in full-scale, length scales grow without restricting spatial bounds, but since the wind tunnel is a bounded domain, the length scales are limited from growing. Additionally, the von K\'arm\'an method assumes a spectral fit, whereas the autocorrelation method is sensitive to noise, probe size, data quality, etc. - this can affect the reliability of the length scales calculated using these methods. 

Industry standards generally consider it acceptable to use length scale values from wind tunnel data that reach at least one-third of the corresponding full-scale measurements \cite{awes2019quality}. This aligns closely with the results obtained in this project, as shown in Figure \ref{fig:Lux_standards}.

Since the 2 spire `Set 1' test closely resembled the `Profile WF' shown by Figure \ref{fig:lowerts_iconic_U} and \ref{fig:lowerts_iconic_iU}, the longitudinal length scale can be calculated at the representative hub height of turbines in North Sea/ English Channel (95 m), using various methods described in \ref{lxmethods}, to compare the results obtained. The values are given in Table \ref{tab:Lux_hubht}, where the maximum difference is observed between the methods y-intercept, and using the spectra peak. An error of about 3\% has been observed in the $L_{u,x}$ estimation given by the von K\'arm\'an fit and the autocorrelation method. 

\newpage
\section{Conclusion and Future Work}

This project investigated the effectiveness of replicating atmospheric boundary layers in the 10'$\times$5' wind tunnel using Irwin-type spires. Systematic testing examined the influence of spire geometry, number of spires, and their spacing, by comparing results against target profiles `Profile 1' (establishing the commonality between various wind standards, via a common condition of 25 m/s at 10 m), and `Profile WF' (profile which is the most suitable for wind farm inflow in the English Channel and the North Sea). For the marine ABLs, the modelling indicated the required roughness to be of very small size, and it was found that the skin friction against the existing floor worked reasonably well already. 

Overall, the Irwin spires proved highly effective in generating realistic and comparable ABL flow profiles. The generated velocity profiles closely matched the target. The spanwise measurements acquired in the lower, and upper test sections demonstrated a consistent spanwise uniformity in the flow (across a majority of the test cases), with the profiles collapsing across different runs and flow speeds. One of the tests also indicated the need to avoid measurements near the tunnel wall in spanwise flow checks, due to the potential distortion from wall boundary layer interactions. Hence, this deduces the practical applicability of Irwin's methodology as being simple, cost-effective, and efficient approach for modelling atmospheric characteristics in a laboratory setting \cite{coun06}. 

Key findings showed that `Set 1' spires aligned well with `Profile 1', making them suitable for replicating the common features between the standards. `Set 2' spires are in some ways a variant of `Set 1', designed and tailored for replicating `Profile WF', profiles approaching wind turbine farms offshore of the Northern Sea or in the English Channel. The recommended downstream distance to simulate a fully developed atmospheric boundary layer is at 6 times the spire height, though measurements taken at approx. $5.5 \cdot h$ (in the upper TS measurements), still showcased a reasonable spanwise uniformity. 

Despite the success story observed in the lower TS, the upper test section exhibited inherently higher turbulence intensities due to the influence of large tunnel fans, particularly at the lower heights. This led to a noticeable, but small, deviation from the target turbulent intensities. Addressing this elevated turbulence remains a potential area for improvement in future experiments. 

Velocity measurements acquired using both the techniques, LDA, and MHP, were cross-validated, to assess their reliability. Though the MHP showcased its limitations in capturing the smaller scales of turbulence, particularly in the $V$ and the $W$ directions near the wall, it still remained as a robust complementary tool to LDA. However, it was confirmed that the larger scales of turbulence that dominate turbulence characteristics and integral length scales, would be retained, via spectral analysis. This justified the use of the MHP in the upper section tests, where LDA measurements are not feasible. Deviations from the expected symmetric and collapsing behaviour in $V$ and $W$ components were recorded, due to the uncertainties that arose from misalignments between the calibration case and the use case of the MHP coordinate axes. Nevertheless, error propagation analysis revealed these deviations fluctuated outside the fluctuations shown, possibly revealing a minimal impact. LDA exhibited smaller misalignment errors, though it is also susceptible if the probe axes are not precisely aligned with the tunnel axes, and if the seeding was not sufficient in terms of quality and quantity.  

This project further compared the variations in length scale from experimental data with standard profiles. Length scale values calculated using the von K\'arm\'an fit, and the autocorrelation methods closely agreed with each other, differing by an error of about 3\%. However, the wind tunnel data values of $L_{u,x}$ are about one-third of the full-scale values, which are often deemed acceptable for industrial applications \cite{awes2019quality}. The fewer assumptions and direct velocity measurement basis followed by the autocorrelation method makes it the preferred approach. Changes in the spire geometry and number was seen to influence the PSD and the turbulence intensity, verifying that the area under the PSD vs $f$ related to the mean square of the velocity fluctuations, and hence the turbulence intensity. 

Regarding spire configuration, the study clarified the spacing issue, whether the $h/2$ spacing should be based on the original spire height or the truncated height. For the unified standards profile, a $h/2$ spacing with the original spire height proved to work best, with the 3 spire `Set 1' tests yielding the closest match to within $\pm10\%$, and the 2 spire `Set 1' also nearly overlapping with the match. 5 spire `Set 1' provided the best results at the upper TS, with the 4 spire `Set 1' also performing well. For the specific wind farm profile, a $h/2$ spacing based on the truncated spire was optimal, with 2 spire `Set 2' best matching for the lower TS, and the corresponding 4 spire `Set 2' in the upper TS. 

For standardised ABL profiles, `Set 1' spires are recommended, while `Set 2' suits site-specific profiles like for wind farm applications. The choice of spacing depends on the target profile application. More importantly, the spire design (set number, in this case) proved to be more important than the number of spires, which would help simplify the experimental setup, since only minor adjustments in the spire count would be necessary, keeping the same fundamental spire design. 

This thesis has thereby demonstrated a solid, low-cost, and effective approach to simulate the ABL and replicate the desired flow characteristics, utilising Irwin spires, even without additional roughness elements (depending on the profile being generated). This provides valuable insights toward standardising ABL generation in wind tunnels, supporting applications such as offshore wind turbine design and control. By advancing coordinated and efficient modelling strategies beyond traditional methods, this work lays a foundation for future aerodynamic optimisation of wind energy systems and many more. 

\subsection{Future Work}

It would be beneficial to integrate the floor roughness elements to complement the spires, as suggested by Irwin, which could better replicate the overall ABL profile. To achieve the intensity of turbulence for the marine boundary layer, a low-impact roughness material like a carpet-textured surface can be installed, to improve the accuracy of the flow being replicated. Further investigation is necessary to understand and reduce the inherently high turbulence intensities observed in the upper test section, majorly influenced by the large tunnel fans. Fidelity of the upper ABL replication can be enhanced if strategies are developed to mitigate these fan-induced flow disturbances. To improve the precision of velocity measurements, a fixed hexagonal mount for the MHP could be developed to ensure consistent and accurate alignment with the flow direction, reducing repeatability errors across tests. This would also help to reduce/ eliminate the anomalies seen in $V$, and $W$, by avoiding misalignment in the coordinate and the calibration axes. Furthermore, a three-dimensional automated traverse system would help to improve spatial coverage, accuracy, and efficiency of the measurements, in both LDA and the MHP. Furthermore, multiple MHPs could acquire measurements simultaneously to facilitate comprehensive three-dimensional flow mapping and improve understanding of the spatial correlations within the boundary layer. More precise installation and measurement tools, with documented tolerances, could benefit future experiments. This can help reduce errors such as probe misalignment, laser positioning offsets, and repeatability issues concerning the traverse. Interpolation errors associated with relying on the lower section measurements can be eliminated, by installing a dedicated pitot tube in the upper test section.

Active control systems, such as motorised flaps and dynamic roughness elements, can be integrated to create adaptive ABL simulations. This would allow real-time modulation to mimic gust fronts, seasonal variations, enabling a more realistic performance assessment of wind turbines. Additionally, a hybrid validation framework linking wind tunnel measurements with LES could be developed through a feedback loop. This would enable high-fidelity tuning and validation of turbulent models using the real ABL data, improving the accuracy of designs for wind turbines, buildings, and so on. These ABL studies can be extended to analyse the interactions between wind farms and other renewable energy sources like solar farms, accounting for microclimate changes due to combined system operation. 

% Future studies can focus on replicating thunderstorm wind profiles, which differ significantly from the conventional BL profiles developed in this project. Thunderstorms and downbursts generate strong, highly turbulent low-level jets where the mean wind speed peaks at low altitudes and then decreases with height, opposite to typical wind tunnel profiles. The complex turbulence during these storms involves multiple scales and varying intensities. To better simulate these conditions, future experiments could use an inclined plate at the top of the tunnel to induce the low-level jet while adjusting floor roughness to control turbulence levels \cite{future}. 

By pursuing these recommendations, future research can significantly advance the precision and applicability of laboratory ABL simulations, supporting improved aerodynamic design and control strategies for wind energy and related fields. In conclusion, this work affirms that while there may be no single method that perfectly reproduces all aspects of a boundary layer, Irwin's approach offers a compelling balance between simplicity, cost, and performance, making it a practical solution for experimental and engineering applications.

\clearpage
\newpage
\addcontentsline{toc}{section}{References}
\pagestyle{empty}  % No page numbers on references (optional)
\bibliography{references}

\clearpage
\newpage

% Start appendix page numbering and style
\pagenumbering{roman}    % lowercase roman page numbers
\setcounter{page}{1}     % start at i

% Set page style to show page numbers (usually 'plain' works)
\pagestyle{plain}        

% Add Appendix heading (unstarred to number it, or star it for unnumbered)
\section*{Appendix}  
\addcontentsline{toc}{section}{Appendix} % add to ToC manually

% Reset subsection counter and label subsections by letters
\setcounter{subsection}{0} 
\renewcommand{\thesubsection}{\Alph{subsection}} 

% \section{Appendix}

\subsection{Simulation of the Urban Profile}

\subsubsection{Profiles from the Standards}
\label{urbanprofiles}
\begin{figure}[H]
    \centering
    \begin{minipage}{.47\linewidth}
        \centering
        \includegraphics[width=1.1\linewidth]{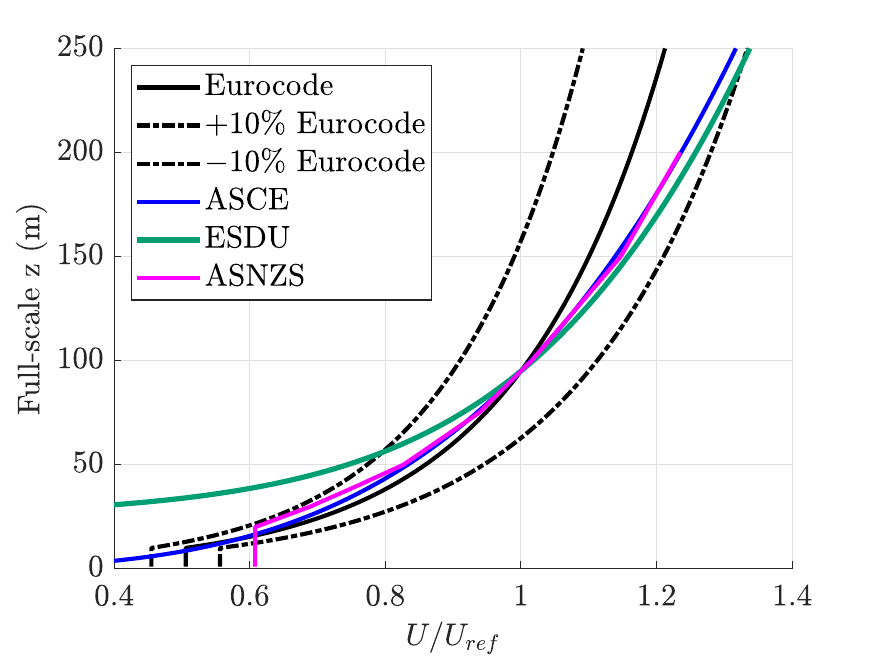}
        \caption{Normalised $U$ profile for the urban standards, wrt 95 m}
        % \label{fig:upperts_iconic_U}
    \end{minipage}\hfill
    \begin{minipage}{.47\linewidth}
        \centering
        \includegraphics[width=1.1\linewidth]{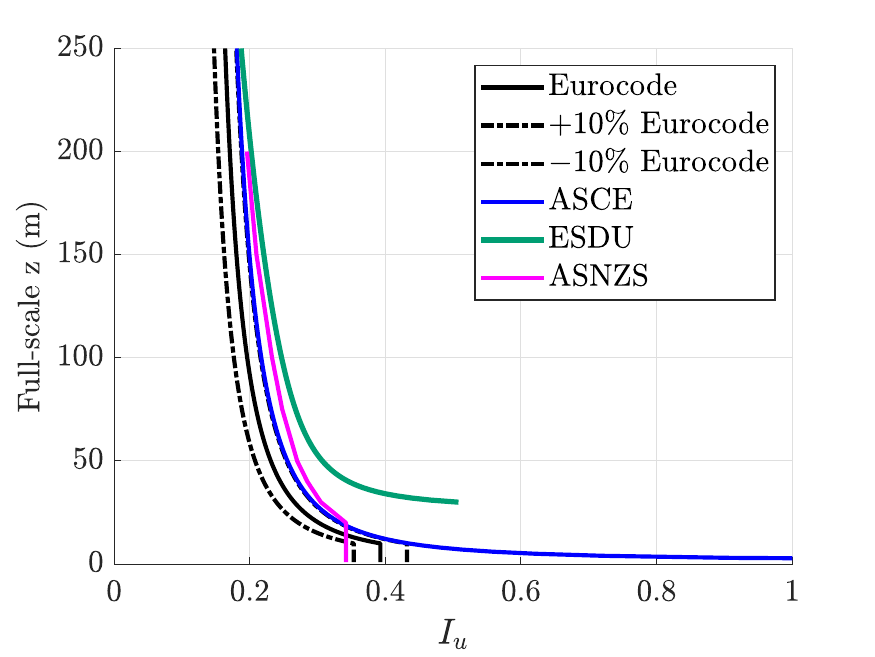}
        \caption{Turbulence intensity for the urban standards}
        % \label{fig:upperts_iconic_iU}
    \end{minipage}
\end{figure}

\subsubsection{Spire and Roughness design}

\begin{table}[H]
\centering
\begin{tabular}{c|c|c|c|c}
\textbf{Parameter} & 
\makecell{\textbf{Lower TS} \\ \textbf{Original}} & 
\makecell{\textbf{Lower TS} \\ \textbf{Truncated}} & 
\makecell{\textbf{Upper TS} \\ \textbf{Original}} & 
\makecell{\textbf{Upper TS} \\ \textbf{Truncated}} \\ \hline \hline

$h$ (m)            & 1.6  & 1.5  & 1.6  & 1.5  \\
$b_1$ (mm)         & 224  & 224  & 210  & 210  \\
$b_2$ (mm)         & N/A  & 9.3  & N/A  & 8.7  \\
$bs$ (mm)          & \multicolumn{2}{c|}{400}  & \multicolumn{2}{c}{400}  \\
Spacing (m) [no.]        & \multicolumn{2}{c|}{
    \makecell{0.8 [3] \\ 0.7 [3]}
} & \multicolumn{2}{c}{
    \makecell{0.8 [6] \\ 0.7 [7]}
} \\
$k$ (mm)           & \multicolumn{2}{c|}{31}   & \multicolumn{2}{c}{31}   \\
$D$ (cm)           & \multicolumn{2}{c|}{20}   & \multicolumn{2}{c}{20}   \\
\end{tabular}
\caption{Spire and roughness elements' design parameters - for simulating urban terrain type, with $\delta=500$ m, and test cale = $1:400$}
\label{tab:spire_params}
\end{table}

\newpage
\subsection{2 Spires `Set 1' - LDA and MHP Measurements}
\label{badlda}
\begin{figure}[H]
    \centering    \includegraphics[width=0.5\linewidth]{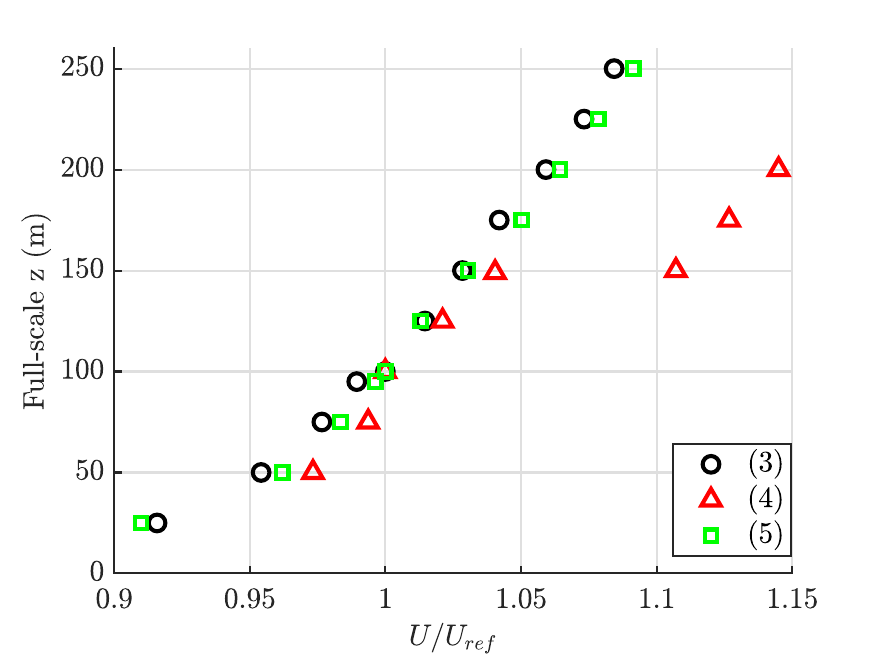}

    \caption{Normalised $U$ of all the LDA spanwise, 2 spire 'set 1'}
    \label{fig:normU_bad}
\end{figure}

\begin{figure}[H]
    \centering    \includegraphics[width=0.5\linewidth]{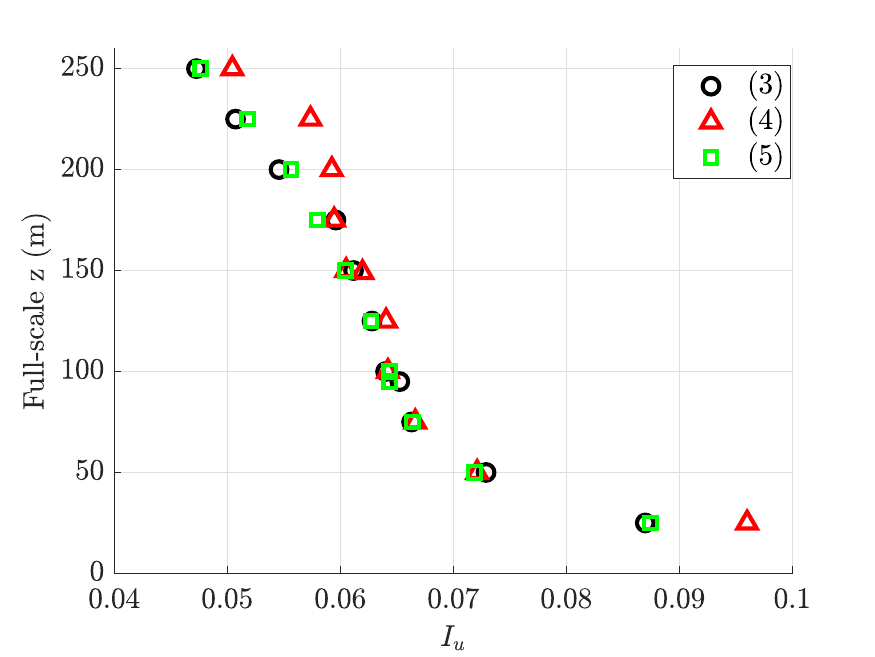}

    \caption{Turbulence intensity in $U$ of all the LDA spanwise, 2 spire 'set 1'}
    \label{fig:iU_bad}
\end{figure}

\begin{figure}[H]
    \centering    \includegraphics[width=0.5\linewidth]{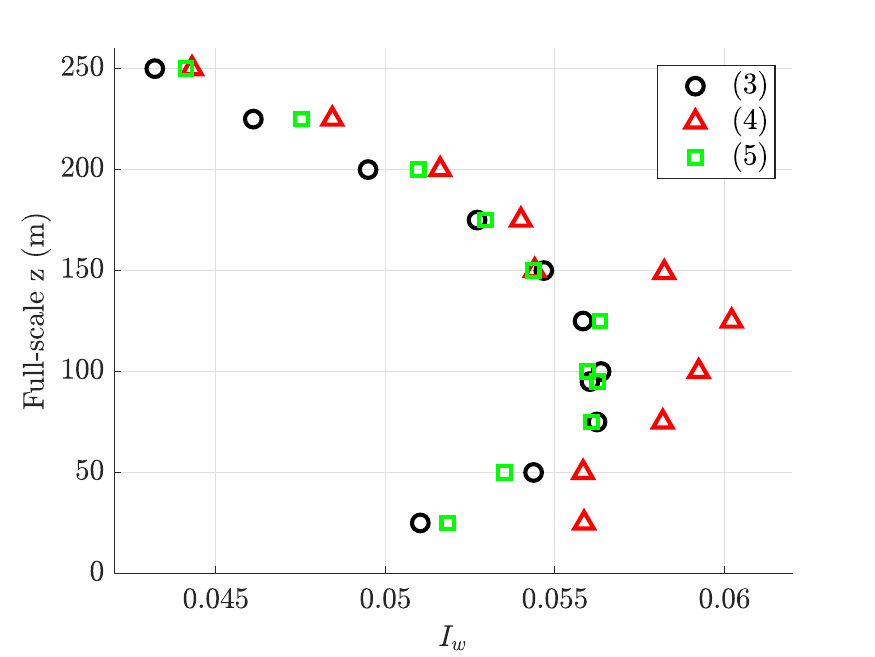}

    \caption{Turbulence intensity in $W$ of all the LDA spanwise, 2 spire 'set 1'}
    \label{fig:iW_bad}
\end{figure}

\newpage
\subsection{Traverse Design}

\subsubsection{Original Traverse}
\label{oldtraverse}
\begin{figure}[H]
    \centering
    \begin{minipage}{.47\linewidth}
        \centering
        \includegraphics[width=0.8\linewidth]{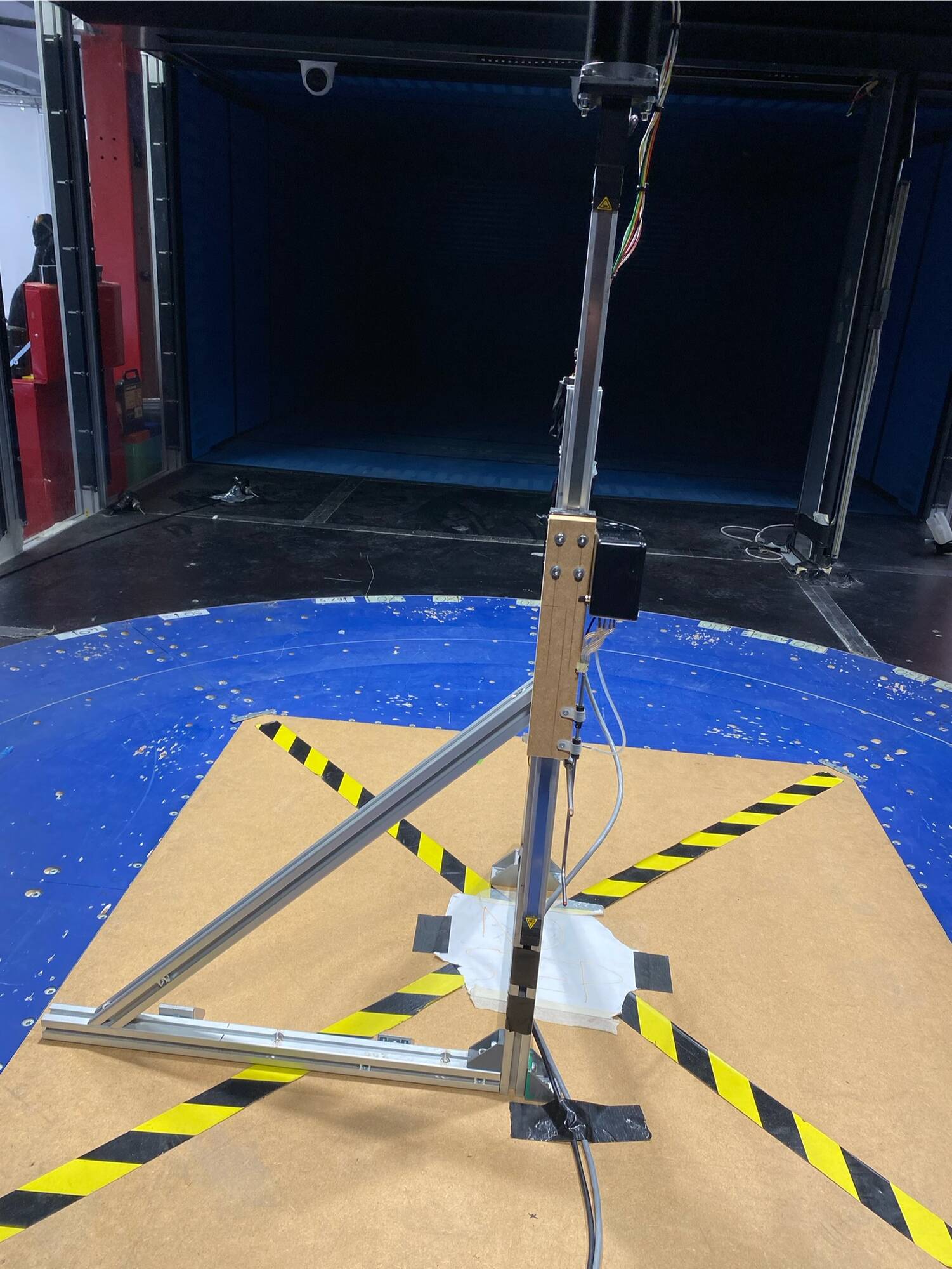}
        \caption{Original traverse design for the MHP}
        % \label{fig:upperts_iconic_U}
    \end{minipage}\hfill
    \begin{minipage}{.47\linewidth}
        \centering
        \rotatebox{-90}{% Negative for clockwise rotation
        \includegraphics[width=1\textwidth]{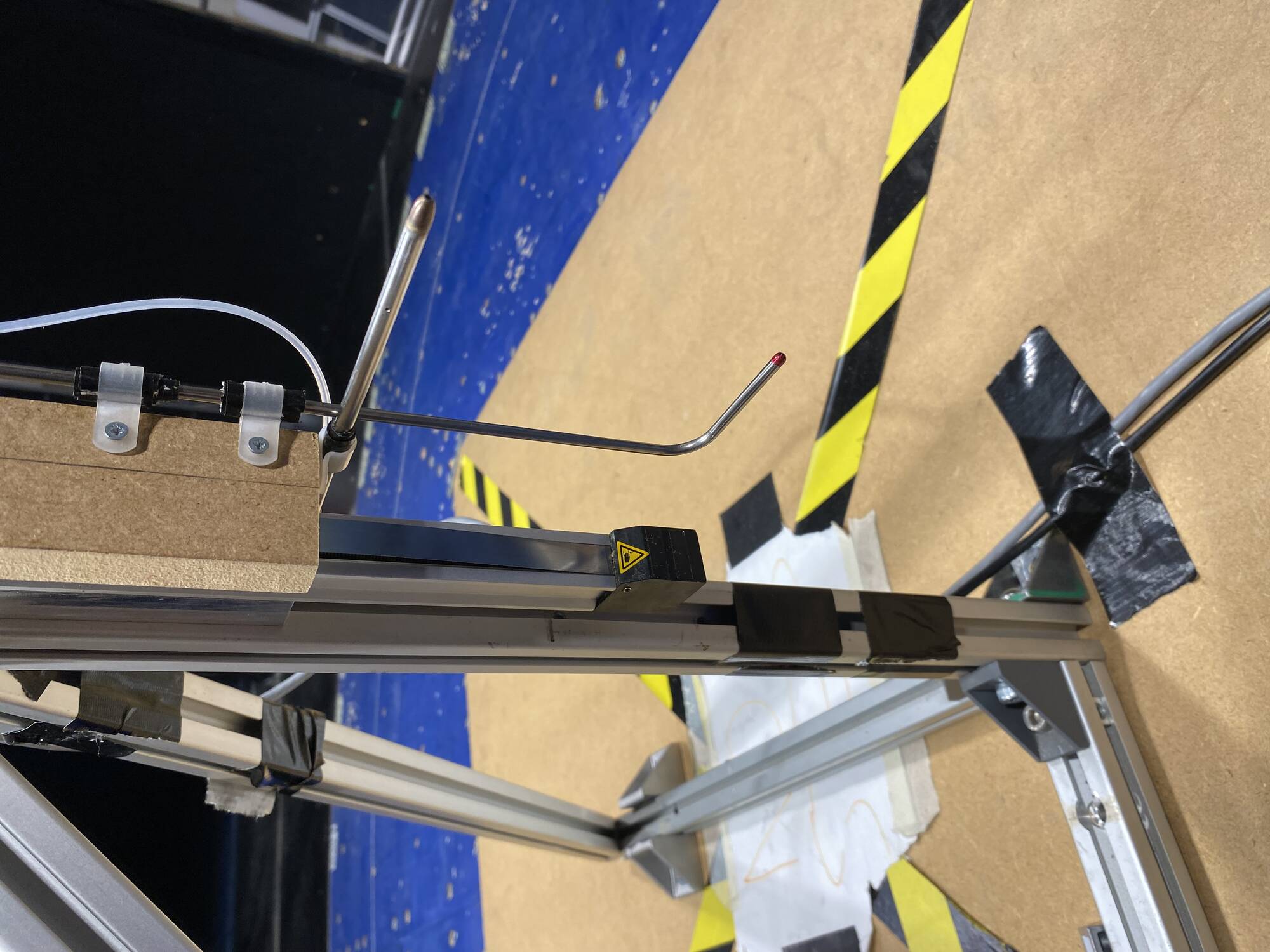}
    }
        \caption{Original traverse design for the MHP - closeup}
        % \label{fig:upperts_iconic_iU}
    \end{minipage}
\end{figure}

\subsubsection{Modified Traverse}
\label{newtraverse}
\begin{figure}[H]
    \centering
    \rotatebox{-90}{% Negative for clockwise rotation
        \includegraphics[width=0.6\textwidth]{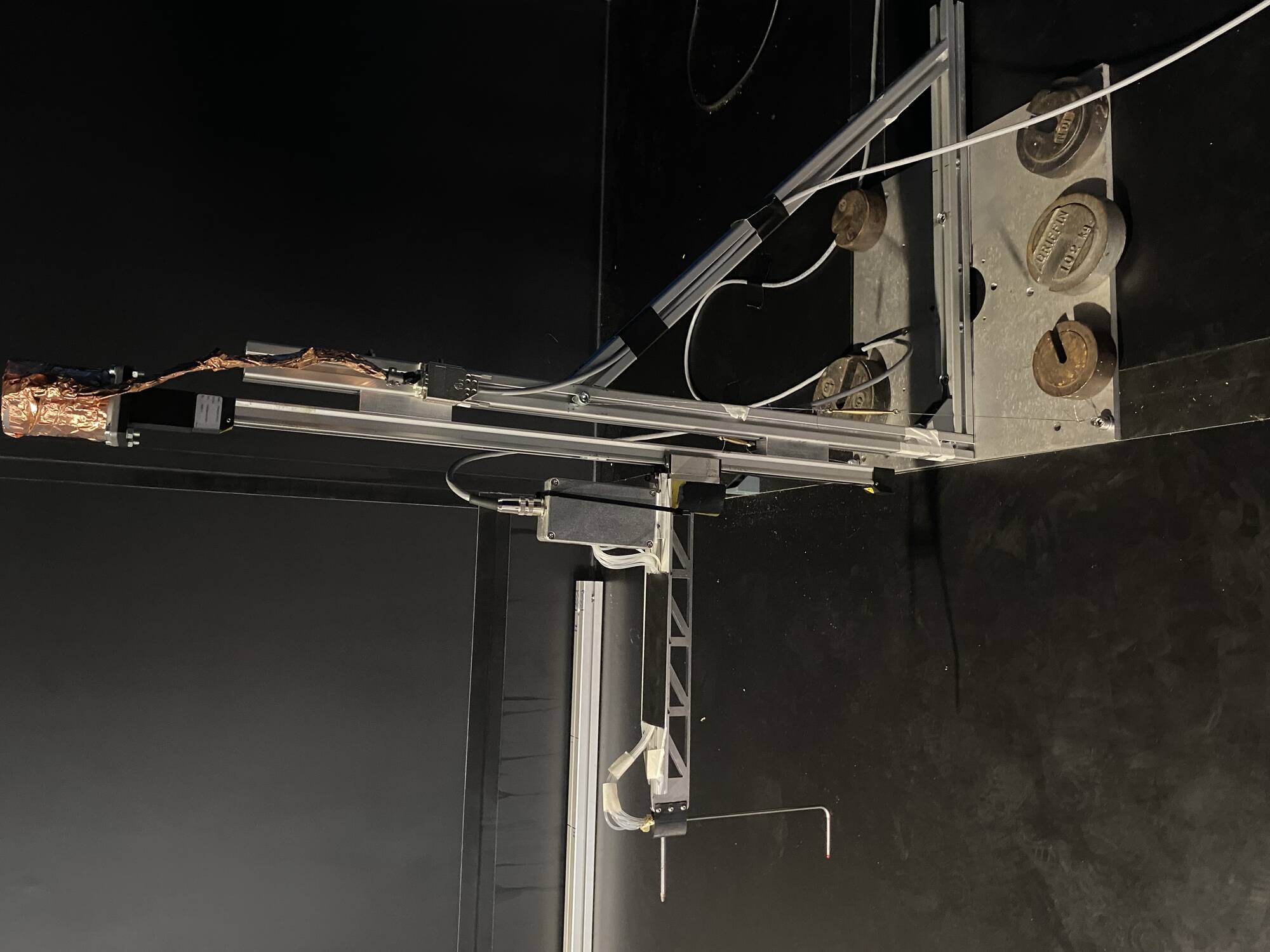}
    }
    \caption{New traverse design for the MHP}
    % \caption{Your rotated figure caption}
    % \label{fig:rotated_figure}
\end{figure}

\subsubsection{Traverse setup}

\label{traverse_setup}
\begin{figure}[H]
    \centering
    \begin{minipage}{.47\linewidth}
        \centering
         \rotatebox{-90}{% Negative for clockwise rotation
        \includegraphics[width=1.1\textwidth]{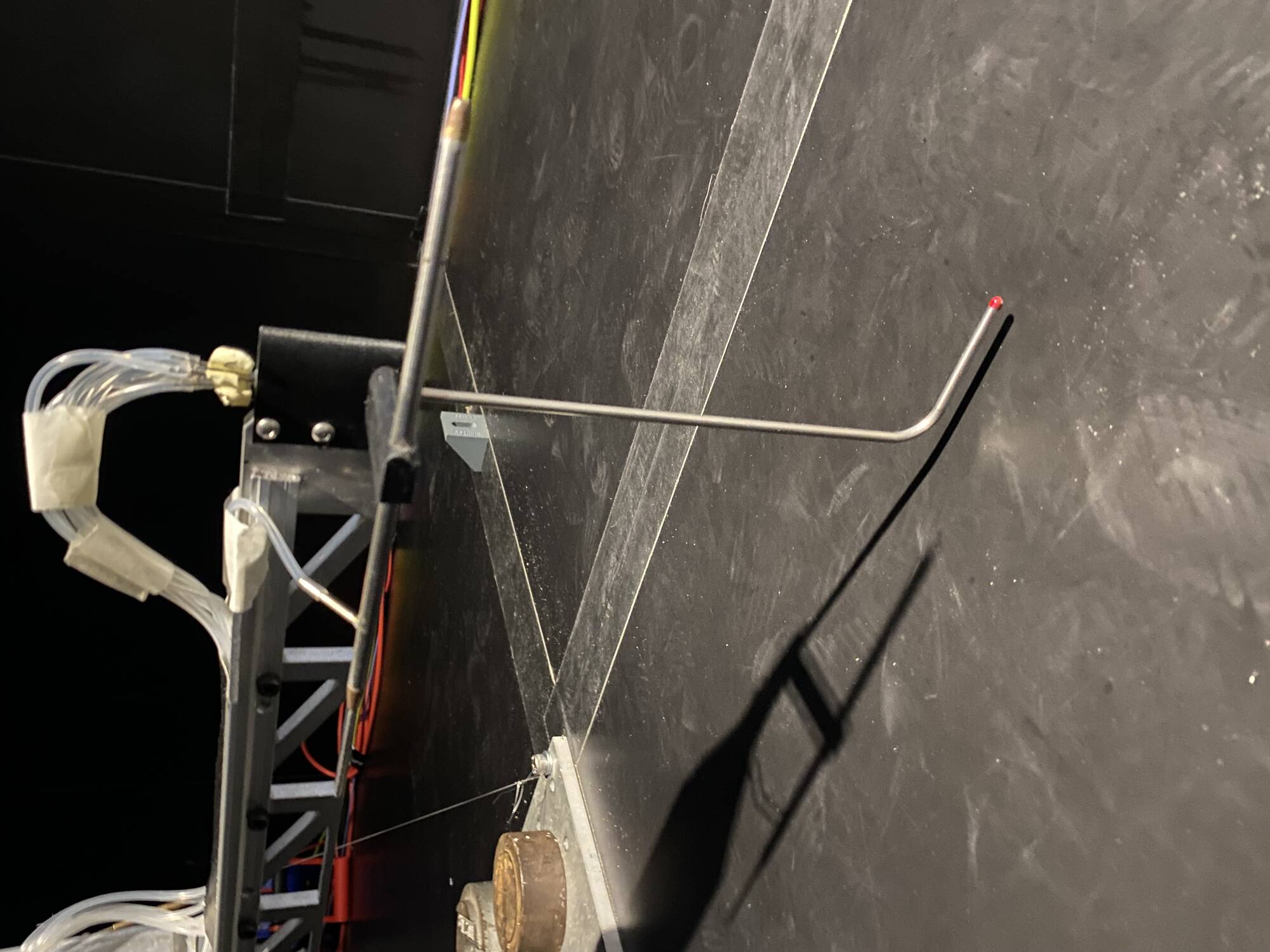}
    }
        \caption{MHP when closest to the floor at 8.13 mm approx.}
        % \label{fig:upperts_iconic_U}
    \end{minipage}\hfill
    \begin{minipage}{.47\linewidth}
        \centering
        \rotatebox{-90}{% Negative for clockwise rotation
        \includegraphics[width=1.1\textwidth]{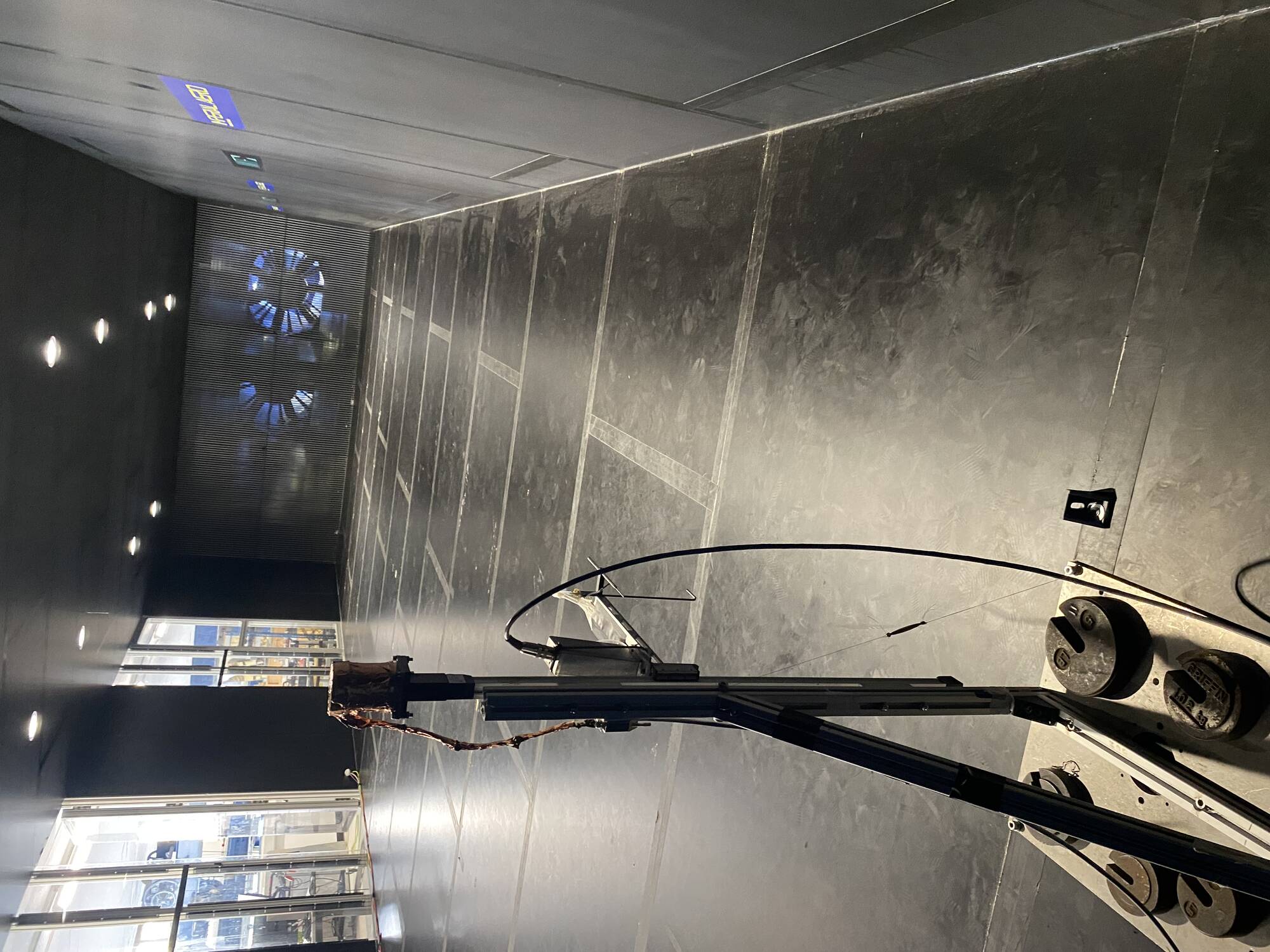}
    }
        \caption{The modified traverse being used in the upper TS}
        % \label{fig:upperts_iconic_iU}
    \end{minipage}
\end{figure}

\begin{figure}[H]
    \centering    \includegraphics[width=0.5\linewidth]{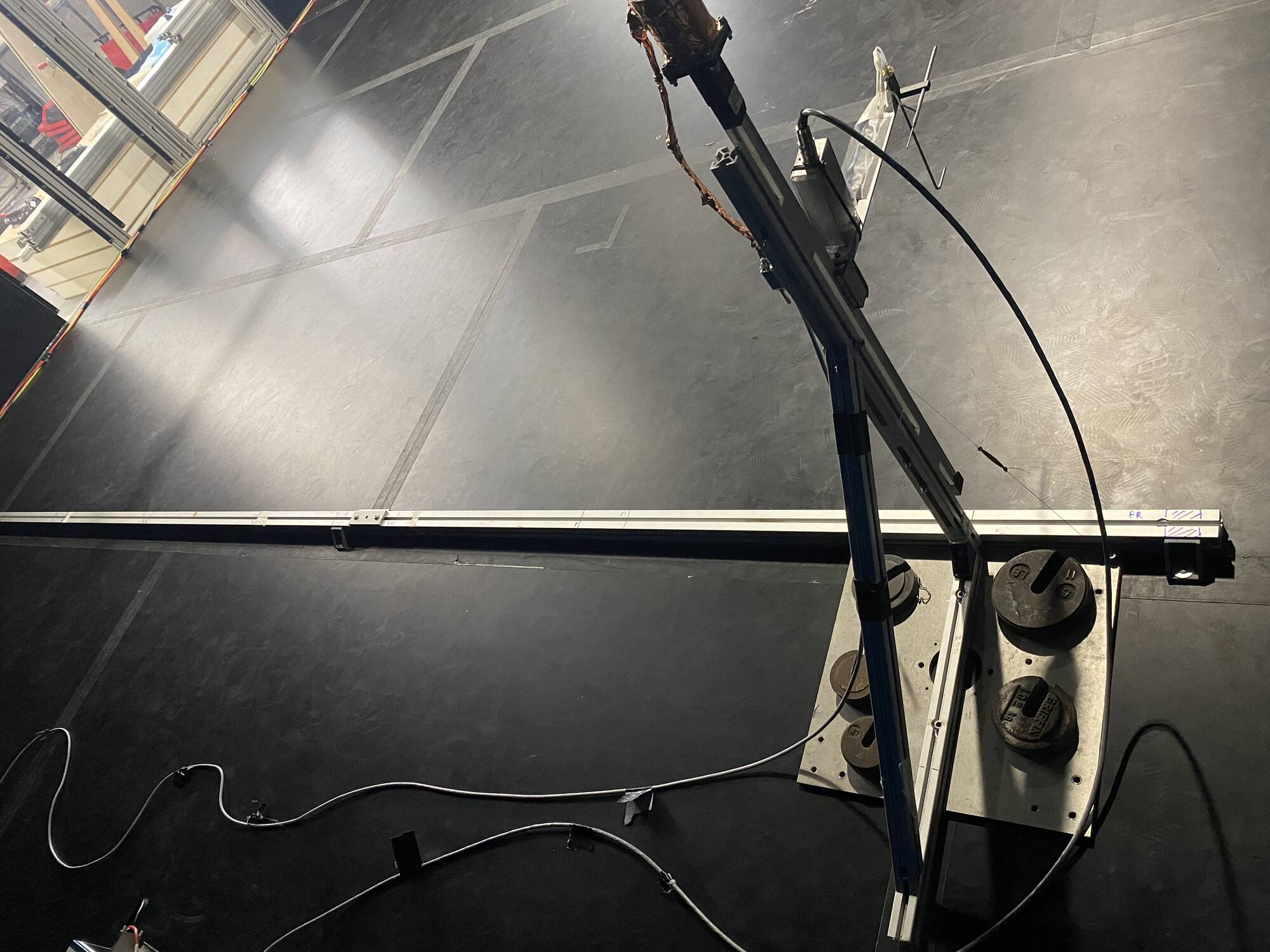}
    \caption{Traverse being aligned spanwise using a minitec (removable) on the tunnel floor}
    \label{fig:iU_bad}
\end{figure}

\subsubsection{MHP Sensor Box}

\begin{figure}[H]
    \centering    \includegraphics[width=0.4\linewidth]{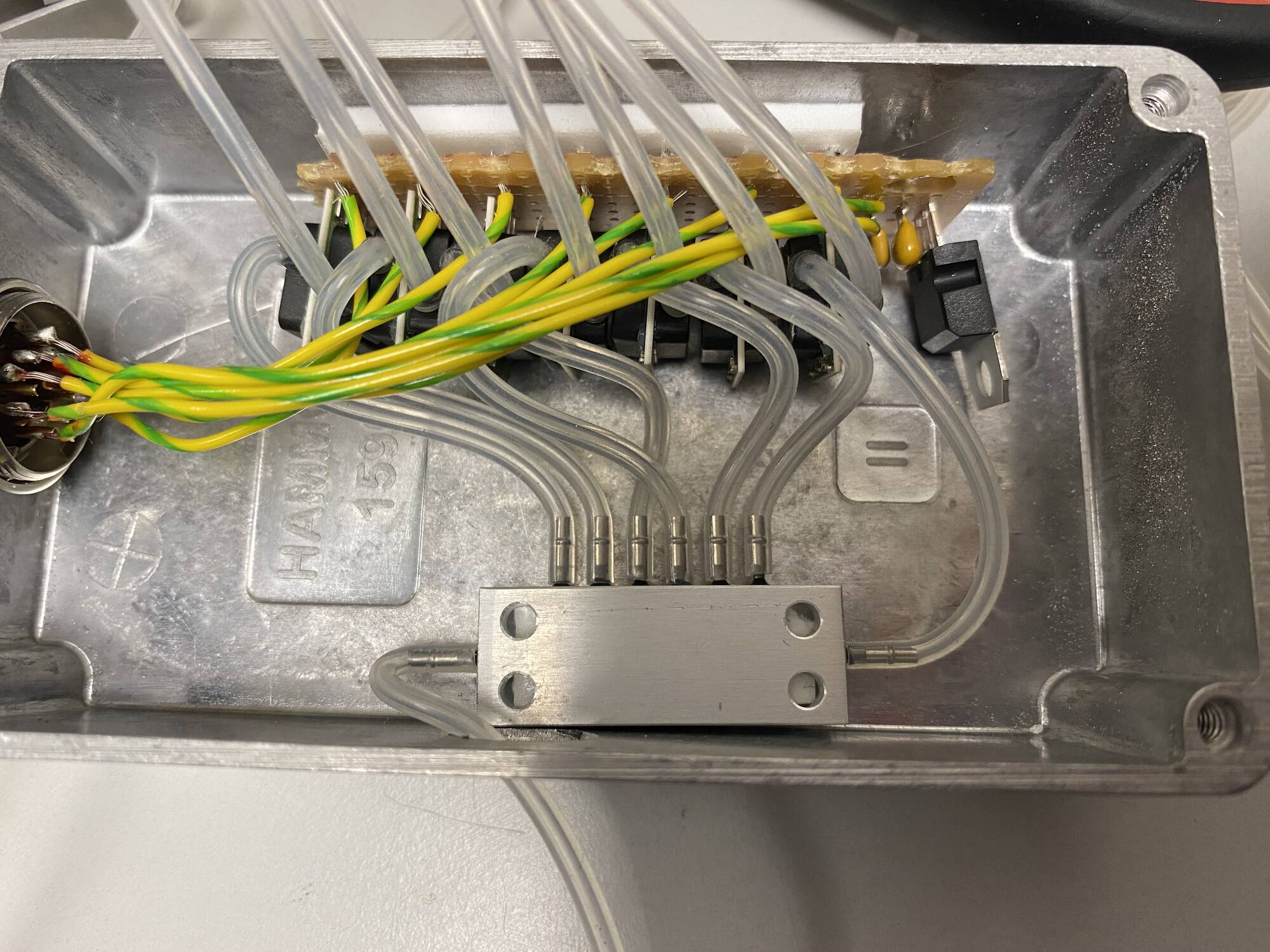}
    \caption{MHP's sensor box}
    \label{fig:iU_bad}
\end{figure}

\subsubsection{CAD designs}
\label{cad}
\textbf{1. MHP and static holder}

\begin{figure}[H]
    \centering    \includegraphics[width=0.7\linewidth]{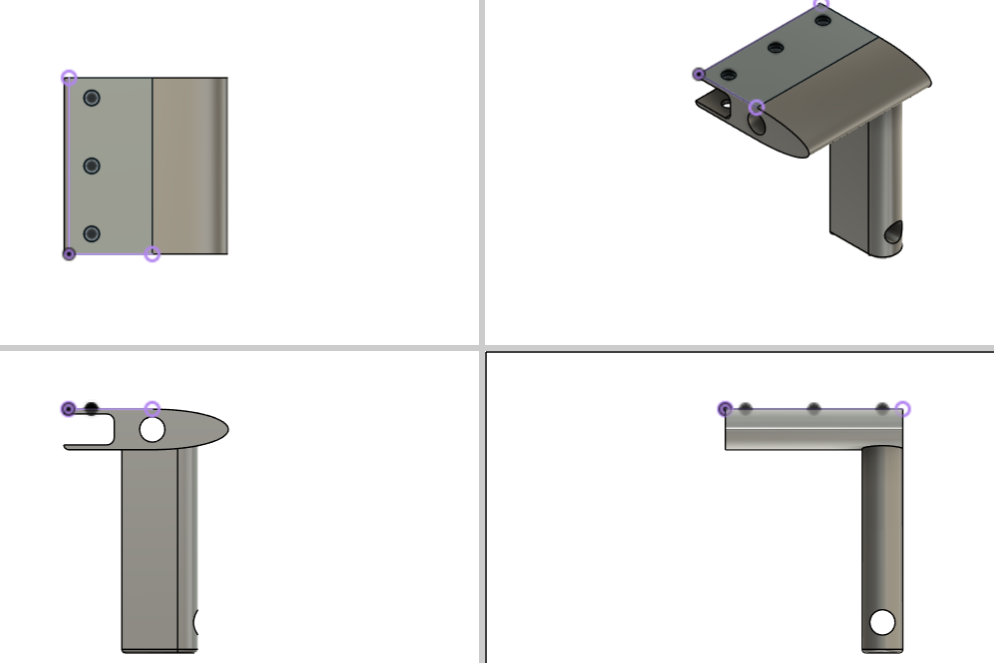}
    \caption{CAD of the holder for MHP and static}
    \label{fig:iU_bad}
\end{figure}

\textbf{2. Aluminium truss structure}

\begin{figure}[H]
    \centering    \includegraphics[width=0.75\linewidth]{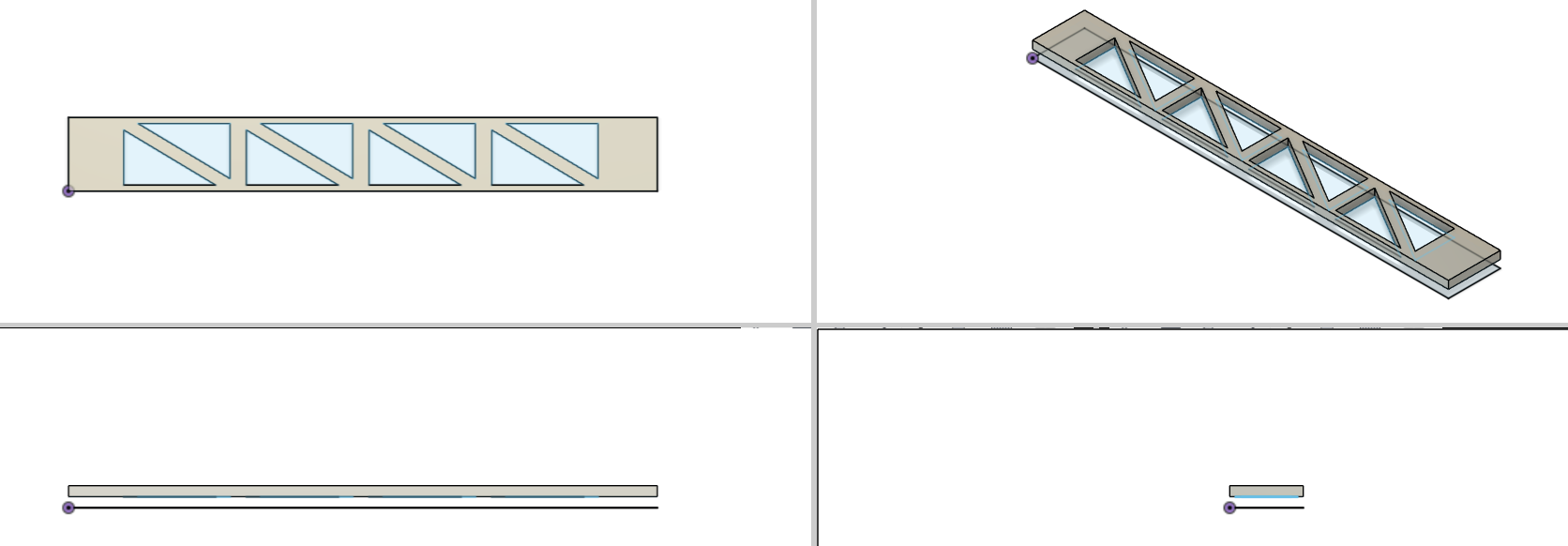}
    \caption{CAD of the Aluminium truss structure for the traverse}
    \label{fig:iU_bad}
\end{figure}

\newpage
\subsection{Wind Tunnel Set-up}

\subsubsection{Green mesh roughness - Lower TS}
\label{greenmesh}
\begin{figure}[H]
    \centering    \includegraphics[width=0.5\linewidth]{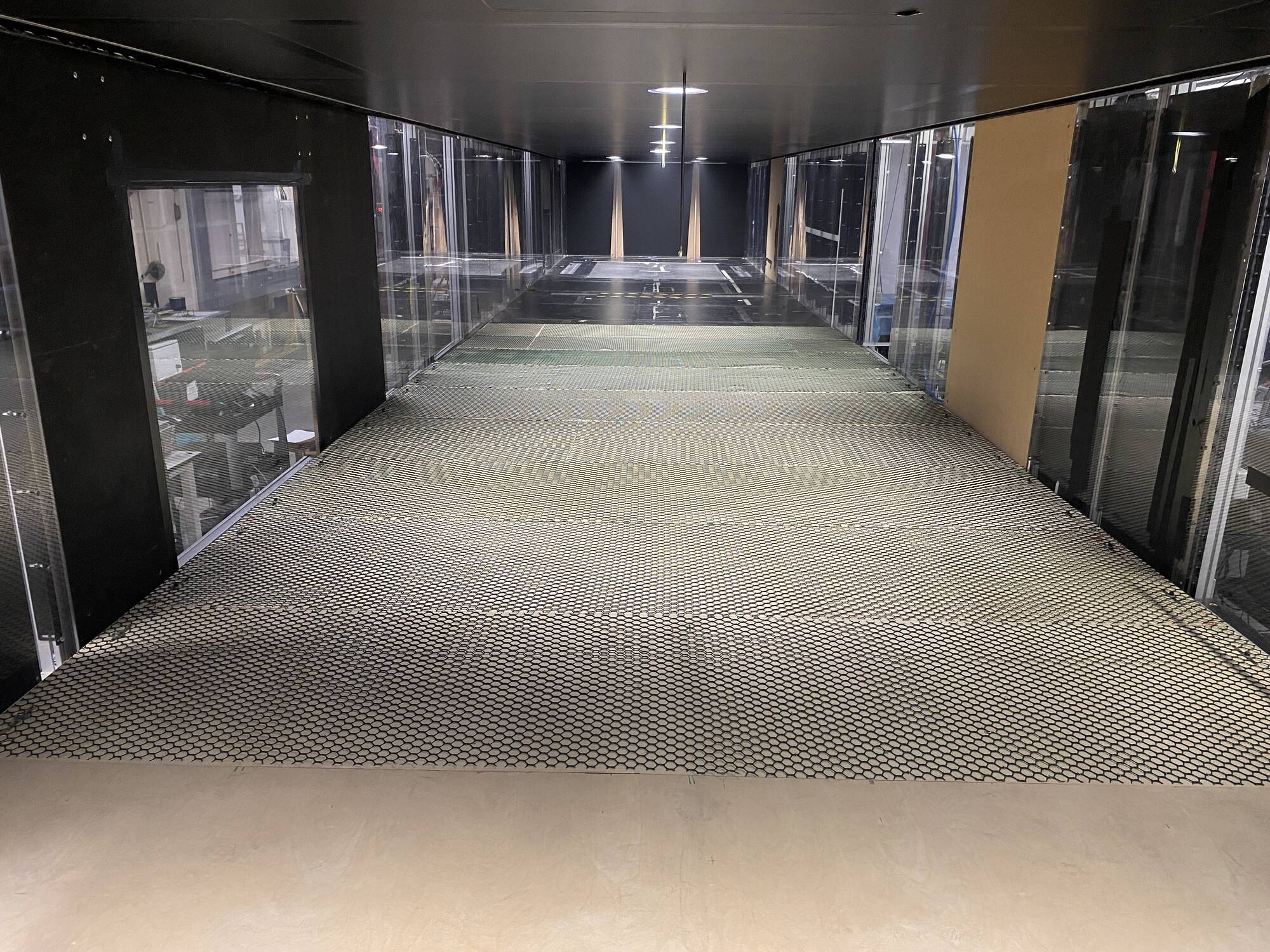}
    \caption{Green mesh roughness with 2 spire `set 1' - lower TS}
    \label{fig:iU_bad}
\end{figure}

\subsubsection{2, 3 spires - Lower TS}
\label{lowerspires}
\begin{figure}[H]
    \centering
    \begin{minipage}{.47\linewidth}
        \centering
        
        \includegraphics[width=1.1\textwidth]{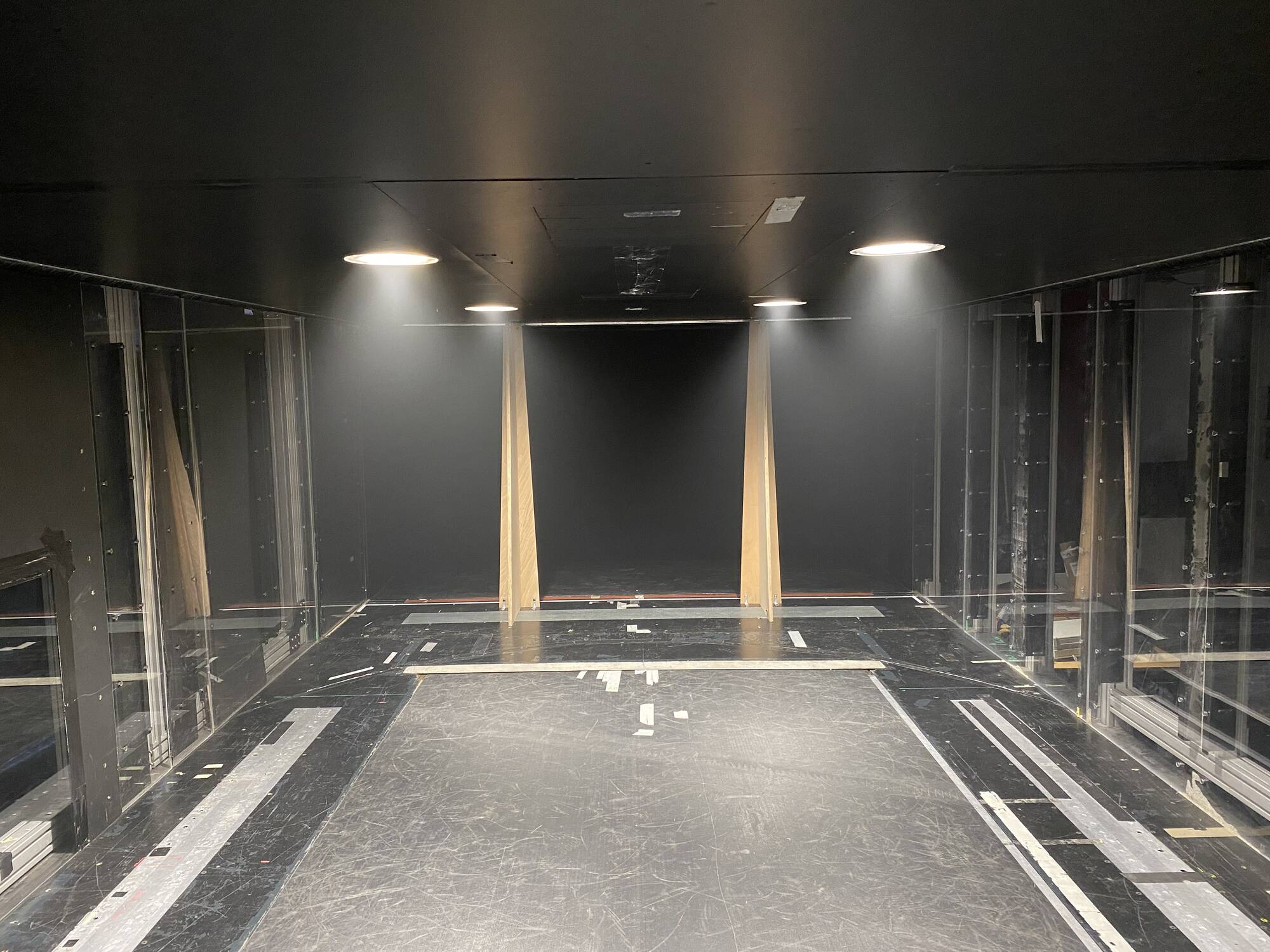}
    
        \caption{2 spires installed in the lower TS}
        % \label{fig:upperts_iconic_U}
    \end{minipage}\hfill
    \begin{minipage}{.47\linewidth}
        \centering
       
        \includegraphics[width=1.1\textwidth]{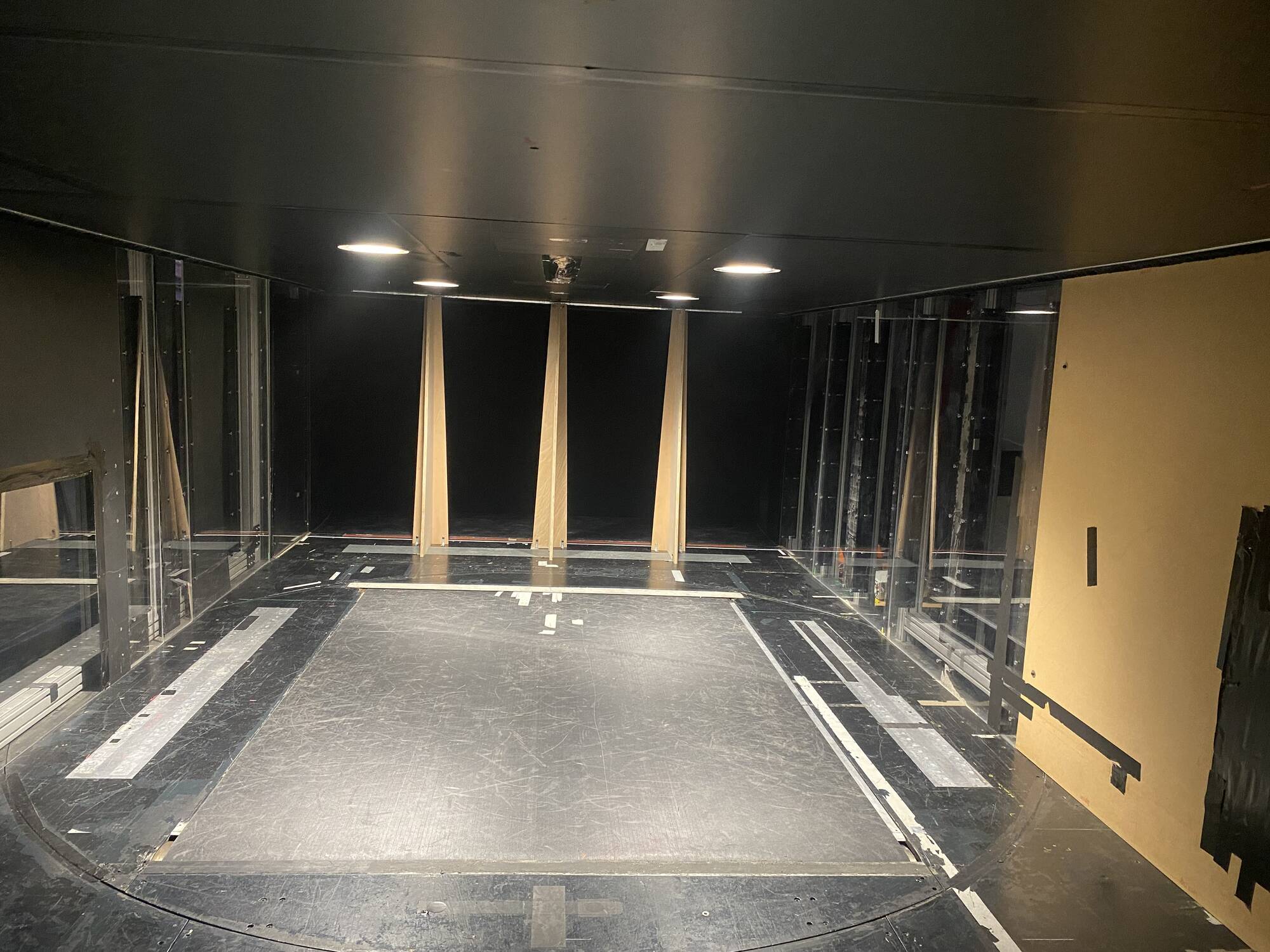}
    
        \caption{3 spires installed in the lower TS}
        % \label{fig:upperts_iconic_iU}
    \end{minipage}
\end{figure}

\subsubsection{4, 5 spires - Upper TS}
\label{upperspires}
\begin{figure}[H]
    \centering
    \begin{minipage}{.47\linewidth}
        \centering
       
        \includegraphics[width=1.1\textwidth]{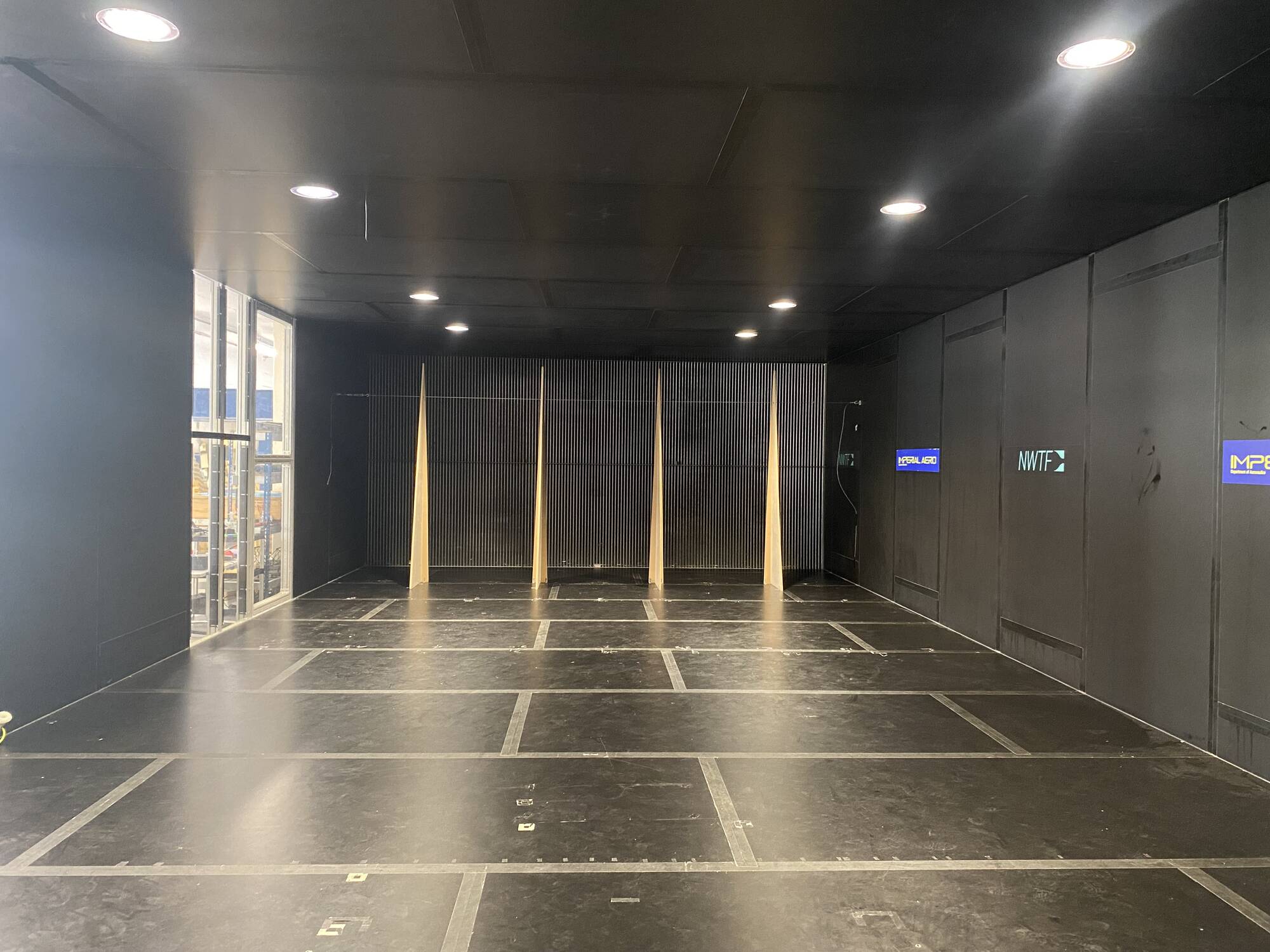}
    
        \caption{4 spires installed in the upper TS}
        % \label{fig:upperts_iconic_U}
    \end{minipage}\hfill
    \begin{minipage}{.47\linewidth}
        \centering
       
        \includegraphics[width=1.1\textwidth]{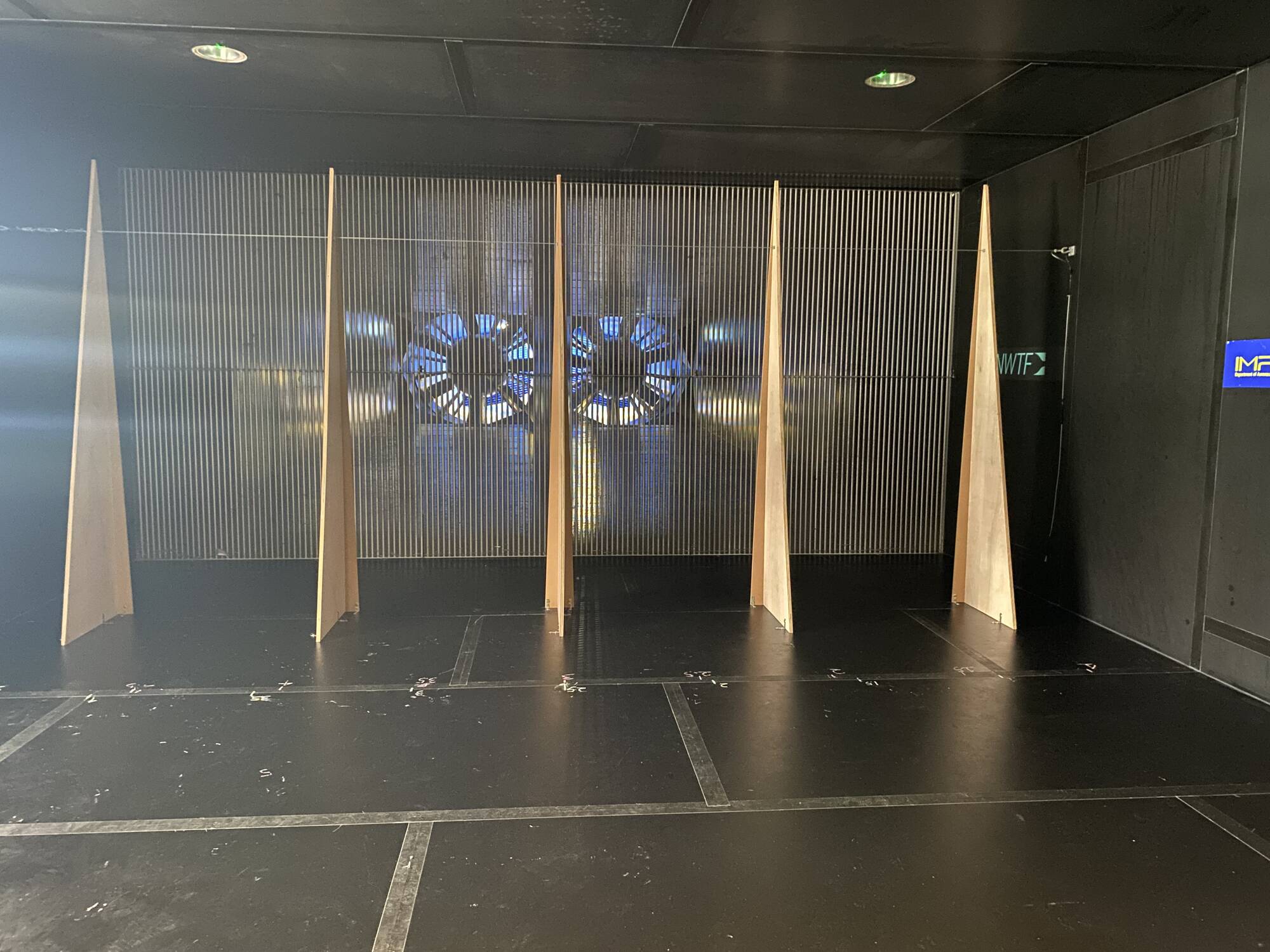}
    
        \caption{5 spires installed in the upper TS}
        % \label{fig:upperts_iconic_iU}
    \end{minipage}
\end{figure}

\subsubsection{LDA - Lower TS}
\label{ldascan}
\begin{figure}[H]
    \centering    
    \rotatebox{-90}{% Negative for clockwise rotation
        \includegraphics[width=0.5\textwidth]{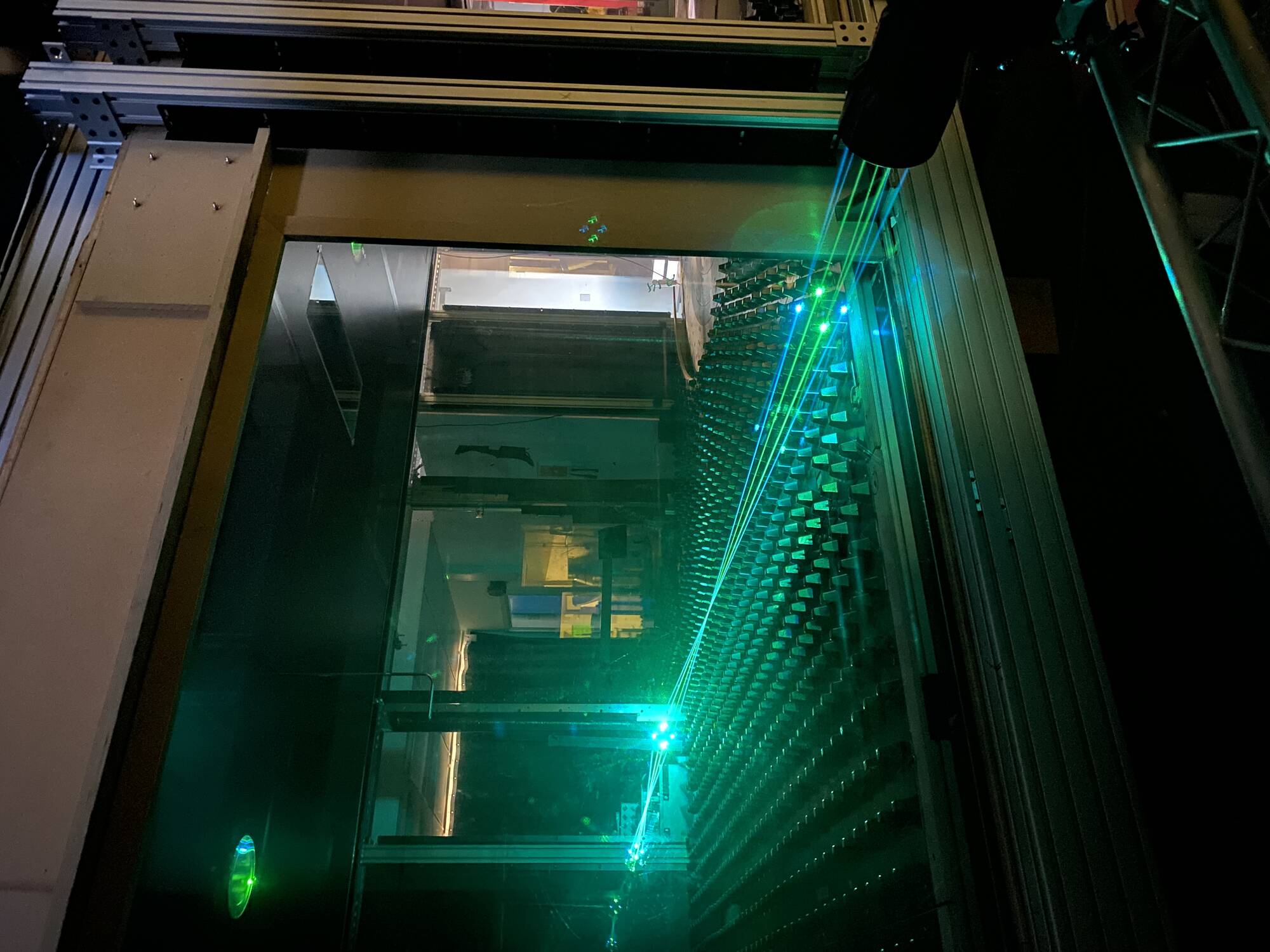}
    }
    \caption{LDA setup - lower TS}
    \label{fig:iU_bad}
\end{figure}

\subsubsection{LDA and the old traverse MHP - Lower TS}
\label{refoldtraverse}
\begin{figure}[H]
    \centering    \includegraphics[width=0.5\linewidth]{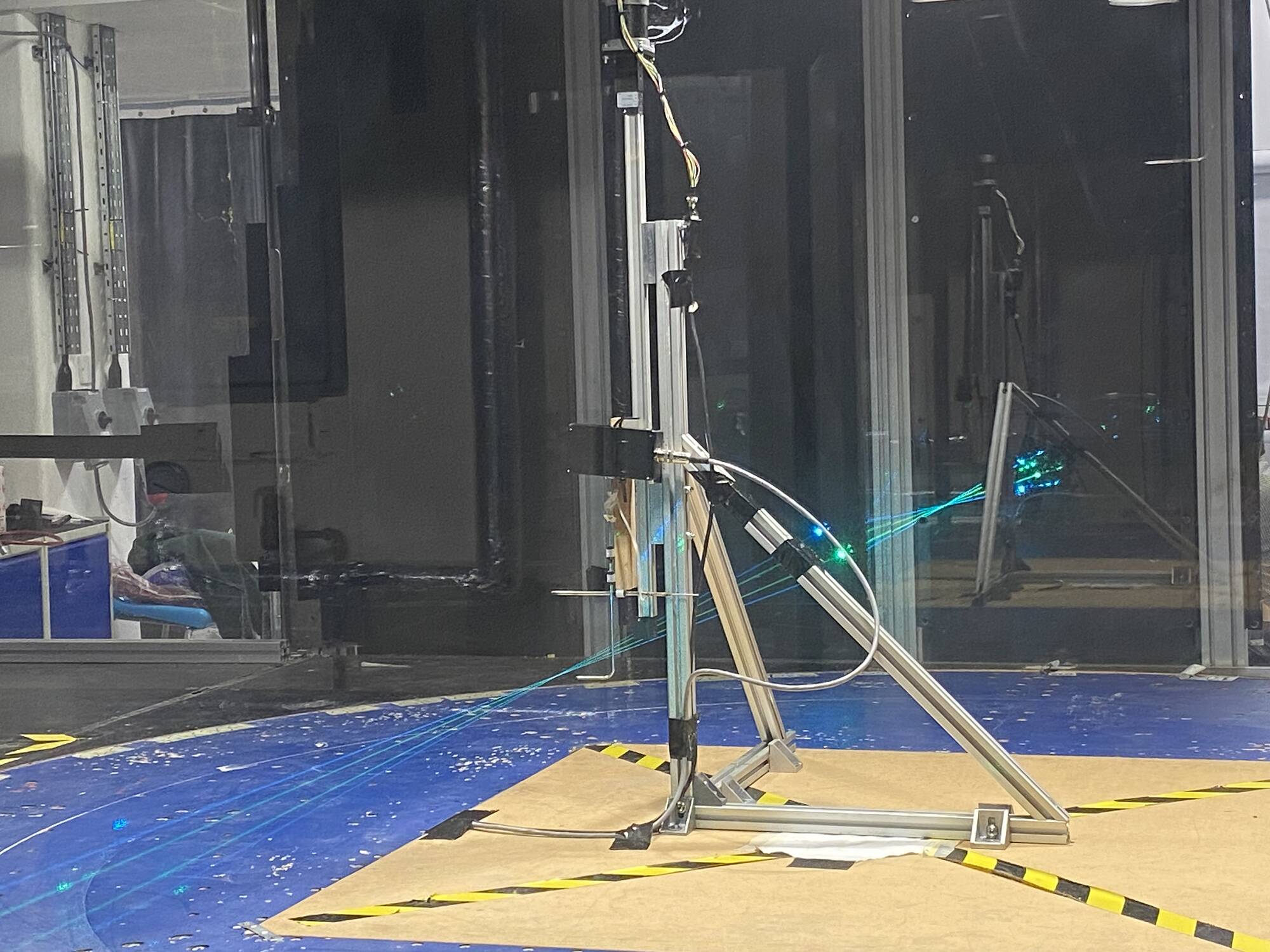}
    \caption{Old MHP traverse system, with LDA - lower TS}
    \label{fig:iU_bad}
\end{figure}

\subsubsection{Spire installation - Upper TS}
\label{upperts_cable}
\begin{figure}[H]
    \centering
    \begin{subfigure}[t]{0.32\textwidth}
        \centering
        \includegraphics[width=\linewidth]{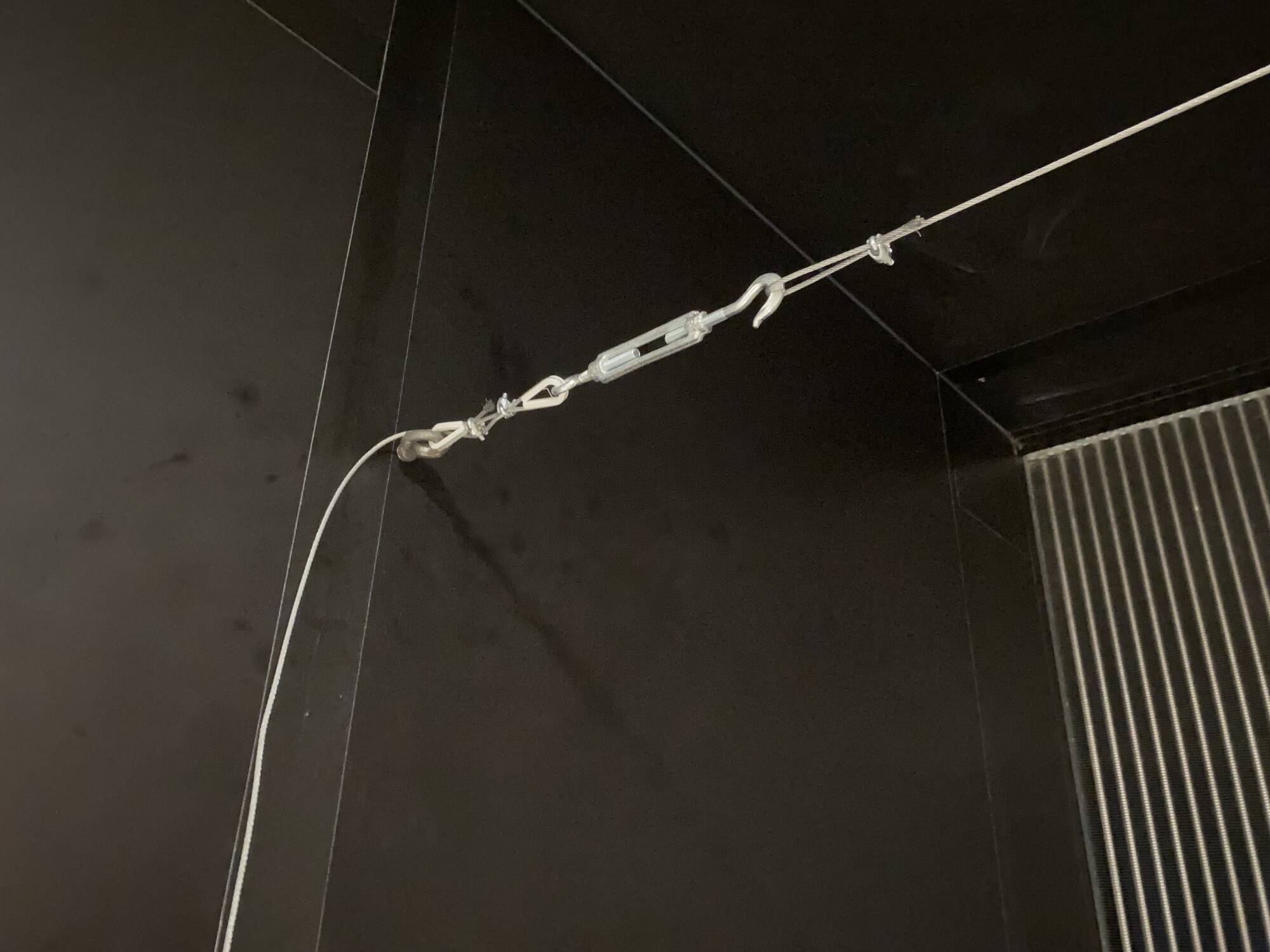}
        \caption{Turnbuckle to keep the cable under tension}
        \label{fig:sub1}
    \end{subfigure}
    \hfill
    \begin{subfigure}[t]{0.32\textwidth}
        \centering
            \rotatebox{-90}{% Negative for clockwise rotation
        \includegraphics[width=\textwidth]{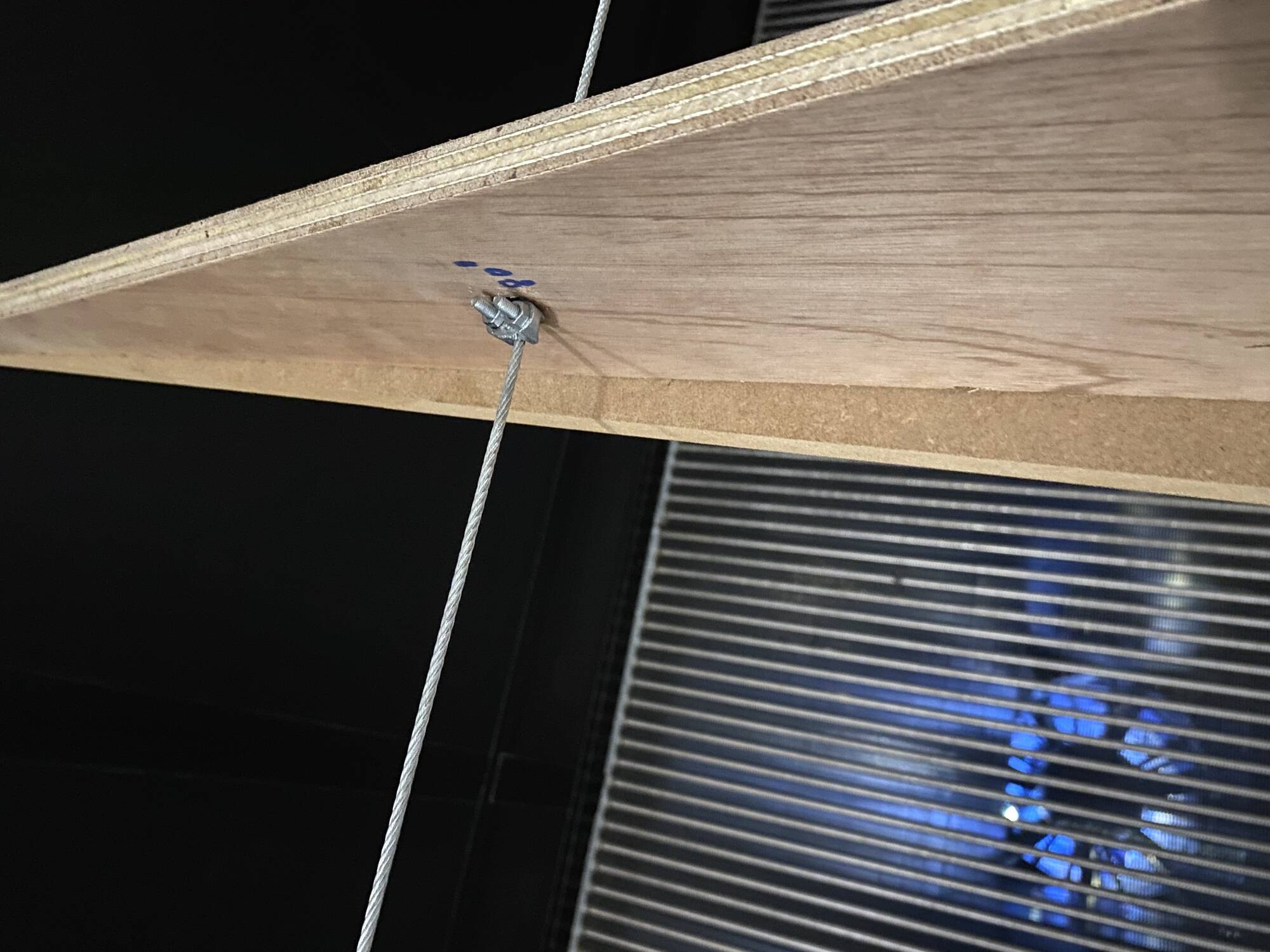}
    }
        \caption{Cable clamp onto the spire side}
        \label{fig:sub2}
    \end{subfigure}
    \hfill
    \begin{subfigure}[t]{0.32\textwidth}
        \centering
        \includegraphics[width=\linewidth]{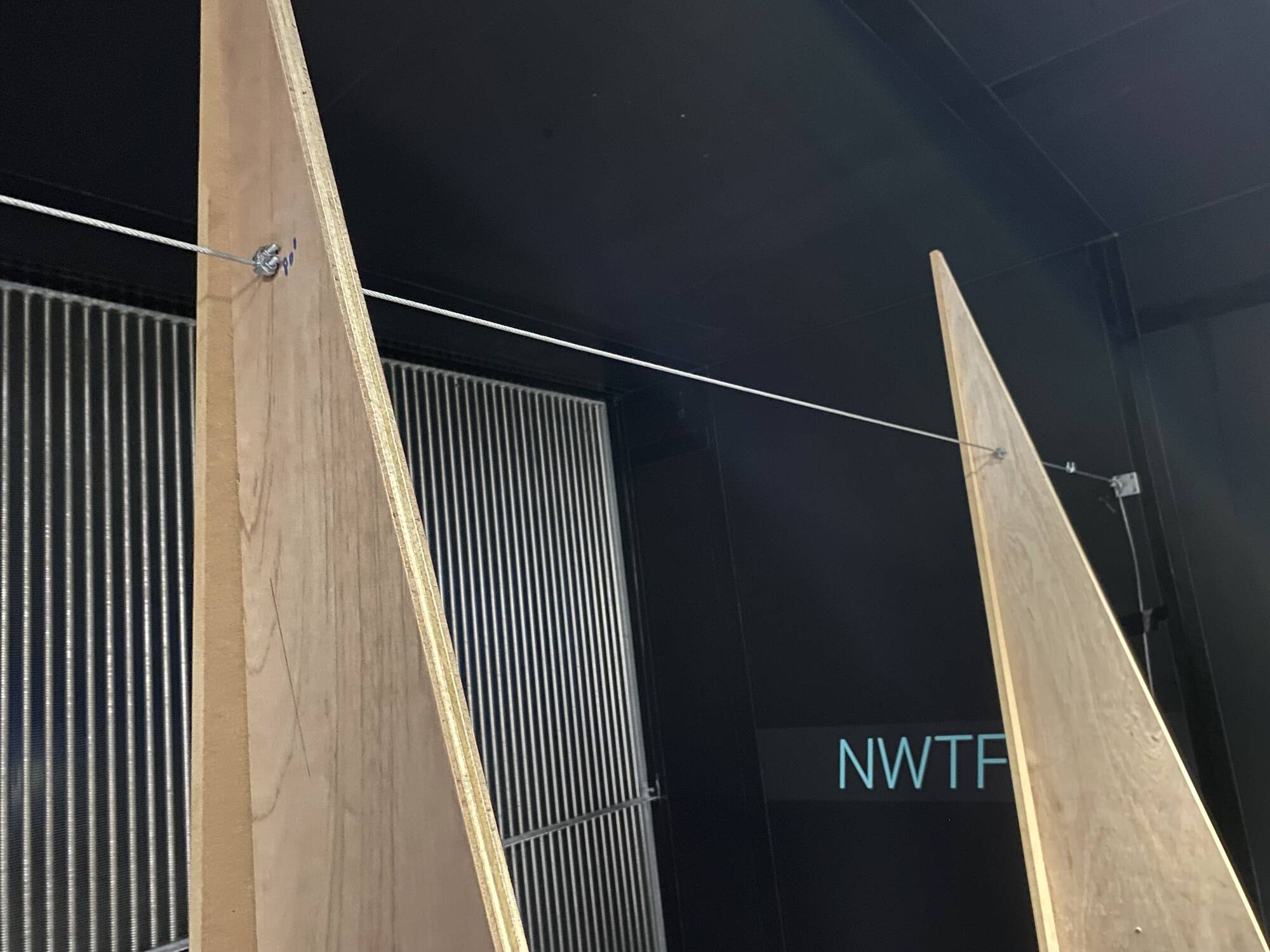}
        \caption{Cable connecting different spires}
        \label{fig:sub3}
    \end{subfigure}
    \caption{Upper TS - spire setup}
    \label{fig:three_subfigs}
\end{figure}
 % your appendix content

\end{document}